\documentclass[twocolumn,trackchanges]{aastex7}
\usepackage[T1]{fontenc}
\usepackage{booktabs}
\usepackage{amsmath}

\begin{document}

\title{A window for water-hydrogen demixing on warm metal-rich sub-Neptunes}

\correspondingauthor{Caroline Piaulet-Ghorayeb}
\author[0000-0002-2875-917X]{Caroline Piaulet-Ghorayeb}
\altaffiliation{E. Margaret Burbidge Postdoctoral Fellow}
\affiliation{Department of Astronomy \& Astrophysics, University of Chicago, 5640 South Ellis Avenue, Chicago, IL 60637, USA}
\email[show]{carolinepiaulet@uchicago.edu}  

\author[0000-0002-5113-8558]{Daniel P. Thorngren}
\affiliation{Department of Physics and Astronomy, Johns Hopkins University}
\email{dpthorngren@jhu.edu}  

\author[0000-0002-1337-9051]{Eliza M.-R.\ Kempton}
\affiliation{Department of Astronomy \& Astrophysics, University of Chicago, 5640 South Ellis Avenue, Chicago, IL 60637, USA}
\affiliation{Department of Astronomy, University of Maryland, College Park, MD 20742, USA}
\email{ekempton@uchicago.edu}  

\author{Justin Lipper}
\affiliation{Institut Trottier de Recherche sur les Exoplanètes, Département de Physique, Université de Montréal, 1375 Avenue Thérèse-Lavoie-Roux, Montreal, H2V 0B3, Canada
}
\email{justin.lipper@umontreal.ca}

\author[0000-0003-0638-3455]{Leslie Rogers}
\affiliation{Department of Astronomy \& Astrophysics, University of Chicago, 5640 South Ellis Avenue, Chicago, IL 60637, USA}
\email{larogers@uchicago.edu}

\author[0000-0002-4365-2258]{Fernanda Correa Horta}
\affiliation{Department of Astronomy \& Astrophysics, University of Chicago, 5640 South Ellis Avenue, Chicago, IL 60637, USA}
\email{fpazcorrea@uchicago.edu}

\author{Shi Lin Sun}
\affiliation{Institut Trottier de Recherche sur les Exoplanètes, Département de Physique, Université de Montréal, 1375 Avenue Thérèse-Lavoie-Roux, Montreal, H2V 0B3, Canada
}
\email{shi.lin.sun@umontreal.ca}





\begin{abstract}

Sub-Neptunes represent the largest exoplanet demographic, yet their bulk compositions remain poorly understood. Recent studies suggested that only very cold planets, such as Uranus and Neptune, could experience stratification of volatiles in their envelopes. Transiting warm sub-Neptunes, with $10^3$ to $10^4$ times more stellar irradiation, were therefore believed to have fully-miscible compositions. Here, we present ATHENAIA, an interior-atmosphere composition inference framework we leverage to assess the potential for water-hydrogen demixing on warm sub-Neptunes and for the 350\,K planet TOI-270\,d as a case study, using radiative-convective atmosphere models coupled to interior models. We find that the higher temperatures at which hydrogen and water demix in water-rich environments open a window for demixing on sub-Neptunes with bulk envelope metallicities of $\sim 150$ to $700\times$ solar, compatible with TOI-270 d. Demixing is easier to achieve on more massive and colder planets, but still broadly affects warm ($\simeq $330 to 450\,K) metal-rich sub-Neptunes. Therefore, combining atmosphere metallicities with models of fully-miscible envelopes may lead to underestimated bulk envelope metallicities and mass fractions. Further, we find that considering the increased greenhouse effect in metal-rich atmospheres in concert with the composition-dependent adiabatic gradient in the convective envelope increases the range of compositions under which molten mantle conditions should be expected on sub-Neptunes. This work encourages a reconsideration of the current paradigm for linking sub-Neptune atmospheres to their interiors and motivates evolutionary modeling describing the onset of metallicity gradients in sub-Neptune envelopes.
\end{abstract}



\section{Introduction} \label{sec:intro}

Linking planetary compositions to formation conditions and evolutionary scenarios is one of the main goals of exoplanetary science. This feat has proven particularly challenging for sub-Neptunes, despite their ubiquity as the most common type of exoplanet discovered by transit surveys \citep{howard_planet_2012,batalha_planetary_2013,fulton_california-kepler_2018}. Sub-Neptune sizes range from approximately 2 to 4 Earth radii, with densities too low for pure-rock compositions implying potentially thick gaseous envelopes. Yet, they are smaller than the solar system ice giants, limiting the prior assumptions that can be made when interpreting their bulk makeup including, at the most basic level, whether their envelopes are H/He-dominated, akin to gas giants, rich in ices deep below the observable atmosphere, similarly to Uranus and Neptune, or even metal-rich all the way to the upper atmosphere \citep{miller-ricci_atmospheric_2009,rogers_framework_2010}. 

An understanding of sub-Neptune interior compositions is critical to deciphering their origins at the population level, particularly when it comes to evaluating the relative importance of formation conditions \citep{kuchner_volatile-rich_2003,alibert_formation_2017,burn_radius_2024}, and evolution through chemical interactions between the gas envelope and the mantle \citep{kite_water_2021,schlichting_chemical_2022} or atmospheric mass-loss driven by stellar irradiation \citep{rogers_photoevaporation_2021}. Specifically, knowledge of sub-Neptunes' envelope mass fractions and compositions will inform their overall metal/ice content, and is critical to interpreting their size and density distribution \citep{luque_density_2022,neil_evaluating_2022,rogers_conclusive_2023} as well as modeling envelope mass-loss \citep{gupta_signatures_2020,owen_evaporation_2017}. Further, envelope extent and metal content both impact the planet's thermal structure, dictating the molten or solid state of the mantle which dictates the efficiency of dissolution/outgassing of volatile species \citep{lichtenberg_bifurcation_2021,werlen_atmospheric_2025}, and sets the planet's cooling rate over its thermal evolution.

Given the vast array of processes shaping sub-Neptune atmospheres and our little grasp on their relative importance from a theoretical standpoint, observational constraints on sub-Neptune envelope properties are crucial. A growing sample of sub-Neptunes now have both mass and radius measurements precise enough to infer envelope mass fractions via interior-atmosphere structure modeling, traditionally assuming either H/He-dominated envelopes, pure-water hydrospheres, or layered structures with water at high pressures underlying the gas-phase hydrogen \citep[e.g.][]{lopez_understanding_2014,zeng_growth_2019,turbet_revised_2020,aguichine_mass-radius_2021,madhusudhan_interior_2020}. The resulting envelope mass fraction estimates are however extremely degenerate with the envelope metallicity, with uncertainties of about an order of magnitude on the mass budget of the envelope between the two extremes of H/He-dominated solar-metallicity gas, and e.g. pure-water envelope scenarios. Atmosphere studies, which traditionally probe the upper layers of planetary envelopes in transmission at $\sim$ mbar pressures, have been proposed as a way to resolve this degeneracy. However, the traditional view of layered structures, with most volatiles confined to the high pressure region of the envelope and inaccessible to observations, limited the information content that transmission spectroscopy could provide as to the metal or water content of the entire envelope. 

Recent theoretical work has shown that what was traditionally thought of as the ``hydrogen-water boundary'' in sub-Neptune envelopes, i.e.\ the pressure and temperature at which the hydrogen-rich upper envelope transitions to a water-rich lower layer, occurs at conditions where water would be in the supercritical phase because of the warming greenhouse effect of the hydrogen-rich gas \citep{madhusudhan_interior_2020,piaulet_evidence_2023}. Only for the coldest planets can water condense out into clouds in the atmosphere, or in even more extreme cases where the hydrogen layer is extremely thin, form a liquid ocean (referred to as ``hycean'' worlds; \citealp{madhusudhan_habitability_2021,innes_runaway_2023}). Since experiments and density-functional-theory (DFT) simulations supported the miscibility of supercritical water with hydrogen \citep{seward_system_1981,bergermann_ab_2024,gupta_miscibility_2024}, a picture has emerged where all warm sub-Neptunes too hot for liquid ocean formation would have fully miscible envelopes with hydrogen and water well-mixed from the supercritical phase all the way to the gas probed at lower pressures in transmission spectroscopy \citep{nixon_new_2024,innes_runaway_2023,pierrehumbert_runaway_2022,benneke_jwst_2024}. This perspective opened up the exciting possibility that the atmospheric conditions measured on sub-Neptunes may reflect the bulk envelope composition, and further facilitate breaking the core mass/envelope mass degeneracy.

However, both experimental data and theoretical predictions relying on numerical calculations support the existence of a ``solvus'', also known as the ``critical curve'', in temperature-pressure space, beyond which water and hydrogen, even at high pressures, are no longer miscible and would form two distinct phases \citep{seward_system_1981,bergermann_ab_2024,gupta_miscibility_2024}. Since the stability of coexisting hydrogen and water also depends on the mixture composition, this can give rise to ``demixing'', a process whereby two fluids cannot thermodynamically mix into a homogeneous phase. Demixing is distinct from a phase transition (such as water condensation), or what is traditionally thought of as stratification via rainout (driven by gravitational separation). 

Demixing has already been shown to impact the interior structures and thermal evolution of Uranus and Neptune in the solar system, leading to envelope compositional gradients due to H/He and H/H$_2$O phase separation \citep{nettelmann_thermal_2011,amoros_h_2-h_2o_2024,arevalo_different_2025}. Demixing could similarly affect the atmospheres of sub-Neptunes and lead to compositional gradients that would break the link between bulk envelope and upper atmosphere metallicity, and impact both the susceptibility of sub-Neptune atmospheres to mass-loss  and their cooling rates.
Recent theoretical predictions extended the range of thermodynamical conditions over which this critical curve is constrained to those relevant for sub-Neptune interiors \citep{gupta_miscibility_2024}, opening the possibility to assess their propensity to demixing at the population level.

\begin{figure}
    \centering
    \includegraphics[width=0.45\textwidth]{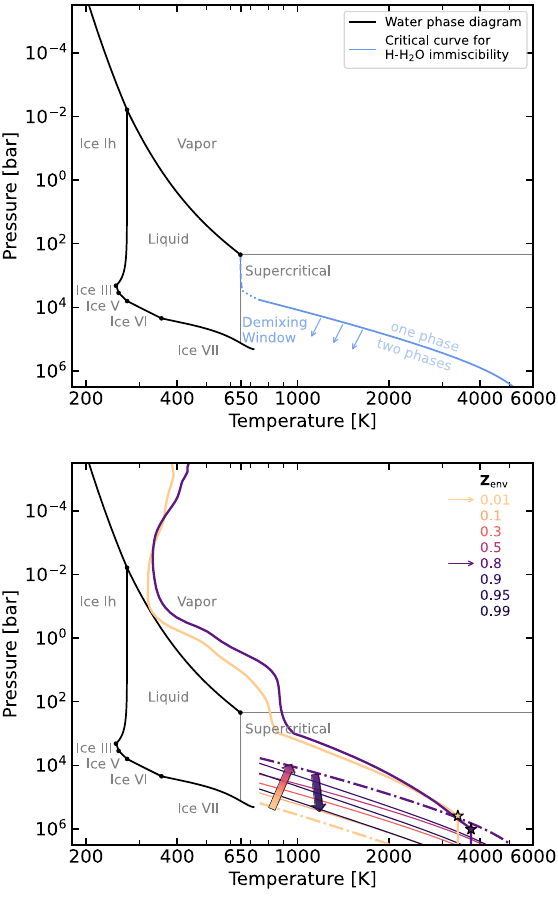}
    \caption{Illustration of the demixing window concept. \textit{Top panel:} Water phase diagram (black lines) with labeled regions; the critical curve (blue) indicates the highest temperature for single-phase H-H$_2$O mixing, compiled from low-pressure data (dashed, \citealt{seward_system_1981}) and high-pressure data (solid, \citealt{gupta_miscibility_2024}), with interpolation (dotted). Envelope T-P profiles below the blue line can experience demixing for some compositions. \textit{Bottom panel:} Demixing window at different metallicities. High-pressure H-H$_2$O coexistence curves (colored), for various envelope metallicities $Z_\mathrm{env}$, show that the  window for immiscible conditions is maximal for metallicities near $Z\sim$0.8, but recedes for lower and higher metallicities (illustrated by colored arrows). Example T-P profiles for a 5.1 $M_\oplus$ planet at T$_\mathrm{eq}=360$\,K, with an envelope mass fraction of 30\% and a metallicity of either 0.01 (1$\times$ solar; orange) or 0.8 ($\sim$300$\times$ solar; purple), are shown, with dash-dotted linestyles for the corresponding coexistence curves. At low metallicity, the envelope profile (down to the mantle-envelope boundary, star marker) does not intersect the coexistence curve, but for $Z_\mathrm{env}=0.8$, it does, implying that under stable conditions, the upper atmosphere metallicity would be lower than that of the bulk envelope. Note that due to the low water abundance (VMR$\sim 10^{-3}$) in the $Z=0.01$ model, it would not experience water condensation despite its proximity to the \textit{pure-water} condensation curve.}

    \label{fig:phase_space_motivation}
\end{figure}

In particular, the small (2.216 $R_\oplus$, \citealp{eylen_masses_2021}) sub-Neptune TOI-270 d, with its low insolation resulting in a zero-albedo equilibrium temperature of about 380\,K, has received considerable scrutiny since the detection of volatiles such as methane, carbon dioxide, as well as potentially large amounts of water in its hydrogen-dominated, but metal-enriched atmosphere \citep{benneke_jwst_2024,holmberg_possible_2024,felix_evidence_2025,nixon_magma_2025,constantinou_atmospheric_2025}. In terms of its interior structure, the scenarios proposed for TOI-270 d range from Hycean conditions \citep{holmberg_possible_2024}, to a fully-miscible envelope \citep{benneke_jwst_2024}. The photospheric temperature measured on TOI-270 d now disfavors the formation of a liquid water ocean except perhaps on the nightside (``dark Hycean world'' scenario, \citealp{madhusudhan_habitability_2021,constantinou_atmospheric_2025,rigby_surface_2025}). 

Although initial estimates claimed that demixing should not occur even on the colder $\sim$255\,K sub-Neptune K2-18 b \citep{gupta_miscibility_2024}, independent modeling exploration suggested that both K2-18 b and TOI-270 d could experience demixing \citep{howard_possibility_2025}. 
However, both were impacted by critical parameter space or model setup limitations. The first exploration of K2-18 b \citep{gupta_miscibility_2024} only considered low-metallicity, H/He-dominated conditions -- neglecting the compositional dependence of planetary thermal structures. The latter work \citep{howard_possibility_2025} described the behavior of water in the interior of the planet using an equation of state \citep{more_new_1988} that has been superseded by more recent prescriptions \citep{mazevet_ab_2019}. The differences in the predicted temperatures in sub-Neptune interiors, using the updated adiabatic gradients, range from hundreds to more than 1000K \citep{aguichine_mass-radius_2021,howard_possibility_2025}. Further, using self-consistent radiative-convective atmosphere models, compared to predictions from profiles combining adiabats with parameterized profiles for the upper radiative regions (as in \citealp{howard_possibility_2025}), was shown to have important impacts on predicting sub-Neptune radii and their thermal evolution, especially for metal-rich compositions \citep{aguichine_evolution_2024}. Yet, no modeling effort has yet used such radiative-convective models for the atmosphere thermal structure coupled with interior models to assess the susceptibility of sub-Neptune envelopes to demixing across the mass-radius diagram. 

\begin{figure*}
    \centering
    \includegraphics[width=0.85\textwidth]{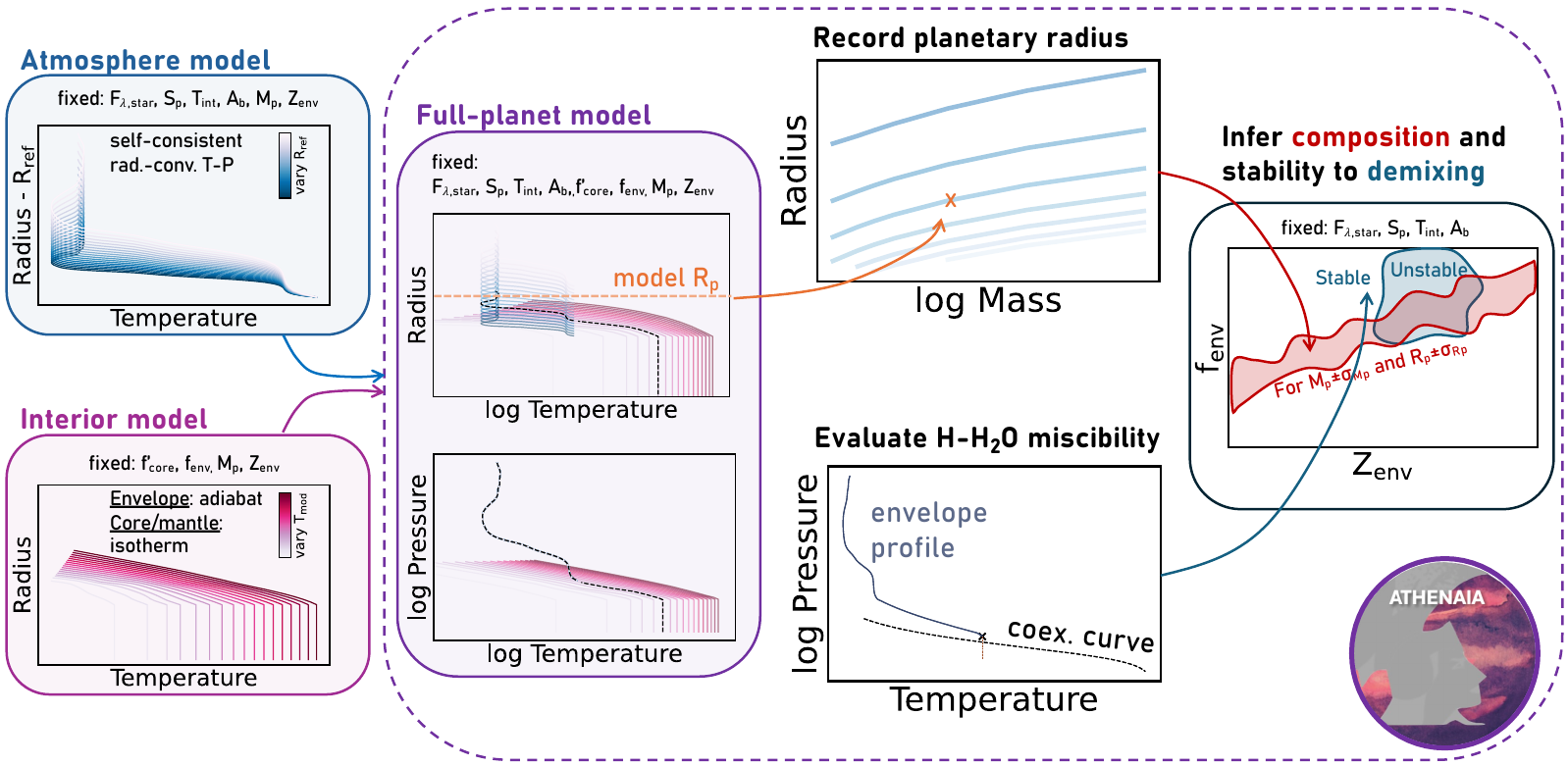}

    \caption{Illustration of the workflow for the construction of coupled interior-atmosphere models with ATHENAIA. For each composition, atmosphere models are calculated with SCARLET (top left) and interior models following \citet{thorngren_connecting_2019}. We couple them with ATHENAIA by finding the $R_\mathrm{ref}$ (atmosphere model) and $T_\mathrm{mod}$ (interior model) that minimize $\delta_\mathrm{TPR}$ (see Eq. \ref{eq:deltaTPR}). The radius at 20 mbar is extracted to construct constant-composition mass-radius curves, while the pressure-temperature profile of the envelope is used to evaluate the stability to demixing by comparing to the corresponding H-H$_2$O coexistence curve (dashed). Finally (right panel), given planetary mass and radius, the range of potential compositions is mapped to $f_\mathrm{env}-Z_\mathrm{env}$ space (red) and compared to the map of conditions where envelopes are (un)stable against demixing (blue).}
    \label{fig:link_int_atm_flowchart}
\end{figure*}

The contrasting predictions ranging from fully-miscible envelopes for warm sub-Neptunes, to demixing-driven compositional gradients for the more massive and much less irradiated (less than 0.3\% of Earth's irradiation) solar system ice giants, suggests the existence of a ``demixing window'' (Fig. \ref{fig:phase_space_motivation}): planetary conditions where envelope metallicity gradients could exist. The evaluation of this compositional transition is essential to assess the extent to which, in particular, water abundances and metallicity estimates from transmission spectroscopy can be translated to bulk envelope water contents. In this study, we apply full-planet models combining interior models with state-of-the-art water EOS, and self-consistent atmosphere models in radiative-convective equilibrium to evaluate the temperature/pressure conditions in sub-Neptune atmospheres and their composition- and irradiation-dependent susceptibility to demixing. The aims of this study are two-fold: (1) provide a framework for the systematic evaluation of whether demixing occurs on a given sub-Neptune planet provided mass, radius, energy budget, and potential atmospheric constraints, and (2) delineate the demixing conditions in mass, radius, and composition space in order to support population-level interpretation of sub-Neptune compositions and evolutionary models.

This paper is organized as follows. In Section \ref{sec:model}, we describe our methods for the atmosphere and interior modeling and the evaluation of hydrogen-water miscibility. In Section \ref{sec:compositional_inference}, we describe our grid of full-planet models and the composition inference algorithm. We discuss our results in terms of the envelope conditions on TOI-270 d in Section \ref{sec:toi_270_d_res_disc}, and implications for the sub-Neptune population in Section \ref{sec:implications}. We compare our results to previous work in Section \ref{sec:comparison_prev}, discuss potential caveats in Section \ref{sec:caveats} and  summarize our findings in Section \ref{sec:conclusion}.

\section{Full-planet modeling}\label{sec:model}

This study aims to determine whether well-mixed envelope compositions are stable on planets with varying bulk properties, compositions, and energy budgets, or whether compositional gradients should be expected throughout the envelope. We construct full-planet models with envelopes that are well-mixed from the mantle-envelope (MEB) boundary up to the top of the atmosphere, before evaluating whether the conditions for H-H$_2$O coexistence in a single phase are met. While previous work assumed either stratified water- and H$_2$-dominated layers (e.g. \citealp{madhusudhan_interior_2020}), or well-mixed envelopes and atmospheres with uniform compositions throughout (e.g. \citealp{nixon_new_2024}), we perform a systematic evaluation of the phase stability of the mix to evaluate the validity of either assumption across expected sub-Neptune conditions. 

We present ATHENAIA, the ATmospHEre-iNterior bAyesian Inference frAmework (Fig. \ref{fig:link_int_atm_flowchart}). The ATHENAIA framework builds up on an earlier version of the code that was tailored to planets with a layered structure where hydrogen and water are stratified \citep{piaulet_evidence_2023,roy_is_2022, benneke_jwst_2024}. ATHENAIA performs three main tasks: (1) self-consistent modeling of full-planet profiles, given atmosphere and interior modeling frameworks and input parameters for the bulk planetary composition, properties, and energy budget; (2) Bayesian inference of bulk planetary compositions, given known stellar and planetary properties; and (3) evaluation of water-hydrogen miscibility throughout the planetary envelope. Using ATHENAIA, we introduce the concept of a ``demixing window'', the region of parameter space where planetary conditions are no longer consistent with fully miscible envelopes. Each of these components are presented in the following sub-sections.

\begin{table}[h]
    \centering
    \renewcommand{\arraystretch}{1.2}
    \begin{tabular}{ll}
        \toprule
        \textbf{Parameter} & \textbf{Value} \\
        \midrule
        \multicolumn{2}{c}{\textbf{Star: TOI-270}} \\
        \midrule
        Spectral Type & M3.0V \\
        Effective temperature $T_{\mathrm{eff}}$ [K] & 3506 $\pm$ 70 \\
        Metallicity [Fe/H] & -0.2 $\pm$ 0.12 \\
        $\log g$ [cm/s$^2$] & 4.872$\pm$0.026 \\
        Radius $R_\star$ [$R_\odot$]& 0.378$\pm$0.011 \\
        \midrule
        \multicolumn{2}{c}{\textbf{Planet: TOI-270 d}} \\
        \midrule
        Radius $R_p$ [$R_\oplus$] & 2.216$^{+0.065}_{-0.064}$ \\
        Mass $M_p$ [$M_\oplus$] & 4.78 $\pm$ 0.43 \\
        Semi-major axis $a_p$ [AU] & 0.0742 $\pm$ 0.0014 \\
        Insolation $S_p$ [$S_\oplus$] & 3.5$\pm$0.4 \\
        Eq. Temperature $T_{\mathrm{eq}}$ ($A_\mathrm{B}$=0.0) [K] & 381 $\pm$ 10 \\
        $T_{\mathrm{eq}}$ ($A_\mathrm{B}$=0.2) [K] & 361 $\pm$ 10 \\
        $T_{\mathrm{eq}}$ ($A_\mathrm{B}$=0.4) [K] & 336 $\pm$ 9 \\
        \bottomrule
    \end{tabular}
    \caption{Stellar and planetary parameters for the TOI-270 system used in this work. The stellar parameters and planet mass are from \citet{eylen_masses_2021}, while the planetary radius and orbital parameters are from the fit to the JWST/NIRISS SOSS data \citep{benneke_jwst_2024}. We report the derived zero-Bond albedo equilibrium temperature and representative equilibrium temperatures for two other albedo values represented in our model grid. We use the median parameters to describe the stellar irradiation, but marginalize over the distribution of possible masses and radii accounting for their uncertainties to infer the range of compositions compatible with TOI-270\,d's parameters.}
    \label{tab:star_planet_params}
\end{table}

\subsection{Interior modeling}

For the interior portion of our full-planet models, we use one-dimensional interior structure models \citep{thorngren_mass-metallicity_2016,thorngren_connecting_2019} in order to derive profiles of temperature, pressure, radius, and density given bulk planetary properties. These models consist of an inert rock/iron interior with an Earth-like fraction $f'_\mathrm{core}=0.325$ of the mass contained in an iron core and the other 2/3 in a silicate mantle, overlaid with a homogeneous convective envelope that has a metallicity $Z_\mathrm{env}$ regulated by the amount of H$_2$O mixed with the H, He, and makes up a fraction $f_\mathrm{env}$ of the total planet mass. The models solve the equations of hydrostatic equilibrium, mass conservation, and the equation of state (EOS). For the latter we use \citet{chabrier_new_2019,chabrier_new_2021} for the H/He mix, \citet{mazevet_ab_2019} for H$_2$O, and \citet{Thompson90} for rock and iron.

\begin{figure}
    \centering
    \includegraphics[width=\linewidth]{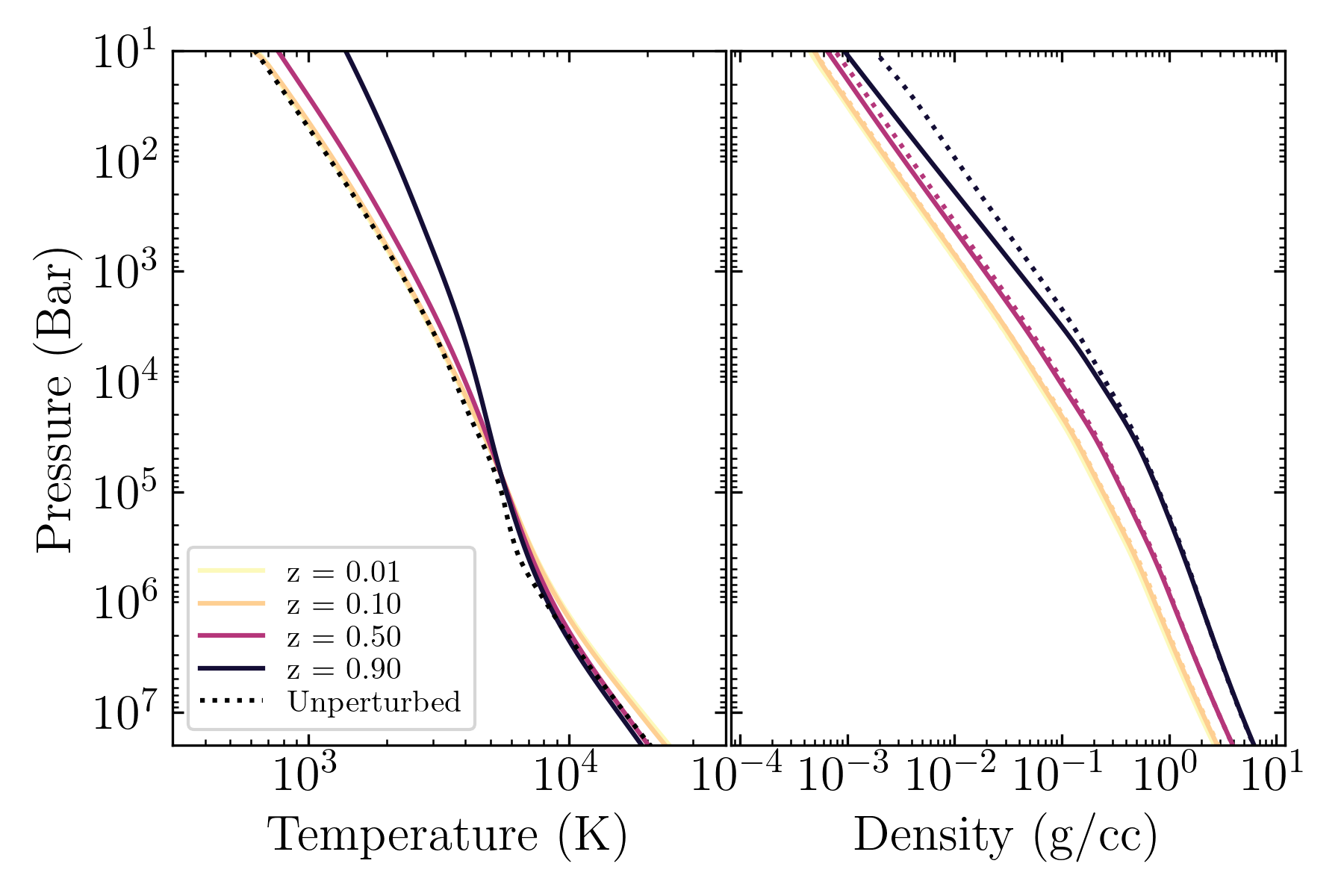}
    \caption{Adiabatic pressure-temperature profiles (left) and corresponding pressure-density profiles (right) used for the water/H/He mixtures in our interior models at various compositions. We compare this approach against the approximation that the P-T profile is the same as for $Z=0$ (pure H/He, dotted lines). The densities only differ significantly at low pressures, but the temperature gradient is significantly different by Z=0.5, which affects the thermal evolution.}
    \label{fig:eos_adiabats}
\end{figure}

The interior temperature-pressure profile is calculated assuming convection throughout the deep H/He/H$_2$O envelope, and is isothermal below the MEB. The temperature-dependence of iron and silicate EOSs, as well as the potential phase changes that may occur at high pressures, were shown to have a sub-percent impact on the radii of rocky planets \citep{valencia_internal_2006,seager_mass-radius_2007,fortney_planetary_2007}. This factors at an even more minor scale into the radii of the gas-enveloped sub-Neptunes modeled in this work with extended radiative atmospheres (see also \citealp{tang_reassessing_2024}), and such effects lie well within current radius measurement uncertainties for sub-Neptunes. 
The water and H/He equations of state were mixed using the additive-volumes rule, with the adiabatic profile obtained from numerical integration \citep[analogous to][]{nettelmann_ab_2008} of the adiabatic gradient, which was in turn computed from the density and internal energy values provided by the original EOS \citep[see][Sec. 1.4.1 for more details]{thorngren_bayesian_2019} as a function of pressure and temperature. This approach is an upgrade from e.g. \citet{thorngren_connecting_2019}, which used H/He adiabats without perturbation by the mixed metals; a comparison is shown in Fig. \ref{fig:eos_adiabats}. 

These standalone interior models do not include atmosphere models to serve as an upper boundary condition in temperature-pressure space, so they are calculated for various reference temperatures $T_\mathrm{mod}$, here taken at 200 bar, which set the conditions for the adiabatic profile in the envelope. This means that the thermal evolution of the adiabat is indexed by the 200-bar potential temperature, rather than the specific entropy $s$ as in e.g. \citet{thorngren_mass-metallicity_2016}, but these approaches are equivalent. The interior models are calculated over a grid of $Z_\mathrm{env}$, planet masses $M_p$, envelope mass fractions $f_\mathrm{env}$, and $T_\mathrm{mod}$. Only the subset of $T_\mathrm{mod}$ values where the temperature at the link point between the atmosphere model and deep envelope (interior model) is compatible with predictions given the planet's irradiation level and internal temperature $T_\mathrm{int}$ (see next subsection) are actually represented in the final models we produce for this work. We use the interior profiles to describe the temperature and density at pressures deeper than those probed by the atmosphere model, or after the onset of convection (see next subsection).

\subsection{Atmosphere modeling}

In parallel with the grid of interior models, we calculate a grid of atmosphere models that describe the behavior of the temperature/pressure/radius profiles at lower pressures ($\lesssim 1$ kbar), accounting for both convective and radiative energy transport. We use a version of the SCARLET atmosphere forward modeling and retrieval framework \citep{benneke_atmospheric_2012,benneke_characterizing_2013,benneke_sub-neptune_2019} adapted for the computation of self-consistent models up to the high metallicities required to describe sub-Neptunes, and amenable for coupling with interior models \citep{piaulet_evidence_2023,piaulet-ghorayeb_jwstniriss_2024}.

The calculation of a self-consistent radiative-convective atmosphere model takes as inputs a stellar spectrum, atmospheric composition, and planetary properties such as mass $M_p$, radius $R_\mathrm{ref}$ at a set reference pressure $p_\mathrm{ref}$, bolometric insolation $S_{p}$, internal temperature $T_\mathrm{int}$ describing the energy deposited in the planet's atmosphere at the lower boundary, i.e. the highest pressure layer, and Bond albedo $A_\mathrm{B}$. 

SCARLET solves iteratively the hydrostatic equilibrium and radiative transfer equations until the energy output at the top of the atmosphere, and the combined energy input from the interior (regulated by  $T_\mathrm{int}$) and incident stellar irradiation (dependent on the stellar spectrum, atmospheric composition, $S_{p}$, and $A_\mathrm{B}$) balance in each layer. We use a version of SCARLET that lifts the assumption of constant gravity throughout the atmosphere \citep{piaulet_evidence_2023} and rather accounts for the mass contained in each atmosphere layer when solving hydrostatic equilibrium. We also implement new methods for the convergence of the radiative-convective profiles and adiabatic gradient prescriptions, appropriate for the high levels of metal enrichment covered in our modeling exploration (see Section \ref{sec:TP_convergence}). In order to couple these atmosphere models with the appropriate interior models given any $f'_\mathrm{core}$ and $f_\mathrm{env}$, we calculate atmosphere models for each $Z_\mathrm{atm}=Z_\mathrm{env}$, $M_p$, $T_\mathrm{int}$, and $A_\mathrm{B}$, over a fine grid of $R_\mathrm{ref}$ (set at $p_\mathrm{ref}=1$ kbar) covering the equivalent 1-kbar radii from the interior model grid. The hydrostatic equilibrium and radiative transfer equations are solved independently for each $R_\mathrm{ref}$.

\subsubsection{Atmospheric composition}

The SCARLET models assume well-mixed compositions, with the volume mixing ratios (VMRs) of the three component gases H$_2$, He, and H$_2$O kept constant throughout the atmosphere. H$_2$O is used as the metallicity proxy, and we translate a chosen $Z_\mathrm{atm}=Z_\mathrm{env}$ to VMRs assuming the protosolar Y/X ratio of $r=Y/X=0.275/0.725$ is maintained \citep{lodders_solar_2010}. Letting
\begin{equation}
    X = \frac{1 - Z_{\mathrm{env}}}{1 + r}
\end{equation}
\noindent and
\begin{equation}
    Y = rX = \frac{r(1 - Z_{\mathrm{env}})}{1 + r},
\end{equation}\label{eq:Y}
\noindent we obtain:
\begin{equation}
    \mathrm{VMR}_{S} = \frac{X_S / m_{\mathrm{S}}}{D}
\end{equation}
\noindent where $S$ is either H$_2$, He, or H$_2$O, substituting for $X_S$: $X$, $Y$, or $Z_\mathrm{env}$, respectively, and
\begin{equation}
    D = X/m_\mathrm{H_2} + Y/m_\mathrm{He} + Z_{\mathrm{env}}/m_\mathrm{H_2O}.
\end{equation}

\subsection{Link between atmosphere and interior models}

In order to obtain full-planet models for a given planetary energy budget, mass, envelope composition $Z_\mathrm{env}$, and relative mass budgets of iron, silicates, and H/He/H$_2$O envelope (set by $f'_\mathrm{core}$ and $f_\mathrm{env}$), we employ an algorithm to optimize the choice of interior and atmosphere models among the set calculated for those parameters. This methods consists of finding $T_\mathrm{mod}^*$ and $R_\mathrm{ref}^*$ that minimize the $\delta_\mathrm{TPR}$ distance, defined as:
\begin{equation}\label{eq:deltaTPR}
    \delta_\mathrm{TPR}=\sqrt{\delta T\big|_{P_\mathrm{link}}^2 + \delta T\big|_{R_\mathrm{link}}^2}
\end{equation}
\noindent with $P_\mathrm{link}$, $R_\mathrm{link}$ the atmosphere model's pressure in bars and radius in Earth radii at the pressure level where the link with the interior model happens. This pressure is either that of the radiative-convective boundary, $P_\mathrm{rcb}$ (if $P_\mathrm{rcb}<1$\,kbar), or 1 kbar otherwise. 
$\delta T\big|_{P_\mathrm{link}}$ is the temperature difference between each of the atmosphere models (for each $R_\mathrm{ref}$), and each of the interior models (for each $T_\mathrm{mod}$). $\delta T\big|_{R_\mathrm{link}}$ is defined similarly by comparing the temperatures of the atmosphere and interior models at the radius where $P=P_\mathrm{link}$ in the atmosphere model. The full planetary profiles for temperature/pressure/radius/density are then set to be the atmosphere profiles of the best-match atmosphere model $R_\mathrm{ref}^*$ at pressures lower than $P_\mathrm{link}$, and the interior profiles of the best interior model $T_\mathrm{mod}^*$ at deeper pressures. The planetary radii at 1, 10, and 20 mbar are recorded for each full-planet composition scenario. When displaying the resulting constant-composition curves in mass-radius space, or comparing to TOI-270 d's radius, we use the 20 mbar radii for easy comparison with previous works \citep{lopez_understanding_2014,benneke_jwst_2024}.

In the rare cases (very low $f_\mathrm{env}$) where the MEB happens at pressures that are shallower than the RCB, or shallower than 1 kbar if the RCB is not reached at 1 kbar, we instead link the interior and atmosphere models that minimize $\delta R_\mathrm{MEB}$, the difference in radius at the MEB pressure (identified in the interior model). In these cases, the high pressures over which the composition-dependent demixing conditions of hydrogen and water referenced in this work are applicable (see Fig. \ref{fig:phase_space_motivation}) are not reached within the envelope, and miscibility cannot be evaluated from the planet model. This does not impact our results, since demixing preferentially occurs at larger $f_\mathrm{env}$ (Fig. \ref{fig:demixing_window_fwd}).

\begin{figure}
    \centering
    \includegraphics[width=0.48\textwidth]{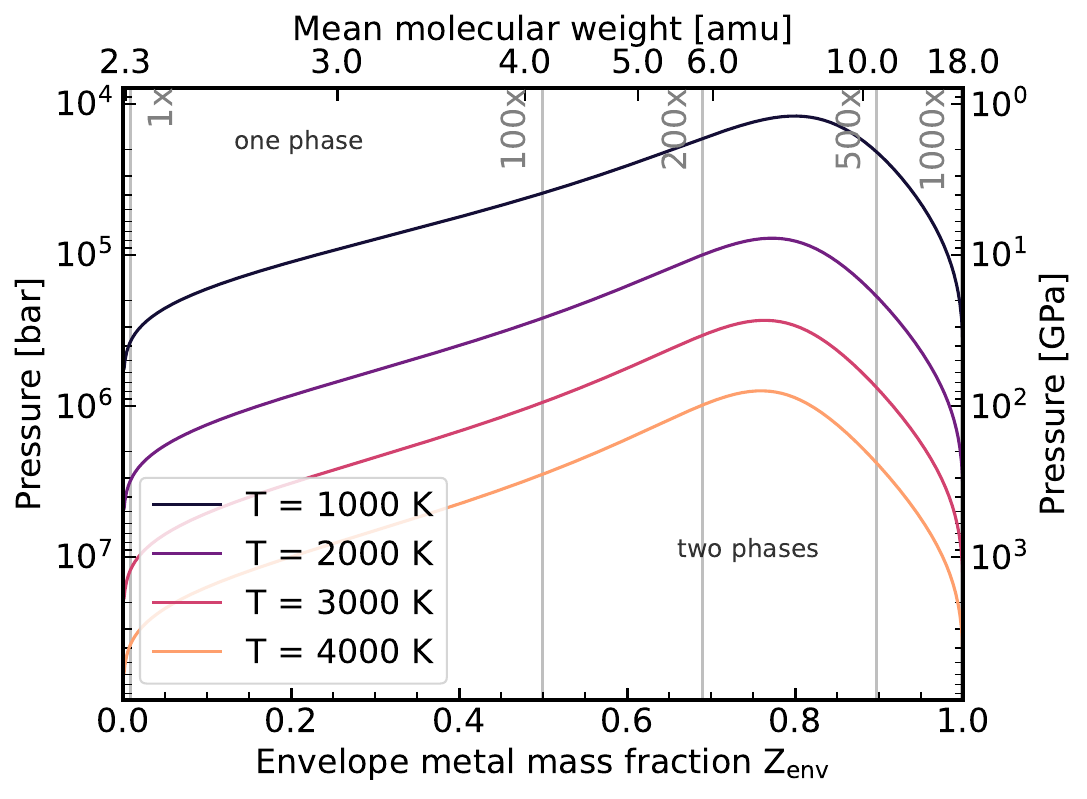}

    \caption{Coexistence curves for hydrogen and water in pressure-metallicity space, for 4 different temperatures. Vertical lines indicate 1 to 1000$\times$ solar metallicity. The corresponding mean molecular weight values are indicated at the top. Demixing is easier to achieve in colder envelopes with moderate metal enrichments.}
    \label{fig:coexistence_vs_composition}
\end{figure}

\begin{figure*}
    \centering
    \includegraphics[width=0.98\textwidth]{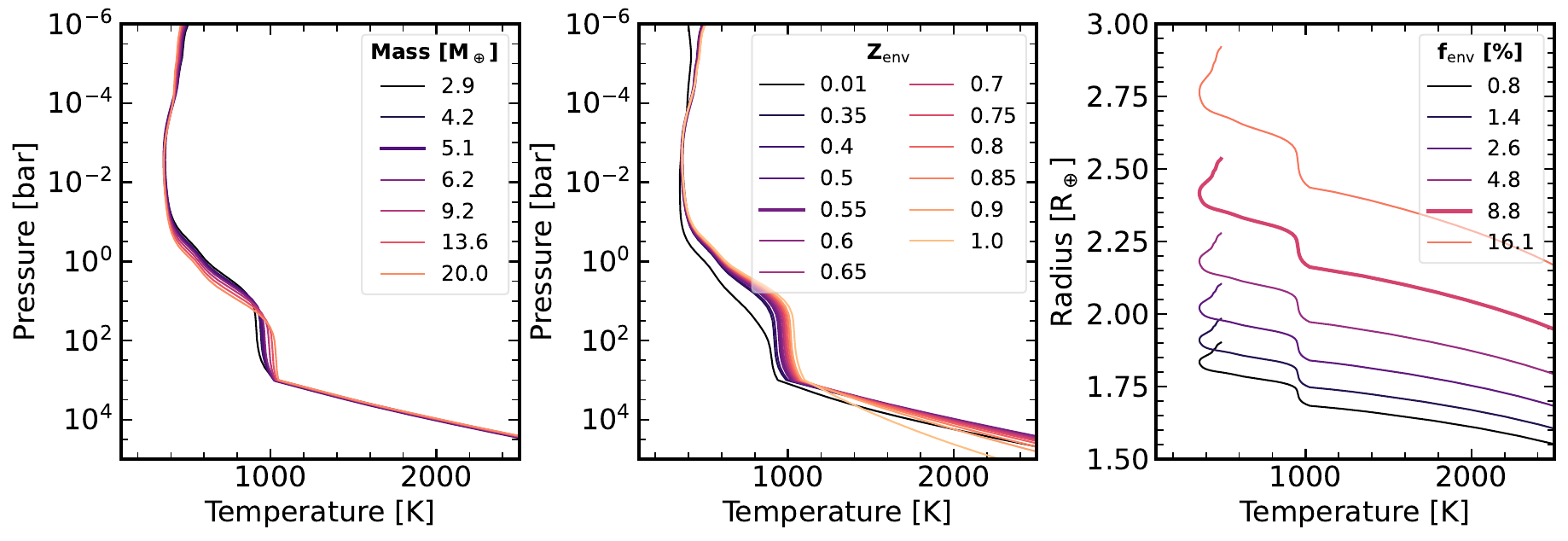}

    \caption{Impact of physical parameters varied across the grid on the pressure-temperature and pressure-radius profiles. The left, middle, and right panel illustrate the impact of varying the planet mass, envelope metallicity, and envelope mass fraction respectively. Aside from the parameter varied in each panel, all other parameters are kept fixed to $T_\mathrm{eq}=381$\,K, $M_p=5.1 M_\oplus$, and $f_\mathrm{env}=8.79$\% (bolded profiles). Decreasing the equilibrium temperature shifts the envelope profiles, as illustrated in Fig. \ref{fig:tp_zenv}. }
    \label{fig:master_grid_varyparam}
\end{figure*}
\subsection{Evaluation of hydrogen-water miscibility}

Water-hydrogen demixing in a planetary envelope occurs when local thermodynamic conditions do not allow for the coexistence of hydrogen and water in a single miscible phase. This typically leads to metallicity gradients, with more metal-rich material in the deeper atmosphere, such that a steady state is reached where the atmosphere composition at each pressure is stable against further demixing. The stability of planetary envelopes to demixing results from a complex interplay between energy budget and composition. Generally, hotter planets will be stable against demixing over vast regions of the parameter space, since the instability occurs when the temperature at a given pressure is \textit{lower} than the water-hydrogen coexistence curve. The coexistence curves themselves are highly compositionally-dependent, with variations in the demixing conditions spanning thousands of K in the deep planetary interior, depending on the exact metallicity value (Figs. \ref{fig:phase_space_motivation} and \ref{fig:master_grid_varyparam}). Meanwhile, although the upper atmospheres of more metal-rich planets will generally be warmer, the adiabatic temperature gradient governing the temperature as a function of depth in the convective envelope becomes \textit{shallower} at higher metallicities (Fig. \ref{fig:eos_adiabats}). Combined with the coexistence curves being pushed to much lower pressures around $Z_\mathrm{env} \in \left[0.6,0.8\right]$ (approximately 140 to 300$\times$ solar metallicity; Fig. \ref{fig:coexistence_vs_composition})\footnote{Throughout this study, metallicities relative to solar are computed from the envelope O/H ratio by number, following the typical prescription in atmosphere models.}, we can qualitatively expect the demixing window presented in this work.

For each combination of model parameters, ATHENAIA assesses the stability of a fully-mixed envelope against hydrogen-water demixing. Within the scope of this study, we aim to evaluate the range of well-mixed envelope compositions where hydrogen and water are miscible all the way to the top of the rocky mantle (MEB pressure). We interpret profiles where the miscibility criterion is not met throughout the entire envelope as unstable against demixing, resulting instead in compositional gradients from a metal-poor upper envelope to the more metal-rich deep envelope. For the demixing criterion, we use the recent DFT calculations from \citet{gupta_miscibility_2024} that extend up to the high pressures and temperatures expected in sub-Neptune interiors.

The criterion for fully miscible envelopes requires that, throughout the entire range of pressures up to the MEB, the temperature profile of the planet lies above the composition-dependent hydrogen-water coexistence curve (Figs. \ref{fig:phase_space_motivation} and \ref{fig:link_int_atm_flowchart}). For the computation of the coexistence curves, we calculate the compositional parameter $x_\mathrm{H_2}$ describing the H$_2$-H$_2$O conditions from the value of $Z_\mathrm{env}$:

\begin{equation}
    x_\mathrm{H_2} = \frac{N_\mathrm{H_2}}{N_\mathrm{H_2} + N_\mathrm{H_2O}}
\end{equation}

\noindent where $N$ refers to the number of particles. In practice, we adopt the following relationships:

\begin{equation}
     x_\mathrm{H_2} = 1- \frac{2\beta}{1 + 2\alpha - 4 \alpha \beta};
\end{equation}

\noindent with:

\begin{equation}    
\alpha = \frac{N_\mathrm{He}}{N_{\rm H_\mathrm{free}}} 
 = \frac{r \cdot m_H}{m_{\rm He}}
\end{equation}

\noindent describing the $Z$-independent ratio (by number) of He over ``free'' H (i.e., not bound in H$_2$O), with $m_i$ defined as the mass of species $i$, and:
\begin{align}
    \beta &= \frac{N_O}{N_{\rm H\mathrm{bound+free}}} \nonumber\\
    &= \frac{\alpha m_{He} + m_H }{m_{H_2O} \left(1/Z_\mathrm{env}-1\right) + 2\left(\alpha m_{He} + m_H\right)}
\end{align}

\noindent describing the ratio (by number) of oxygen over total H  (i.e., both free and bound in H$_2$O), which encodes the $Z$-dependent oxygen-to-hydrogen abundance of the H/He/H$_2$O mixture. Given $x_\mathrm{H_2}(Z_\mathrm{env})$, we calculate the temperature-dependent quantities $W_v$ (describing the pressure-dependent term in the Gibbs free energy calculation) and $\lambda$. We define $\lambda$ following \citet{gupta_miscibility_2024} in order to capture the asymmetry in the coexistence curves, and compute it following their Eq. (6-7), to finally compute the variable $y$:

\begin{equation}
    y=\frac{x_\mathrm{H_2}}{x_\mathrm{H_2} + \lambda \left(1-x_\mathrm{H_2}\right)}
\end{equation}

\noindent which is then translated into the pressure level of the coexistence curve at each queried temperature over the range of validity (750 to 6000\,K) following Eq. (8) in \citet{gupta_miscibility_2024}. We sample the parameters $W_{v,1}$, $W_{v,2}$, $\lambda_1$, $\lambda_2$, $W_H$ and $W_S$ using Gaussian distributions centered at the 50th percentile fitted values in their Table 1, with a standard deviation adopted to be the largest of the quoted $\pm 1\sigma$. For the final evaluation, we assess whether over the range of pressures spanned by both the envelope (above the MEB) and the coexistence curve within its 95\% C.I., there exists any point where the temperature in the envelope drops below the coexistence curve, which defines a profile as unstable against demixing. In practice, we record for each profile across the full parameter grid explored the minimum distance $D_\mathrm{min}$ (in units of K) between the envelope temperature profile and the coexistence curve and draw the contours of the ``demixing window'' in $f_\mathrm{env}$-$Z_\mathrm{env}$ space for each combination of planet mass and Bond albedo as interpolated contours of $D_\mathrm{min}=0$ (Figs. \ref{fig:met_fenv_posterior} and \ref{fig:demixing_window_fwd}).

\section{Planet composition inference}\label{sec:compositional_inference}

Our planetary composition inference framework relies on the computation of a grid of coupled interior-atmosphere models that predict planetary temperature, density, and radius as a function of pressure. Within the context of this work, the grid setup is tailored to the exploration of the potential for envelope compositional gradients even on warm sub-Neptunes, as described below. Beyond this broad parameter space exploration, we evaluate the specific bulk parameters consistent with TOI-270 d's known properties and whether these could lead to potential demixing, in order to assess claims of a fully miscible envelope.

\subsection{Grid setup}

Our model grid focuses on planets with irradiation levels comparable to TOI-270 d, warm enough to avoid both water condensation at the surface and in the atmosphere under conditions compatible with the JWST-inferred compositional constraints except in high-albedo scenarios \citep{holmberg_possible_2024,benneke_jwst_2024,felix_evidence_2025,constantinou_atmospheric_2025}, further supported by our modeling (Fig. \ref{fig:met_fenv_posterior}).

For the core and mantle, we fix the composition to an Earth-like mass fraction $f'_\mathrm{core}=0.325$ (following the terminology in \citealp{piaulet-ghorayeb_jwstniriss_2024}) of the core and mantle underlying any of the envelopes modeled for this work. We set the orbital distance and incident stellar spectrum to those of TOI-270 d (Table \ref{tab:star_planet_params}) except for the $T_\mathrm{eq}=400$\,K and 450\,K smaller model grids (Table \ref{tab:model_grid}) where we adopt instead $S_p=4.3$ and 6.8$S_\oplus$, respectively. In our nominal model grid, we vary $T_\mathrm{eq}$ via the planetary Bond albedo between 0 and 0.4, with the upper limit of 0.4 informed by the range of photosphere temperatures of $\sim 300-350$\,K retrieved by transmission studies \citep{benneke_jwst_2024,felix_evidence_2025,constantinou_atmospheric_2025}. For the stellar irradiation, we use the closest stellar model in the PHOENIX grid; \citealp{husser_new_2013}) and adopt $T_\mathrm{int}=25$\,K as the lower boundary condition for atmosphere models (discussed in Section \ref{sec:caveats}). 
Envelope compositions are varied from solar-metallicity H/He-dominated all the way to pure-water envelopes. Finally, the range of envelope mass fractions and planetary masses considered span the full range of expected surface gravities across the sub-Neptune regime (see Table \ref{tab:model_grid}). For each model, we follow the steps outlined in Section \ref{sec:model} (Fig. \ref{fig:link_int_atm_flowchart}) to record the full-planet profiles (Fig. \ref{fig:master_grid_varyparam}) and planetary radii to construct mass-radius relationships.



\begin{table}[h]
    \centering
    \renewcommand{\arraystretch}{1.2}
    \begin{tabular}{ccccc}
        \toprule
        $T_\mathrm{eq}$ [K] & M$_p$ [$M_\oplus$] & $Z_\mathrm{env}$ & H$_2$O VMR & $f_\mathrm{env}$ [\%] \\
        \midrule
        336  & 2.87  & 0.01   & 0.0013   & 0.0010  \\
        361  & 4.23  & 0.10   & 0.0142    & 0.0018  \\
        381  & 5.14  & 0.35   & 0.0652     & 0.0034  \\
        400*   & 6.24  & 0.40   & 0.0795    & 0.0061  \\
        450*  & 9.20  & 0.50   & 0.1147    & 0.0113  \\
             & 13.56 & 0.55   & 0.1367    & 0.0206  \\
             & 20.00 & 0.60   & 0.1627    & 0.0378  \\
             &       & 0.65   & 0.1940    & 0.0693    \\
             &       & 0.70   & 0.2322    & 0.1269    \\
             &       & 0.75   & 0.2799    & 0.2325    \\
             &       & 0.80   & 0.3414    & 0.4259    \\
             &       & 0.85   & 0.4234    & 0.7802    \\
             &       & 0.90   & 0.5384    & 1.4293    \\
             &       & 1.00   & 1.0000    & 2.6186    \\
             &       &        &           & 4.7973    \\
             &       &        &           & 8.7887    \\
             &       &        &           & 16.101    \\
             &       &        &           & 29.497    \\
             &       &        &           & 54.039    \\
             &       &        &           & 99.000    \\
        \bottomrule
    \end{tabular}
    \caption{Planetary parameters spanned by the model grid. The first three values of $T_\mathrm{eq}$ were chosen to correspond to a Bond albedo of 0.0, 0.2, and 0.4 for TOI-270\,d. $^*$The last two $T_\mathrm{eq}$ dimensions of the grid are only computed for $M_p=5.14\,M_\oplus$ and $M_p=20\,M_\oplus$ (see Appendix Section \ref{sec:dmx_limits}). H$_2$O VMR values are derived from each model's $Z_\mathrm{env}$. The envelope mass fractions span 10$^{-5}$ to 1 in log space. For all models, we use $T_\mathrm{int}=25$\,K and $f'_\mathrm{core}=0.325$.}
    \label{tab:model_grid}
\end{table}

\subsection{Metallicity-Envelope Mass Fraction Inference}

We use ATHENAIA to perform inference of the range of envelope mass fractions and metal contents consistent with TOI-270 d's observed mass, radius, and irradiation to evaluate its compatibility with a fully-miscible envelope. ATHENAIA inherits its Bayesian inference framework for envelope and bulk properties from previous work (\citealp{piaulet_evidence_2023}, see applications in \citealp{roy_is_2022,benneke_jwst_2024}). The method is briefly summarized below.

One grid of planetary radii is generated for each $T_\mathrm{eq}$ value we explore, spanning the dimensions $\left(M_p; \log_{10} f_\mathrm{env}; Z_\mathrm{env}\right)$. From each grid, an interpolator is created to infer planetary radii in-between grid points via linear interpolation. We first construct a likelihood grid by finely sampling all three dimensions, and calculating for each the Gaussian likelihood value knowing the planetary radius and its uncertainty (Table \ref{tab:star_planet_params}). Then, we compute the prior which is uniform with $\log_{10} f_\mathrm{env}$ and $Z_\mathrm{env}$, but Gaussian with $M_p$ in order to marginalize over the mass uncertainty (Table \ref{tab:star_planet_params}). Finally, we extract the coordinates in the 2D $f_\mathrm{env}$-$Z_\mathrm{env}$ space corresponding to the 68.2\% and 95.4\% confidence intervals from the Bayesian posterior distribution (Fig. \ref{fig:met_fenv_posterior}).






\section{Application to TOI-270 d}\label{sec:toi_270_d_res_disc}

We demonstrate on TOI-270 d our approach for evaluating the potential for demixing in warm sub-Neptune interiors using ATHENAIA. We  combine information from mass, radius, and atmosphere characterization with constraints on the range of interior structures where fully miscible envelopes would be unstable.

\subsection{Inferred envelope properties}

\begin{figure}
    \centering
    \includegraphics[width=0.48\textwidth]{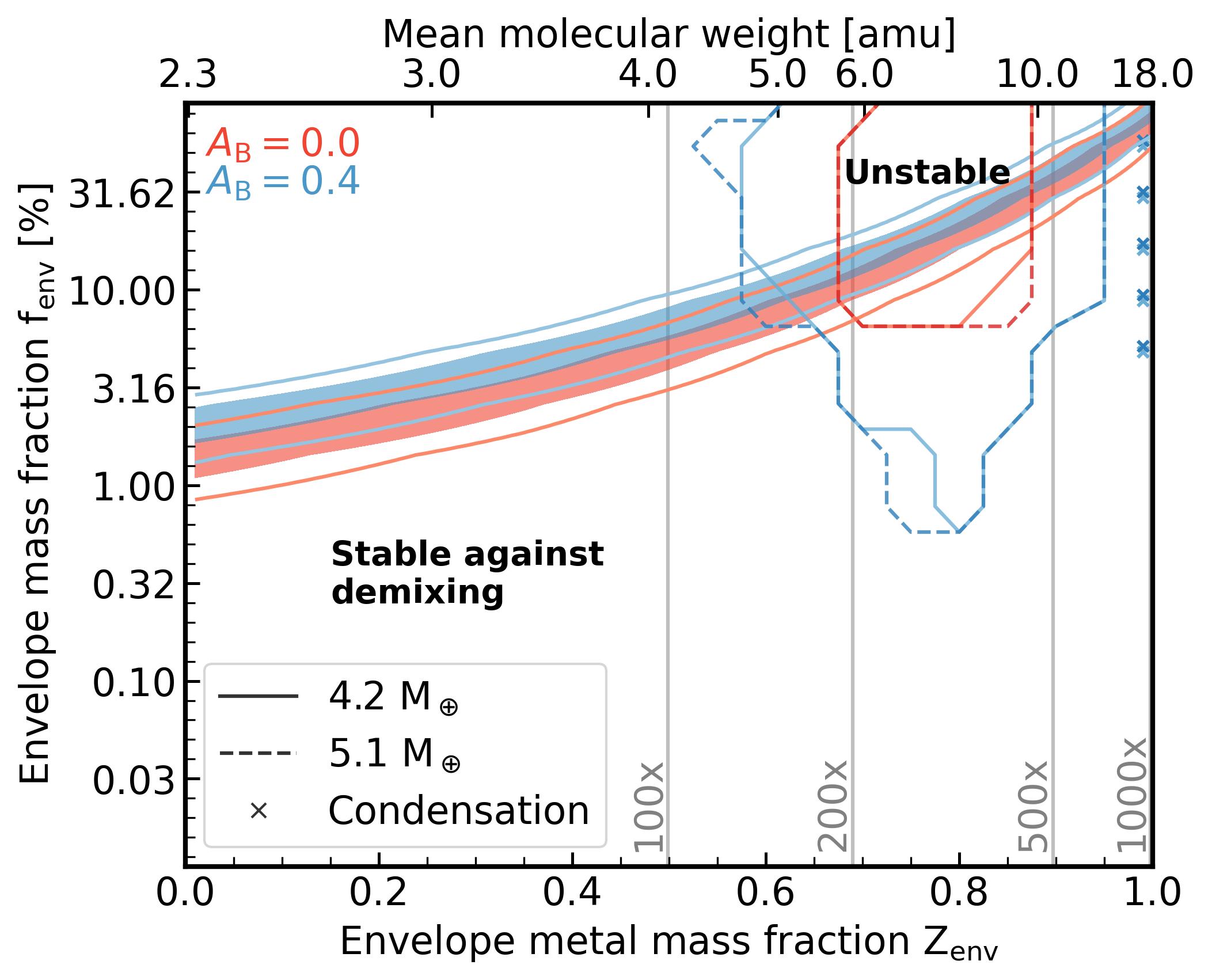}

    \caption{Joint constraints on the envelope mass fraction and envelope metal content of TOI-270 d. The red (blue) contours spanning the full range of potential envelope metallicities represent the 1 and 2$\sigma$ composition constraints obtained with ATHENAIA assuming TOI-270\,d's insolation and a Bond albedo of 0 (0.4). The contours in the top right corner outline the region of the parameter space where fully-mixed compositions are unstable (red for $A_\mathrm{B}=0$, blue for $A_\mathrm{B}=0.4$, line style encodes planet masses bracketing TOI-270 d). Over the range of atmosphere metal mass fractions allowed by the transmission spectrum, the $f_\mathrm{env}$-$Z_\mathrm{env}$ posterior distribution overlaps with the demixing window, where envelope compositional gradients can occur. Vertical grey lines show the locations of various metallicity enrichments relative to solar (labeled). Crosses (slightly offset to the left for clarity) indicate the models where the water condensation curve is crossed: they lie beyond the demixing window and do not affect the conclusions.}
    \label{fig:met_fenv_posterior}
\end{figure}

We obtain statistical constraints on the range of envelope mass fractions and envelope metallicities that are compatible with TOI-270 d's bulk properties and irradiation assuming a well-mixed envelope (Fig. \ref{fig:met_fenv_posterior}). The posterior distribution in this 2D space exhibits the well-known sub-Neptune compositional degeneracy: the planet can be explained either by a 1--3 percent-by-mass (wt\%) H/He-rich solar-metallicity atmosphere, or at the other extreme by a 30--60 wt\% pure-water atmosphere. 
Assuming that the atmosphere metal mass fraction $Z_\mathrm{atm}$ of 200-300$\times$ solar measured with JWST is representative of the envelope metallicity, we reproduce the previously-reported $\sim 10$\% envelope mass fraction \citep{benneke_jwst_2024}. We checked that our results were not impacted by the choice of 20 mbar rather than 1 mbar as the reference pressure for extracting planetary radii: the maximum probability set of $(Z_\mathrm{env};f_\mathrm{env})$ obtained as the best match for TOI-270\,d's mass and radius using 1 mbar as the reference pressure falls fully within the 68\% C.I. shown on Fig. \ref{fig:met_fenv_posterior}.

\begin{figure}
    \centering
    \includegraphics[width=0.48\textwidth]{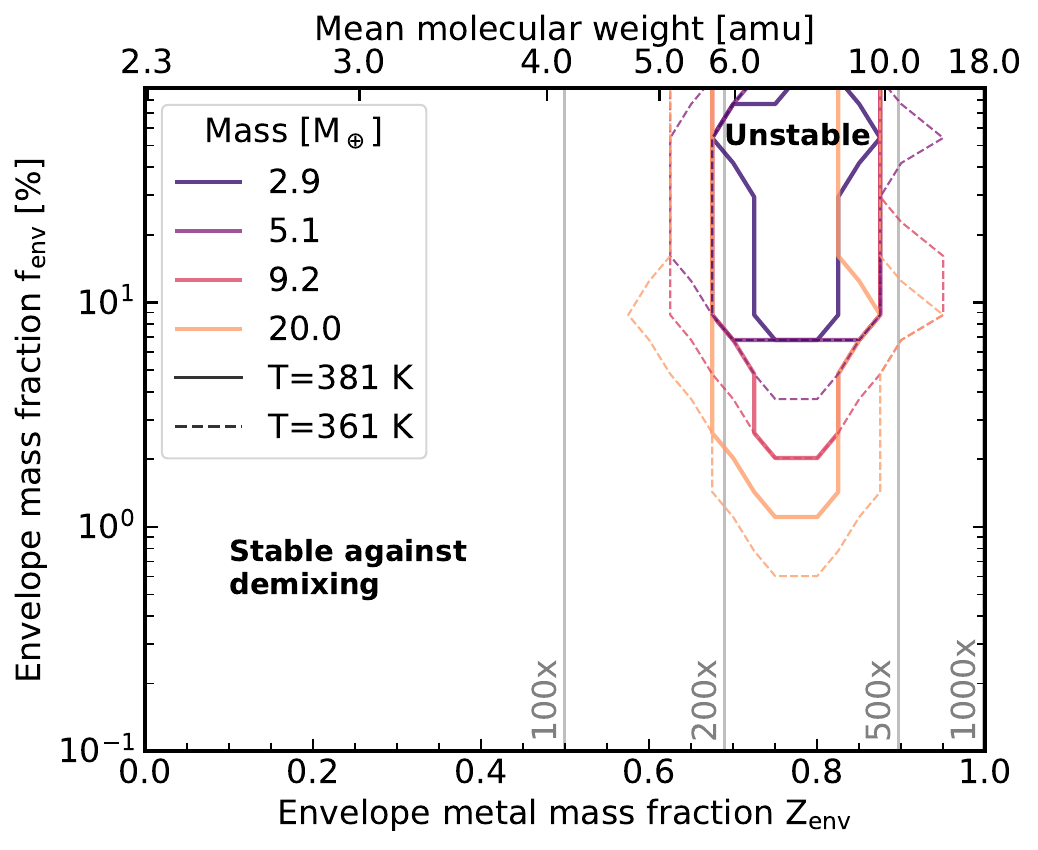}

    \caption{Contours delineating the region of parameter space where demixing occurs for different planet masses and planetary equilibrium temperatures. The color encodes the planet mass, and the line style encode the equilibrium temperature. Vertical gray lines indicate different atmosphere metallicities (labeled). None of the models shown cross the water condensation curve.}
    \label{fig:demixing_window_fwd}
\end{figure}





\subsection{Potential for compositional gradients}

We map in the same 2D $f_\mathrm{env}$-$Z_\mathrm{env}$ space the demixing window inferred for the two masses in our grid closest to TOI-270 d: 4.2 and 5.1 $M_\oplus$ (compared to TOI-270 d's $M_p=4.78 \pm 0.43 M_\oplus$). We find that in these conditions, demixing can happen on TOI-270 d for $Z_\mathrm{env} \in \left[0.55-0.95\right]$ (120 to 700$\times$ solar metallicity), with the exact bounds depending on the assumption on the planetary equilibrium temperature (Fig. \ref{fig:met_fenv_posterior}).

Given the overlap in this demixing space with the observationally-constrained upper atmosphere metallicity of 200 to 300$\times$ solar (\citealp{benneke_jwst_2024}, Fig. \ref{fig:met_fenv_posterior}), our modeling suggests that TOI-270 d's envelope may exhibit compositional stratification.
In this case, TOI-270 d's bulk envelope metallicity would be higher than the estimate from the upper envelope due to demixing. Our zeroth-order attempt at mapping the extent of the compositional gradient required to observe any given photospheric $Z_\mathrm{atm}$ suggests a true envelope bulk metallicity of up to $Z_\mathrm{env}\sim 0.85$ (see Appendix Section \ref{sec:stable_Z}, Figure \ref{fig:stable_comp}). This would imply a less massive planetary core than previously inferred, with an envelope up to twice as massive. Alternatively, if the true $Z_\mathrm{atm}< 0.6$, or if the thermal evolutionary state of TOI-270 d maintains its interior hotter than the $T_\mathrm{int}=25$\,K assumed here, the fully-miscible envelope scenario may still hold (e.g. \citealp{tejada_arevalo_sub-neptune_2025}).

\subsection{Sensitivity to chemical composition}\label{sec:sensitivity_ch4_vs_h2o}

We find that the T-P profiles produced by our water-hydrogen mixtures match the conditions predicted by models incorporating carbon-bearing molecules into the atmosphere's radiative-convective calculation, for metallicities comparable to that of TOI-270\,d. While our temperature profiles assume envelopes made of a mix of hydrogen, helium, and water, other molecules such as methane and carbon dioxide have been detected on the warm sub-Neptune TOI-270\,d \citep{benneke_jwst_2024,holmberg_possible_2024}. Although CH$_4$ is predicted to be underabundant by a factor of $\sim 4$ relative to H$_2$O at the photosphere on TOI-270\,d \citep{benneke_jwst_2024}, the presence of additional infrared absorbers should enhance the greenhouse effect and heat up the lower atmosphere. Water-hydrogen demixing in our simulations primarily depends on the temperature conditions reached at $\sim$GPa pressures (Fig. \ref{fig:tp_zenv}), where our models reproduce temperatures compatible with those predicted by radiative-convective calculations accounting for the full set of retrieved abundances for TOI-270\,d and metallicities of 100 to 300$\times$solar \citep{rigby_surface_2025}. 

\begin{figure}
    \centering
    \includegraphics[width=0.48\textwidth]{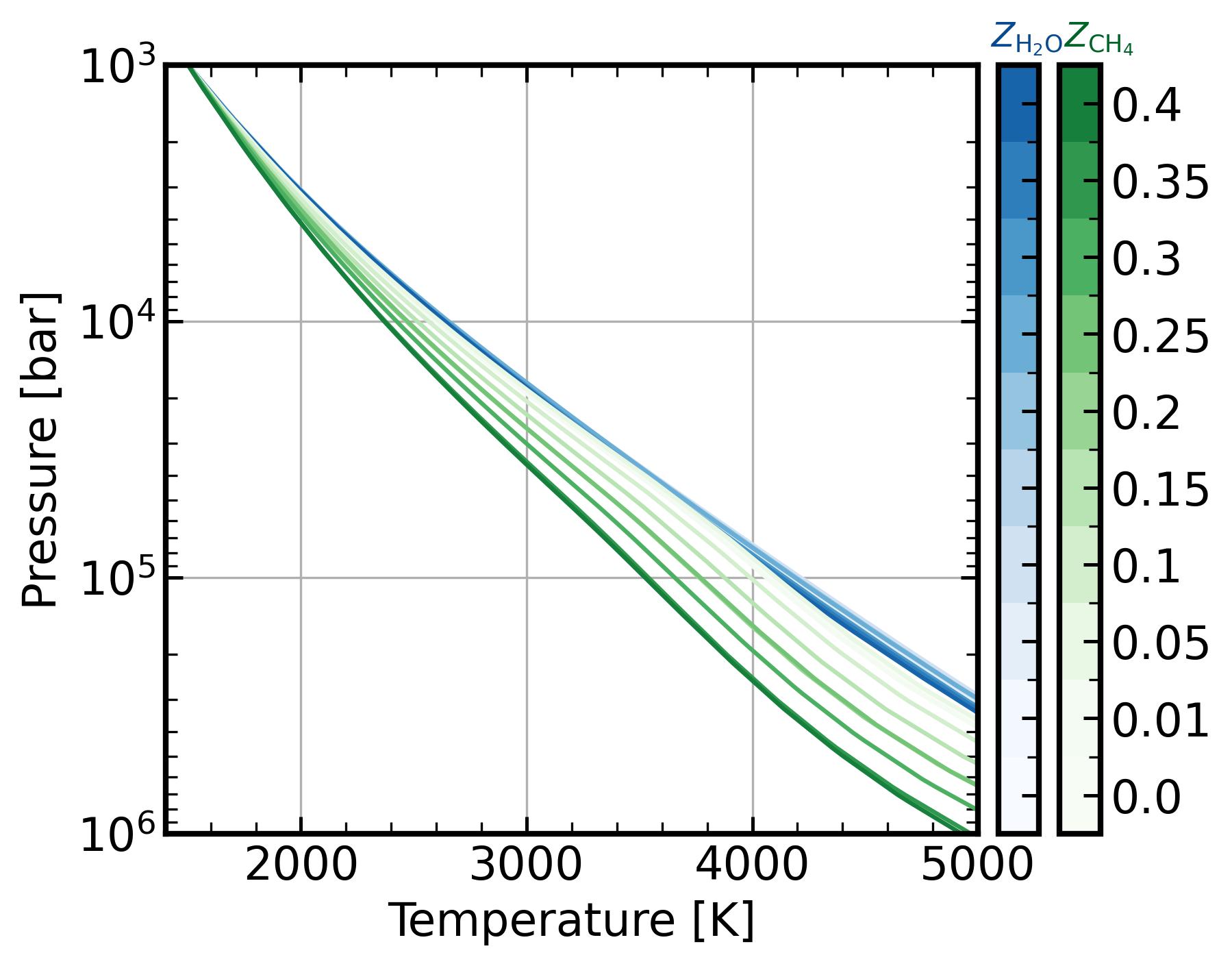}

    \caption{ Illustration of the compositional dependence of the deep-envelope temperature gradients for H-H$_2$O mixtures (blue) or H-CH$_4$ mixtures (green). We show constant-composition adiabats for models with a temperature of 1500\,K at 1 kbar, varying the envelope metallicity $Z$ from 0.001 to 0.4. Increasing the metallicity leads to shallower adiabatic gradients in both cases, but the effect is substantially more pronounced for the CH$_4$-rich scenarios. Since colder interiors are more susceptible to water-hydrogen demixing, the presence of abundant methane mixed with water in metal-rich sub-Neptune envelopes will expand the range of irradiation conditions where envelope compositional gradients are to be expected.}
    \label{fig:h2o_ch4_profiles}
\end{figure}

Our calculation of adiabatic profiles for mixtures including methane in the convective envelope suggests that the presence of methane is likely to make the deep envelopes of sub-Neptunes colder and more susceptible to water-hydrogen demixing (Fig. \ref{fig:h2o_ch4_profiles}). The presence of abundant methane, rather than only water, may impact the compositional dependence of envelope temperature profiles across the critical 0.1-1,000 GPa range over which DFT calculations predict the transition between profiles (un)stable to water-hydrogen demixing (Fig. \ref{fig:phase_space_motivation}). We explore this possibility by performing the first comparison between isentropes calculated for H/He/H$_2$O vs. H/He/CH$_4$ mixtures, across a broad range of compositions ($Z_\mathrm{CH_4/H_2O} \in [0.,0.4]$; Fig. \ref{fig:h2o_ch4_profiles}), using the EOS from \citet{nettelmann_uranus_2016} and \citet{bethkenhagen_planetary_2017}. We find that increasing the methane abundance has a much more substantial effect on the adiabatic profiles than an equivalent increase in the water enrichment, with methane-rich envelopes exhibiting substantially shallower temperature gradients. This results in temperatures hundreds of Kelvin cooler at 100 kbar, even for the same atmosphere boundary conditions at 1 kbar (Fig. \ref{fig:h2o_ch4_profiles}). 

While a full exploration of the propensity for water-hydrogen demixing in profiles perturbed by abundant methane is beyond the scope of this work, our initial exploration reveals that methane may play a crucial role in expanding the propensity of sub-Neptune envelopes to compositional stratification. 
Further study is needed to resolve the interplay between the shallower adiabatic gradients and potential atmospheric heating sources in the presence of carbon species e.g. through graphite precipitation, depending on how the C/O ratio varies throughout the atmosphere \citep{peng_puffy_2024}. Just like water, carbon may demix in the envelope and partition into solid phases in a planet's interior (see \citealp{li_soot_2026} and references therein). The demixing conditions for carbon, and the phases of carbon that would be found in a carbon-rich planetary interiors are poorly known, motivating future work on this topic.

\subsection{Plausibility of mantle melting}

We additionally apply our ATHENAIA models to an evaluation of the physical conditions at the MEB within the context of the fully-mixed envelope assumption inherent to our model grid. We find that the consideration of the greenhouse effect's composition-dependent impact on atmospheric temperature structures critically restricts the range of sub-Neptune compositions where solid compositions are possible compared to previous estimates, making the envelope mass fraction the leading parameter controlling mantle state.

The molten or solid state of the planetary mantle is of paramount importance because of its impact on the atmospheric composition. In the presence of a molten mantle \citep{kite_water_2021,schlichting_chemical_2021}, the relative solubilities of the volatiles present \citep{lichtenberg_redox_2021} and the mantle composition dictate how volatiles will be distributed between the mantle and the overlying gas. Water, in particular, has a solubility orders of magnitude larger than most carbon-bearing species, impacting the atmospheric C/O ratios expected on sub-Neptunes with vs. without magma oceans at the envelope-atmosphere boundary \citep{werlen_sub-neptunes_2025, werlen_atmospheric_2025}. 

Atmosphere studies of TOI-270 d offered mantle-atmosphere chemical equilibration as a potential explanation for the measured molecular abundances \citep{benneke_jwst_2024,nixon_magma_2025}, inferring that the high atmospheric metallicity can be explained without the need for initial accretion of abundant volatiles. In parallel, interior modeling work using isothermal temperature profiles with $T=T_\mathrm{eq}$ to describe the atmosphere suggested that, instead, TOI-270 d might \textit{not} have a molten surface given its high inferred mean molecular weight \citep{breza_not_2025}. Within that study's framework, theoretical lower limits on the mean molecular weight required for solidification range from $\mu \sim 2.7$ to 6 amu, depending on the 1-100 bar pressure range assumed for the radiative-convective boundary where the model transitions to an adiabatic temperature profile. Their composition-independent description of the temperature in the radiative atmosphere made the adiabatic profile -- gradually shallower when moving from solar to increasingly super-solar metallicities -- the only factor dictating the state of silicate at the MEB.

Our modeling rather suggests that the use of radiative-convective temperature profiles to describe the atmosphere weakens the composition dependence of solidification conditions, making envelope mass fraction the leading factor determining the mantle state for a MgSiO$_3$ composition. For TOI-270\,d, this translates into a more restrictive lower limit on the amount of metal enrichment required to allow for solid mantle conditions, specifically a mean molecular weight of $\gtrsim$5 amu corresponding to an inferred envelope mass fraction higher than about 7\%. 

When coupling self-consistent atmosphere models to deep-envelope adiabats, the phase of silicate at the MEB is the result of two competing effects: while the adiabat gets shallower with increasing $Z_\mathrm{env}$ (see e.g. Fig. \ref{fig:tp_zenv}), the increase in infrared opacity strengthens the greenhouse effect in the atmosphere model, leading to a warmer atmosphere over the $\sim 1$\,bar to 1\,kbar pressure range (Fig. \ref{fig:melt_illustrate_profiles}). Only for metallicities greater than $Z_\mathrm{env}\sim 0.6$ does the decreasing gradient of interior profiles ``win over'' to the extent that solid mantle conditions are predicted for masses similar to TOI-270\,d (Fig. \ref{fig:melt_vs_Teq}). Following the intuitively-expected behaviour, the range of compositions where mantle melting is expected expands at higher equilibrium temperatures. 

For low-mass planets, there is both a lower and an upper limit to the envelope mass fractions where a molten mantle is expected for a given metallicity. At the low-$f_\mathrm{env}$ end, MEBs occur in the lower-pressure phase space where silicate rock is in the bridgmanite phase, but do not reach temperatures hotter than the high-pressure solidus of rock, while at higher $f_\mathrm{env}$, silicates are rather in the post-perovskite phase which is solid up to much higher temperatures (Fig. \ref{fig:melt_illustrate_profiles}; \citealp{oganov_theoretical_2004,sakai_experimental_2016}). For higher-mass planets, the MEB reaches higher pressures overall for a given envelope mass fraction, which limits the maximal $f_\mathrm{env}$ allowing for molten conditions (Figs. \ref{fig:melt_illustrate_profiles} and \ref{fig:melt_vs_Teq}). 
 Given the joint constraints on the composition and envelope mass we obtain for TOI-270\,d, our models predict molten mantle conditions at least up to an atmosphere metallicity of $\sim 150\times$solar (Fig. \ref{fig:melt_vs_Teq}). 
Finally, we note that planetary interiors likely form much hotter than the conditions captured in our models, which can lead to both a higher $T_\mathrm{int}$ than covered by our grids, and substantial mantle hydration pushing the solidus to lower temperatures than predicted in this pure-MgSiO$_3$ framework (e.g., \citealp{katz_new_2003}), in which case the parameter space for molten mantle conditions expands further.




\section{Impact for sub-Neptune studies} \label{sec:implications}

Our modeling reveals that water-hydrogen demixing is predicted over a wide range of masses and radii even at the 330--450\,K temperatures we explored, warm enough to avoid water condensation in the atmosphere except for the lowest temperatures and highest metallicities ($T_\mathrm{eq}=336$\,K and $Z_\mathrm{env}\simeq 1.0$). We explore the range of conditions where demixing should be expected on sub-Neptunes as well as the implications for sub-Neptune composition and thermal evolution modeling. We provide data products and open-source scripts to reproduce the contours of the demixing window as well as the mass-radius curves computed by ATHENAIA.

\subsection{Demixing conditions} 

The ATHENAIA model grids are calculated for well-mixed envelopes (composition constant with pressure), and reveal that the calculated temperature profiles reach conditions of hydrogen-water immiscibility over a broad range of envelope mass fractions and envelope metal contents. 

While demixing does not occur in the extreme water-only composition case ($Z_\mathrm{env}=1$) due to the steep increase in the pressure of the coexistence curve near $Z_\mathrm{env}=0.9$ (Fig. \ref{fig:coexistence_vs_composition}), the window in $f_\mathrm{env}-Z_\mathrm{env}$ space where demixing is predicted systematically covers substantially metal-enriched compositions, regardless of planet mass and temperature (Figs. \ref{fig:demixing_window_fwd} and \ref{fig:tp_zenv}). The range of envelope mass fractions and metallicities where demixing occurs expands to lower values both with decreasing temperatures and increasing planet mass, reaching metallicities of $<100\times$ solar for a 20\,$M_\oplus$ planet.

Visualizing our constant-composition curves in mass-radius space reveals the importance of demixing at the population level for sub-Neptunes (Fig. \ref{fig:mass_radius}). While not predicted for solar metallicity envelopes, as suggested by the deep pressures of the coexistence curves at low $Z_\mathrm{env}$ (Figs. \ref{fig:phase_space_motivation} and \ref{fig:coexistence_vs_composition}), we find that $\gtrsim 150\times$ solar metal enrichments can give rise to demixing for planets larger than $2.2\ R_\oplus$ and with masses $>4\ M_\oplus$ in the $T_p=336$\,K case ($A_\mathrm{B}=0.4$ for TOI-270\,d). We download the updated catalog of well-characterized exoplanets from the NASA Exoplanet Archive \citep{NasaExoArchive2013,NasaExoArchive2025}, and map our demixing flags onto the observed population using the calculated radius for each model. Demixing conditions cover the entire mass-radius range of sub-Neptune conditions in the more metal-rich 310$\times$ solar metallicity ($Z_\mathrm{env}=0.8$) case (Fig. \ref{fig:mass_radius}). This suggests that the $\gtrsim 150\times$ solar atmosphere metal enrichments typically inferred on sub-Neptunes \citep{hu_water-rich_2025,benneke_jwst_2024,holmberg_possible_2024,felix_evidence_2025,piaulet-ghorayeb_jwstniriss_2024,ahrer_escaping_2025,teske_jwst_2025,wallack_jwst_2024} might instead reflect even more metal-enriched interiors.

\subsection{Impact on composition and evolution} 

We find that rather than fully-mixed or layered-structure models, the envelopes of warm metal-rich sub-Neptunes are to be described by a continuum of compositional gradient structures, with lower metallicities at low pressures and higher metallicities deeper down in the envelope (as proposed for the solar system ice giants; \citealp{amoros_h_2-h_2o_2024}). Specifically, while envelopes with $\lesssim 150\times$ solar metallicity may be aptly described by fully-mixed models, our simple zeroth-order estimate of the amount of metal depletion to be expected in the upper atmosphere (as detailed in Sec. \ref{sec:stable_Z}; see Fig. \ref{fig:stable_comp}) implies that the metallicities measured via transmission spectroscopy are underestimating the bulk envelope metal mass content over the demixing window. Compositional stratification should have the strongest impact on the upper-atmosphere metallicities of colder and higher-mass planets with metal-rich envelopes, with predicted metal depletions of over 50\% by mass predicted even at masses as low as $\sim 6 M_\oplus$. Meanwhile, planets with bulk envelope metallicities of $\sim 500\times$solar and masses greater than $\sim 10 M_\oplus$ may be well represented by fully layered structures of near-solar-metallicity gas on top of a water-rich interior.
This effect also impacts low-mass sub-Neptunes: the upper-atmosphere metal mass content can be reduced by 20\% even in the case of the lowest planetary mass ($2.87M_\oplus$) and at the zero-albedo equilibrium temperature of TOI-270\,d.

\begin{figure*}
    \centering
    \includegraphics[width=0.98\textwidth]{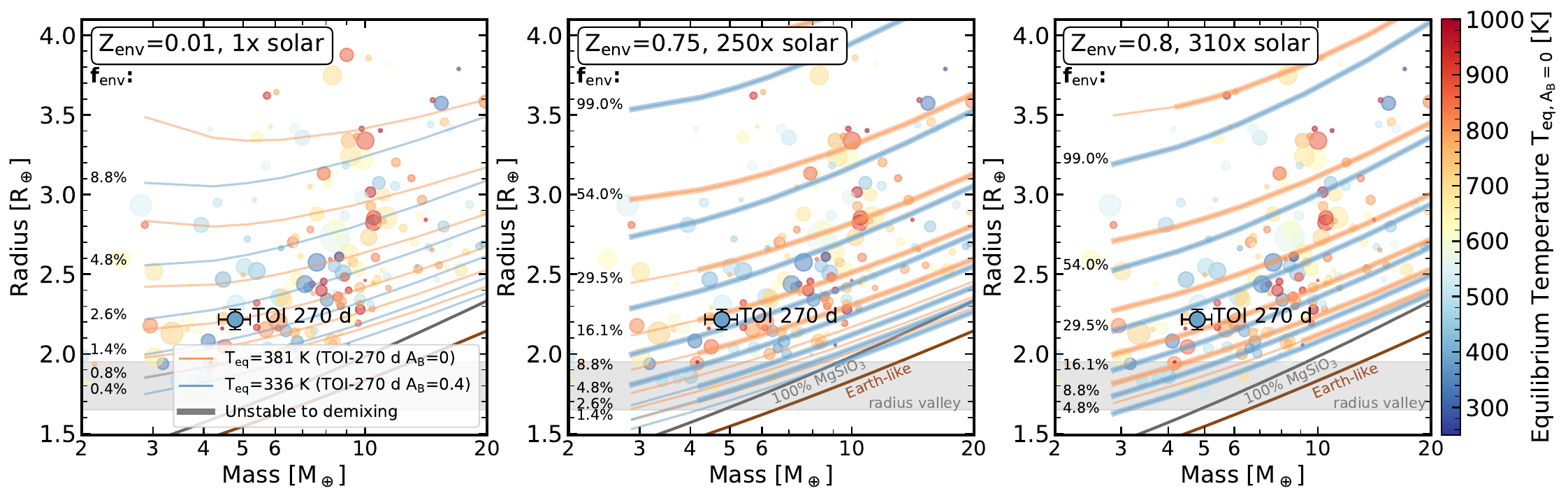}

    \caption{Mass-radius diagram of planets across the sub-Neptune size range (circles, color encodes equilibrium temperature) and their susceptibility to demixing for $Z_\mathrm{env}=0.01$ (left, $1\times$ solar metallicity), 0.75 (middle, $250\times$ solar metallicity), and 0.8 (right, $310\times$ solar metallicity). Circle sizes encode the favorability of planets to transmission spectroscopy via the transmission spectroscopy metric \citep{kempton_framework_2018}. The position of TOI-270\,d in mass-radius space is indicated by the opaque black-outlined circle, and constant-composition curves for an Earth-like and a 100\% MgSiO$_3$ rocky composition \citep{zeng_mass-radius_2016} are shown in grey and brown. The approximate position of the ``radius valley'' \citep{fulton_california-kepler_2017} separating sub-Neptunes from the smaller rocky super-Earths is shown as the grey shaded area. For models of sub-Neptunes with gas envelopes, we show the ATHENAIA mass-radius relationships for various envelope mass fractions (labeled in each panel), for each envelope metallicity, assuming $T_\mathrm{eq}=336$\,K (blue) or 381\,K (orange). Thicker curves indicate conditions unstable to demixing. Over the 330-380\,K temperature regime explored, demixing impacts sub-Neptunes with metal-rich envelopes across their entire range of masses and radii.}
    \label{fig:mass_radius}
\end{figure*}

Overall, using upper-atmosphere metallicities to represent the bulk envelope would result in an \textit{underestimate} of the envelope mass fraction, and therefore overestimated sub-Neptune core masses, over the range of conditions susceptible to demixing. In such a scenario, the upper atmosphere metallicity can no longer be used as a reasonable metric to break the mass-radius degeneracy between low-metallicity envelopes with lower mass fractions, and higher-metallicity envelopes with larger mass fractions (Fig. \ref{fig:mass_radius}). Instead, interior-atmosphere modeling frameworks should be employed to model self-consistently the distribution of metals across the envelope and the corresponding temperature structure that result in structures fully stable against demixing. We expect a degeneracy in the measured atmosphere metallicities from transmission spectroscopy between intrinsically metal-poor envelopes (stable against demixing) and metal-enriched envelopes (unstable against demixing) which would have metal-depleted upper-atmospheres. 

We explore the limits of demixing conditions along the planetary insolation axis by calculating models for $T_\mathrm{eq}=400$\,K and 450\,K for a subset of planet masses (see Figure \ref{fig:dmx_vs_Teq}). We observe a gradual ``closing'' of the demixing window for increasing insolations across the planetary mass range covered by our simulations, which indicates that demixing can affect sub-Neptune interiors up to $T_\mathrm{eq}\sim 450$\,K.

Water-hydrogen demixing can have profound implications on the radius and thermal evolution of sub-Neptunes. The distribution of metals across the envelope would alter the cooling rate for a given internal entropy as well as impact the planet radius, and therefore the cross-section it offers to photoevaporative mass-loss. Further, mass-loss rates themselves would be higher for the same bulk envelope metallicities if compositional gradients exist, as a hydrogen-rich upper atmosphere is more easily lost due to its lighter particle weight and less efficient metal cooling in the absence of abundant water (\citealp{piaulet-ghorayeb_jwstniriss_2024}; Kubyshkina, Egger \& Piaulet-Ghorayeb, subm.). This work therefore motivates further exploration of the ramifications of sub-Neptune demixing in terms of the predictions of mass-loss and envelope thermal evolution models (e.g. \citealp{tang_reassessing_2024,piaulet_wasp-107bs_2021}).

\section{Comparison to previous studies}\label{sec:comparison_prev}

We find that our our modeling results are compatible with the current literature, and that the use of radiative-convective atmosphere models is essential to map the demixing window in $f_\mathrm{env}-Z_\mathrm{env}$ space.

First, our models reproduce the expected qualitative behavior of decreasing planet radius when increasing the envelope metallicity for a given planet mass and envelope mass fraction (Fig. \ref{fig:mass_radius}). The range of envelope mass fractions predicted by ATHENAIA over the sub-Neptune size regime overall, and for TOI-270 d in particular, also matches expectations from models both at the H/He-rich \citep{lopez_understanding_2014} and H$_2$O-rich \citep{aguichine_mass-radius_2021} ends of the compositional range. 

Our model setup for exploration of the susceptibility of planetary envelopes to demixing goes beyond previous work that adopted either fully adiabatic envelopes, or parameterized temperature profiles \citep{guillot_radiative_2010,amoros_h_2-h_2o_2024,howard_possibility_2025}, as we calculate composition-dependent radiative-convective temperature profiles as an upper boundary condition to the adiabatic interior models. This approach has several advantages. First, it allows for the self-consistent determination of the radiative-convective boundary (RCB) where the adiabatic profiles start, which we find to be deep and even $>1$\,kbar in some of our models, rather than the assumed, and often much lower, RCB pressures adopted ad-hoc when combining interior adiabats with parameterized functions in the upper atmosphere \citep{guillot_radiative_2010}. Models with deeper radiative envelopes will have colder interiors, typically more conducive to demixing. Second, compared to using adiabatic profiles, our modeling approach accounts for the (often major; \citealp{tang_reassessing_2024}) contribution of the radiative atmosphere to the planetary radius -- which is the key observable for accurate composition estimation when linking planet observations to interior-atmosphere models. 

We expanded upon previous single-planet, single-composition studies of the propensity of envelopes to demixing by instead systematically exploring its onset throughout the mass-radius diagram as a function of water content and over the range of equilibrium temperatures from the onset of atmosphere water condensation (at the cold end) to that of stable fully-mixed envelopes (at the hot end). We also proposed a new method to statistically assess the impact of demixing on the envelope composition of individual planets by leveraging mass, radius, irradiation, and atmospheric constraints. Taken together, this modeling exploration provides a first assessment of the impact of this process at the population level. 

\section{Model Limitations}\label{sec:caveats}

The modeling work presented here has limitations related to the chosen prescription for demixing, envelope composition, and interior model assumptions.

First, while water condensation may, through the release of latent heat, heat up the planetary atmosphere -- resulting in warmer interiors than we predict, and potentially impacting the diagnosis of hydrogen-water (im-)miscibility, we find that our temperature profiles only cross the water condensation curves in the coldest models with $T_\mathrm{eq}=336$\,K and with high $Z_\mathrm{env}$ (Fig. \ref{fig:stable_comp}). These metallicities are so high that the conditions do not overlap with the demixing window for masses lower than 13.5 $M_\oplus$, and should not impact our conclusions for most sub-Neptunes. 

Beyond our choice of H/He/H$_2$O composition (discussed in Section \ref{sec:sensitivity_ch4_vs_h2o}), the main limitation of the atmosphere model is the choice of $T_\mathrm{int}$. We adopted $T_\mathrm{int}=25$\,K, which is represented in the sub-Neptune literature \citep{madhusudhan_interior_2020,kempton_where_2023,nixon_new_2024,rigby_surface_2025}, although the exact value will be dependent on each planet's thermal evolution, and therefore on the planet mass, atmosphere metallicity, envelope thickness, and any planetary eccentricity (see e.g. \citealp{lopez_understanding_2014,thorngren_bayesian_2018,tang_reassessing_2024}). For the purposes of this work, we choose this value because it is low enough to reveal the \textit{potential} for envelope demixing under realistic conditions in sub-Neptune atmospheres, but high enough to remain realistic. We also calculated limited model grids of planets with well-mixed H/He/H$_2$O envelopes for 5.1\,$M_\oplus$ and 20\,$M_\oplus$ planets with $T_\mathrm{int}=50$\,K and 75\,K and found that in these two cases, the deep-envelope temperatures are sytematically $\gtrsim 110$\,K warmer than the conditions required for immiscibility. This may indicate that water-hydrogen demixing occurs late in sub-Neptune evolution, although the nuances related to the drastic impact of compositional diversity on envelope temperature gradients (see e.g. Fig. \ref{fig:h2o_ch4_profiles}) may allow for this process to impact sub-Neptunes at earlier ages.
While beyond the scope of this study, future work will evaluate the dependence of the demixing window on the compositionally-dependent thermal evolution, and whether the metal-rich sub-Neptune interiors that we proposed as demixing candidates ever get cool enough to match the assumptions made here.


Since we recalculate mixed adiabats from each EOS, our models do not account for non-ideal mixing of H and He covered in \citet{chabrier_new_2021} incorporating the entropy data from \citet{militzer_ab_2013}. This effect should become important at megabar pressures, and for H/He-dominated compositions. We expect that this limitation has minimal impact on our conclusions: our results are dominated by the lower-pressure end of the EOS, and the envelope profiles where demixing conditions are identified have metal-rich, H/He-poor compositions. Non-ideal mixing of hydrogen and water (e.g. \citealp{werlen_effects_2026}), however, may impact the interior temperature profiles. The required DFT calculations and their incorporation in our modeling framework are beyond the scope of this paper.

As mentioned in a few instances in the discussion, some models do not reach the RCB even at 1 kbar, which is a known issue in linking atmosphere and interior models \citep{kempton_where_2023,nixon_new_2024,aguichine_evolution_2024,selsis_steam_2024,cmiel_characterizing_2025}, and is observed in other self-consistent atmosphere models for TOI-270 d presented in the literature \citep{nixon_magma_2025,rigby_surface_2025}. However, we do not expect that this would impact our conclusion that demixing is possible on sub-Neptunes: instead, a deeper RCB would lead to an even colder interior in our models, which would be more prone to demixing.

For the prescription adopted to calculate the coexistence curves, we used the results from \citet{gupta_miscibility_2024}. Although other studies have explored hydrogen-water miscibility using either a parameterized force field instead of DFT, or DFT calculations but determining the solvus using free energies rather than phase coexistence, the reference we adopt provides a better match to experimental data and spans a wider range of temperatures down to 750\,K \citep{bergermann_gibbs-ensemble_2021,bergermann_ab_2024,soubiran_miscibility_2015}. Even if demixing is predicted by our models, we note that convective overshoot (e.g. \citealp{chabrier_heat_2007,korre_convective_2019,anders_convective_2023,tulekeyev_constraints_2024,sur_apple_2024}) may partially counteract it and transport metal-rich material to lower pressures. Future models should explore whether the depth of the metal-rich region and the mixing length could sustain envelope compositional homogeneity even when a H$_2$-H$_2$O mixed phase is thermodynamically disfavored.

In terms of the interior model, we only explore Earth-like ratios of iron and silicates in the interior, but the range of compositions consistent with demixing might extend further throughout the parameter space if studied over a wider range of $f'_\mathrm{core}$ values. Such diversity is motivated by pebble accretion models, which predict a time-varying partitioning of silicates between the envelope and the core over the course of a planet's evolution \citep{vazan_rocky_2023,vazan_how_2024}. On the other hand, the interior we model does not account for silicate-hydrogen miscibility which could lower the density of the mantle \citep{rogers_redefining_2025,young_differentiation_2025} and reduce the range of envelope compositions that reach the deep pressures required for demixing above the MEB. Besides, one of the largest sources of uncertainty in interior models lies in the assumed material properties and the choice of EOS, and our conclusions may be impacted by future improvements to e.g. the hydrogen or water EOS, if they substantially impact the densities and adiabatic gradients over the T-P range we consider.

\section{Summary and Conclusions} \label{sec:conclusion}


In this work, we performed an exploration of the extent to which water-hydrogen demixing can impact the compositions of sub-Neptune envelopes, for planets too warm for water condensation. We used the ATHENAIA framework introduced in this work to explore compositions ranging from solar-metallicity, H$_2$/He-dominated all the way to pure-H$_2$O using coupled atmosphere-interior models, and evaluated the compatibility of the envelope thermodynamic conditions with fully-mixed compositions for irradiation levels comparable to TOI-270 d. 

We find that even if more metal-rich atmospheres are hotter, their shallower adiabatic gradients in the deep envelope, combined with the shift of the hydrogen-water coexistence curve to higher temperatures when metallicities approach $200\times$ solar, open a window for demixing even if water is in the supercritical state on warm, metal-rich sub-Neptunes. 

We use our coupled interior-atmosphere models to infer the joint constraints on the envelope mass fractions and metallicities compatible with TOI-270 d's bulk properties, and find substantial overlap between the inferred properties of TOI-270 d's envelope and the demixing window, which also coincide with the measured atmosphere metallicity from JWST observations \citep{benneke_jwst_2024,holmberg_possible_2024}. Therefore, unless the true upper atmosphere metallicity is close to the lower end allowed by atmosphere constraints (near 100$\times$ solar), TOI-270 d's envelope structure could exhibit a compositional gradient rather than having a fully miscible envelope as previously proposed. This would imply an even larger bulk envelope metallicity than measured in the upper atmosphere. 

Overall, demixing affects less irradiated planets as well as more massive planets over a wider range of envelope mass fractions and metallicities, although it is predicted even on $\lesssim 3$\,$M_\oplus$ planets. We find that warm sub-Neptunes with equilibrium temperatures between approximately 330\,K and 450\,K are impacted throughout the mass-radius diagram, and increasingly so for high $\gtrsim 200 \times$ solar metallicities. The presence of methane in the envelope leads to shallower adiabatic gradients, thus potentially further extending the range of conditions where water-hydrogen demixing is possible.

Our work assumes the stellar host to be an M dwarf, which should lead to shallower radiative zones -- and therefore hotter interior conditions -- in H/He/H$_2$O sub-Neptune atmospheres compared to a solar-type incident stellar spectrum  \citep{kempton_where_2023}. Within the context of water-hydrogen demixing, this suggests that a broader range of atmospheric conditions may lead to stratified envelopes on sub-Neptunes around earlier-type stars. 

Applying our models to the evaluation of mantle-envelope boundary conditions on 5--20$M_\oplus$ sub-Neptunes, we find that using self-consistent atmosphere models as a boundary condition for deep-interior adiabats substantially reduces the composition dependence of silicate mantle solidification conditions on sub-Neptunes. Instead, envelope mass fraction or surface pressure emerges as the leading variable up to $Z\sim 0.8$, with lower envelope mass fractions required for mantle solidification for higher planet masses. Using these constraints, we revise to about 5 amu the minimum atmosphere mean molecular weight required for TOI-270\,d to have a solid mantle given its mass, radius, and temperature.

We provide a new set of open-source constant-composition mass-radius curves calculated with ATHENAIA, the first one to span the full range of compositions from H/He-dominated to H$_2$O-dominated. We also share the $Z_\mathrm{stable}/Z_\mathrm{env}$ metric (Fig. \ref{fig:stable_comp}) that encodes the envelope's stability to demixing throughout our model grid, as well as a script that can be used to draw the demixing contours on the $f_\mathrm{env}-Z_\mathrm{env}$ plane.

Although we identified demixing as a process that can impact even warm sub-Neptunes, thermal evolution calculations are necessary to track how the mass- and metallicity-dependence of the cooling rate in exoplanetary atmosphere changes this picture after Gyr of evolution. Identifying sub-Neptunes with truly fully-miscible envelopes, or quantifying the extent of the mismatch between upper atmosphere and bulk metallicities, will be crucial to interpret observations of sub-Neptune atmospheres in terms of their formation building blocks.

\section{Data Availability} 

The mass-radius curves, demixing contours, and stable metallicity values calculated in this work can be accessed and downloaded on Zenodo\footnote{\url{https://doi.org/10.5281/zenodo.18937844}}.

\paragraph{Acknowledgments}
We thank the anonymous referee for their insightful comments. C. P.-G. thanks Hamish Innes, Raymond Pierrehumbert, Thaddeus Komacek, and Joseph Schools for helpful conversations on the topic of hydrogen-water miscibility and mantle melting, and M. Nixon for providing the phase boundaries of silicate. This work is based on observations with the NASA/ESA/CSA James Webb Space Telescope, obtained at the Space Telescope Science Institute (STScI) operated by AURA, Inc. This research has made use of the NASA Exoplanet Archive, which is operated by Caltech, under contract with the National Aeronautics and Space Administration under the Exoplanet Exploration Program. C.P.-G. acknowledges support from the E. Margaret Burbidge Prize Postdoctoral Fellowship from the Brinson Foundation, and from the Suzuki Postdoctoral Fellowship.

\software{PHOENIX \citep{husser_new_2013},
          \texttt{astropy}\footnote{\url{https://www.astropy.org/}} \citep{astropy_collaboration_astropy_2013,astropy_collaboration_astropy_2018,astropy_collaboration_astropy_2022},
        \texttt{numpy}\footnote{\url{https://github.com/numpy/numpy}} \citep{harris_array_2020},       \texttt{matplotlib}\footnote{\url{https://github.com/matplotlib/matplotlib}} \citep{hunter_matplotlib_2007}, \texttt{scipy}\footnote{\url{https://github.com/scipy/scipy}} \citep{virtanen_scipy_2020}
\texttt{astropy}\footnote{\url{https://github.com/astropy/astropy}} \citep{astropy_collaboration_astropy_2013}
\texttt{pytorch}\footnote{\url{https://github.com/pytorch/pytorch/}} \citep{paszke_pytorch_2019}
          }







\appendix
\setcounter{figure}{0}
\renewcommand{\thefigure}{A\arabic{figure}}
\setcounter{table}{0}
\renewcommand{\thetable}{A\arabic{table}}

\section{Atmosphere temperature profile determination}\label{sec:TP_convergence}

To compute the atmospheric temperature profile in radiative-convective equilibrium with SCARLET, we employ an iterative solver following the \citet{toon_rapid_1989} linearization approach, adapted for non-gray opacities and flexible convection handling. 
The method alternates between radiative flux optimization and convective adjustment, and we implemented \texttt{pytorch}-based GPU acceleration \citep{paszke_pytorch_2019} to employ vectorization for the most computationally-expensive steps.

\paragraph{Radiative Equilibrium Iteration}

Starting from an initial-guess temperature-pressure (T-P) profile, and repeating after each new T-P profile iteration, we perform the hydrostatic equilibrium calculation and compute the opacity terms. We calculate the upward and downward thermal fluxes by solving two-stream, multi-scattering radiative transfer by adapting the routine implemented in \textsc{PICASO} \citep{batalha_exoplanet_2019} for the \citet{toon_rapid_1989} method. 

The resulting radiative fluxes are used to calculate local heating (i.e., flux divergence). The linearization step is performed by constructing the Jacobian of net fluxes with respect to temperature, exploiting automatic differentiation with \texttt{pytorch} for efficient computation of the Jacobian matrix. The temperature correction is then obtained by solving the resulting linear system, with adaptive temperature step size reduction for stability. Radiative equilibrium is considered achieved when per-layer temperature changes and the number of unconverged layers meet  thresholds of proposed temperature changes of less than 10$^{-3}$\,K. 

\paragraph{Convective Adjustment}
Superadiabatic layers are identified after the radiative-equilibrium T-P profile update. We compare the temperature gradient to the adiabatic gradient, $\nabla_\mathrm{ad}$, computed at the same composition by the interior modeling code  and interpolated at the atmosphere model layer's temperature and pressure (linear interpolation in the temperature dimension, and log interpolation in the pressure dimension).

The temperature profile is iteratively modified with temperatures in the superadiabatic layers are recursively adjusted to the environmental lapse rate, since superadiabatic temperature gradients are not expected in the absence of compositional gradients. The process is repeated until no superadiabatic layer remains, and we iterate over the radiative transfer and hydrostatic equilibrium steps, adjusting the temperature in the non-convective layers until  both radiative and convective processes self-consistently reach equilibrium.

\setcounter{figure}{0}
\renewcommand{\thefigure}{B\arabic{figure}}
\setcounter{table}{0}
\renewcommand{\thetable}{B\arabic{table}}

\section{Additional model grid products}

\subsection{Envelope temperature profiles}

\begin{figure}
    \centering
    \includegraphics[width=0.95\textwidth]{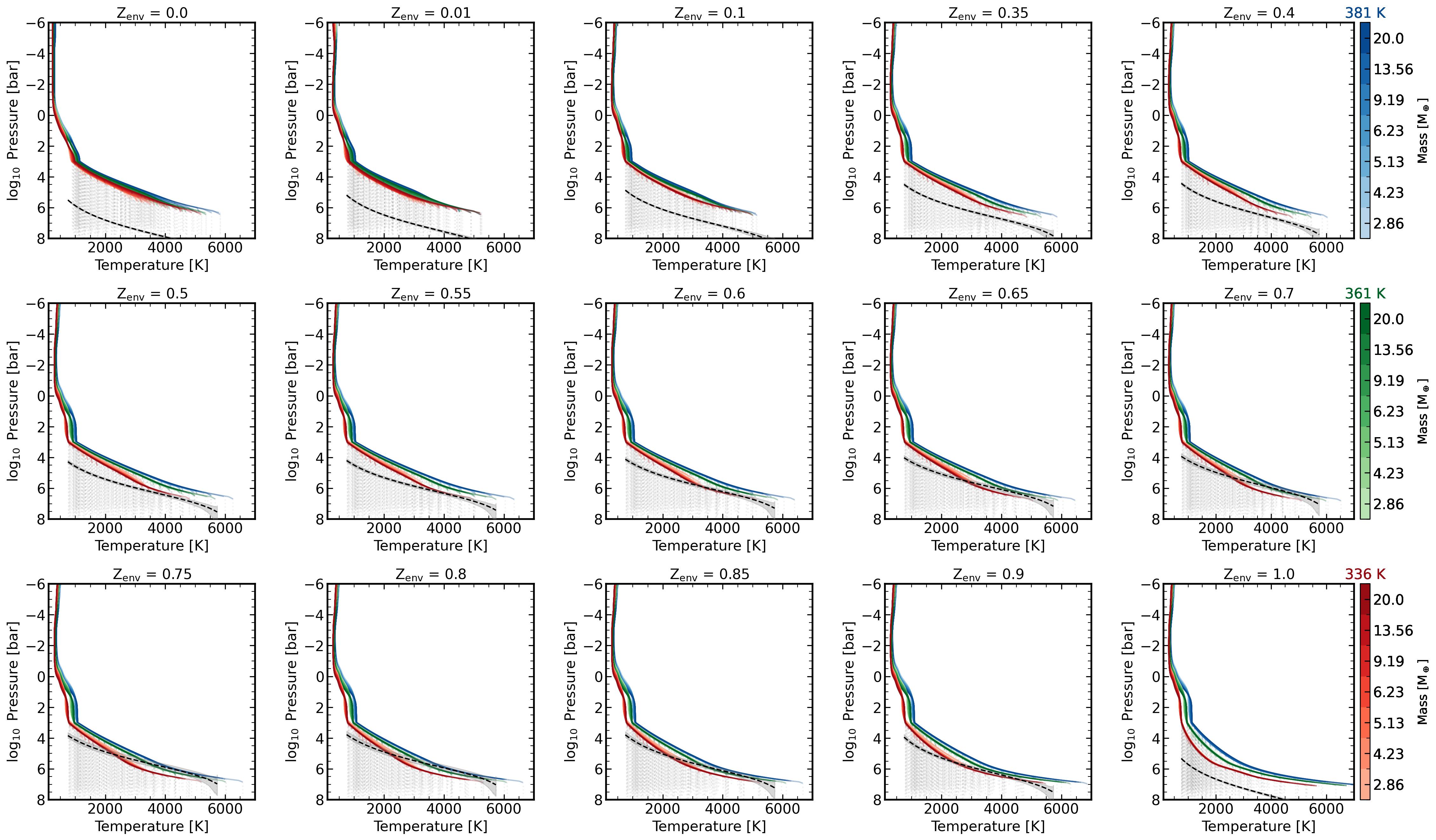}

    \caption{Temperature-pressure profile for the envelope (colors) and interior (dotted, gray) across all conditions in the grid, compared to the coexistence curves (median is black dashed, and 95\% C.I. shown as gray shading), for different envelope metallicities (labeled for each panel). The color shadings encode the planet mass for the three temperatures explored in the grid (shades of blue for $T_\mathrm{eq}=381$\,K, green for $T_\mathrm{eq}=361$\,K, red for $T_\mathrm{eq}=336$\,K). The vertical and horizontal span in temperature-pressure space is kept constant throughout all panels to illustrate the compositional dependence of the demixing window.
}
    \label{fig:tp_zenv}
\end{figure}

We illustrate in Fig. \ref{fig:tp_zenv} the range of temperature-pressure conditions spanned by our models across the grid for all masses, metallicities, albedos, and envelope mass fractions.

\subsection{Estimate of stable upper-atmosphere metallicities}\label{sec:stable_Z}

\begin{figure}
    \centering
    \includegraphics[width=0.98\textwidth]{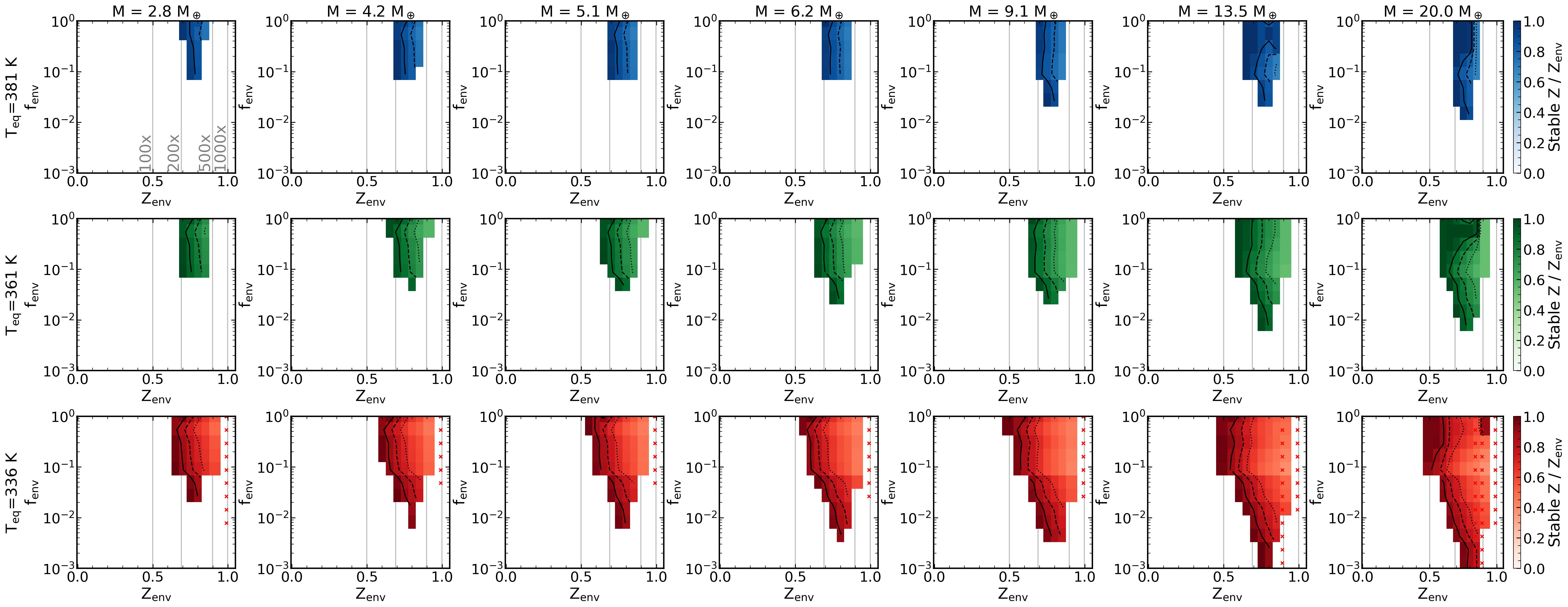}

    \caption{Metal depletion in the upper envelope relative to the bulk envelope metallicity, throughout the demixing window (where the stable Z is lower than the bulk Z), as a function of equilibrium temperature (different rows, corresponding to a Bond albedo of 0., 0.2, 0.4 at TOI-270 d's orbital distance) and planet mass (different columns). Vertical grey lines are added to guide interpretation in terms of the envelope metallicity relative to solar (labeled in the first panel). Not only does the demixing window expand in both dimensions of envelope mass fraction and envelope metallicity as the insolation decreases and planetary mass increases, but the extent of expected metal depletion in the upper envelope simultaneously becomes more severe. Black contours indicate upper envelope metal depletion (in units of Z) of 10\% (solid), 20\% (dashed), and 30\% (dotted). Profiles where water condensation is predicted in the upper atmosphere are indicated with crosses. The area where no demixing is predicted in our models is left in white.
}
    \label{fig:stable_comp}
\end{figure}

We calculate a simple estimate of the extent of demixing in the upper atmosphere by computing the value of $Z$ where the temperature at the first unstable point from the top of the atmosphere (pressure level where the coexistence curve is crossed) would instead remain stable against demixing. We visualize this ``stable $Z$'' value ($Z_\mathrm{stable}$) for the upper envelope over the parameter space where our model predicts demixing (Fig. \ref{fig:stable_comp}).

Although this informs the magnitude of upper-atmosphere demixing, in reality, the temperature profile and ambient metallicity should be iterated over until the metals have been spread across the envelope in a way that stability against demixing is achieved in each atmosphere layer \citep{amoros_h_2-h_2o_2024}, and this exploration should only be considered as illustrative rather than a quantitative assessment, which is beyond the scope of the present study. 









\subsection{Exploring the limits of the demixing window}\label{sec:dmx_limits}

\begin{figure}
    \centering
    \includegraphics[width=0.98\textwidth]{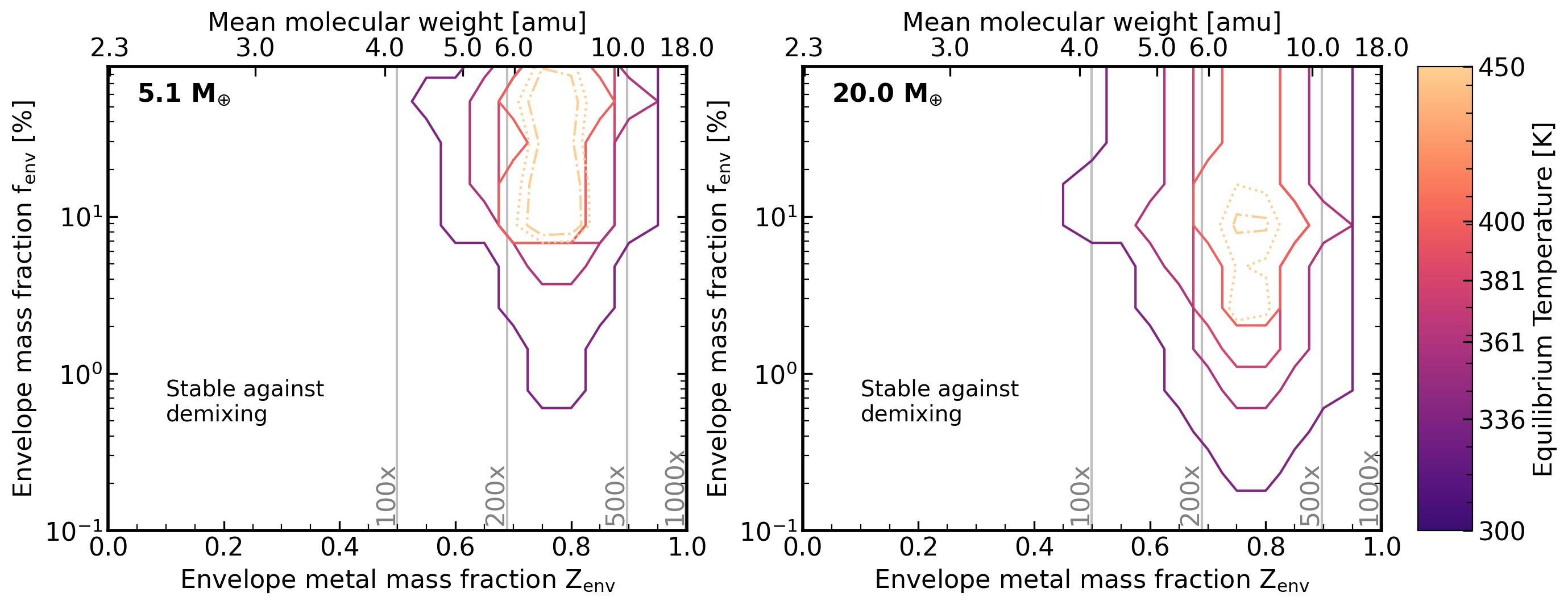}

    \caption{Envelope stability to demixing in the $f_\mathrm{env}-Z_\mathrm{env}$ plane for varying planetary equilibrium temperatures (colors) for a planet mass of 5.1 $M_\oplus$ (left) or 20 $M_\oplus$ (right). The solid contours delineate the region of parameter space where fully-mixed compositions are unstable. For $T_\mathrm{eq}=450\,K$, we instead delineate contours within 250\,K (150\,K) of the H-H$_2$O composition-dependent coexistence curve with dotted (dash-dotted) contours. Such temperature differences are smaller than the typical uncertainty on the temperature of the coexistence curve at a fixed pressure from molecular dynamics simulations (see Fig. \ref{fig:tp_zenv}).
}
    \label{fig:dmx_vs_Teq}
\end{figure}

For two planet masses (5.1$M_\oplus$ and 20$M_\oplus$), we calculate the full planet grid (across the dimensions of $Z_\mathrm{env}$, $f_\mathrm{env}$) for $T_{eq}$ of 400\,K and 450\,K. The results from these two additional model grids are shown on Figure \ref{fig:dmx_vs_Teq}.

\newpage
\setcounter{figure}{0}
\renewcommand{\thefigure}{C\arabic{figure}}
\setcounter{table}{0}
\renewcommand{\thetable}{C\arabic{table}}

\section{Phase of silicates at the mantle-envelope boundary }\label{sec:melt_limits}

\begin{figure*}
    \centering
    \includegraphics[width=0.85\textwidth]{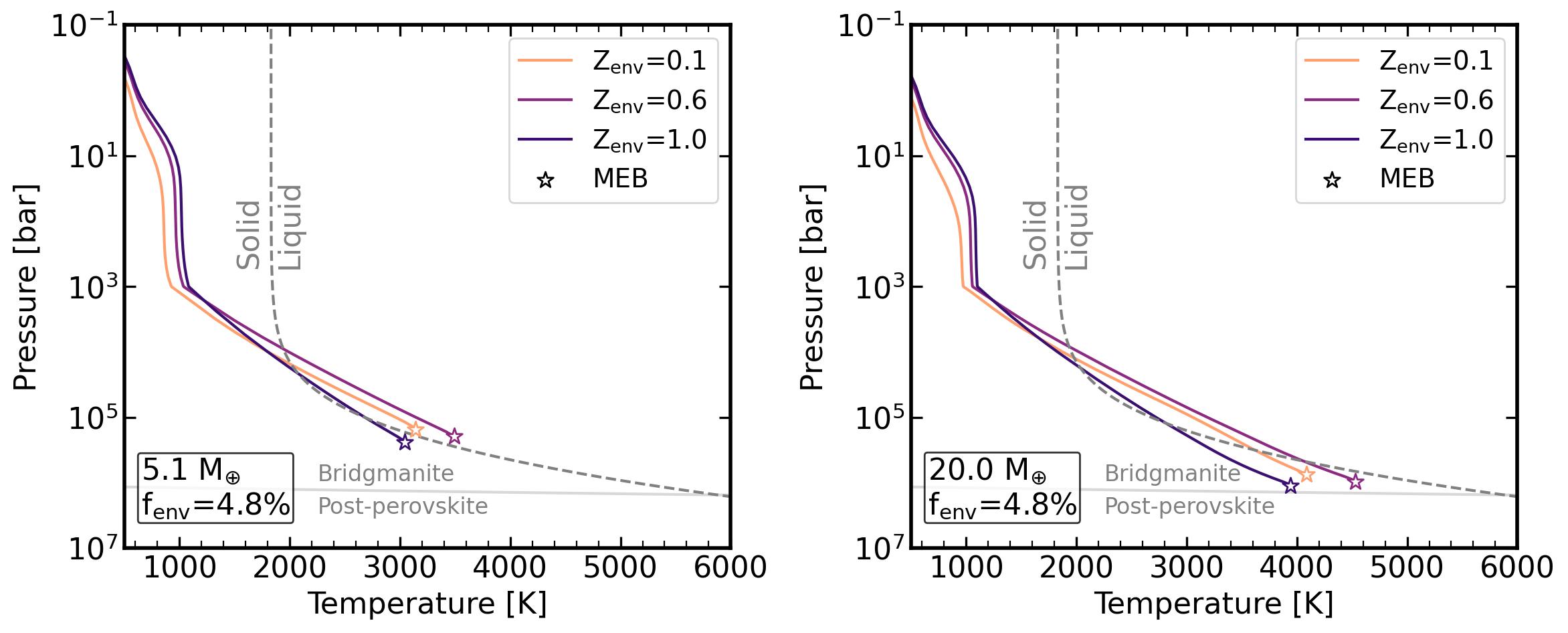}
\caption{Illustration of the competition between increasing atmospheric greenhouse effect and decreasing adiabatic temperature gradients for increasing envelope metallicity, and its impact on the phase expected for the silicate mantle at the boundary. Envelope temperature profiles are shown down to the MEB (star marker) calculated at the zero-Bond albedo equilibrium temperature of TOI-270\,d (381\,K) for two planet masses (left: 5.1\,$M_\oplus$, right: 20\,$M_\oplus$), a fixed envelope mass fraction of 4.8\%, and three metallicities (colors). The phases of SiO$_3$ are delineated in gray: the position of each star marker in pressure-temperature space dictates the inferred mantle conditions at the base of the envelope. Going from $Z_\mathrm{env}=0.1$ to 0.6, the MEB occurs at a higher temperature primarily because of the more pronounced greenhouse heating of the atmosphere (down to 1 kbar), which maintains it within the range of conditions where a molten mantle is expected for the 5.1\,$M_\oplus$ case. Increasing further to a pure-water composition of $Z_\mathrm{env}=1.0$, the higher heating of the deep atmosphere is compensated by a much shallower slope of the deep-envelope adiabat, leading to a prediction of solid bridgmanite. For the 20\,$M_\oplus$ case, we predict higher pressures at the MEB which result in a solid state, illustrating how molten conditions are progressively restricted to lower envelope fractions with increasing planet mass (see Fig. \ref{fig:melt_vs_Teq}).
}
    
    \label{fig:melt_illustrate_profiles}
\end{figure*}

\begin{figure}
    \centering
    \includegraphics[width=0.98\textwidth]{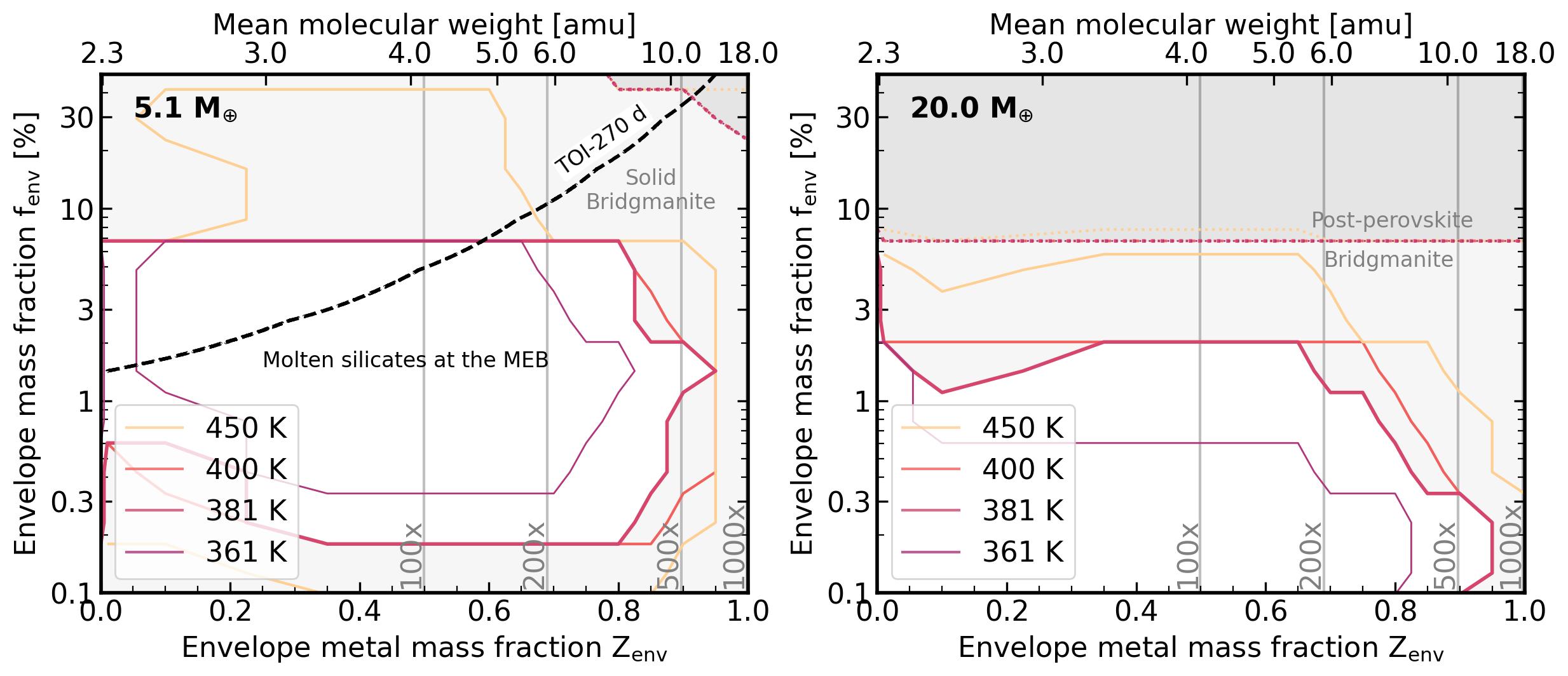}

\caption{Expected phase of MgSiO$_3$ at the mantle-envelope boundary (MEB) across ATHENAIA model grids with either liquid (colored contours encoding the equilibrium temperature) or solid state (gray area shows the $T_\mathrm{eq}=381$\,K limits). The left (right) panel corresponds to a planet mass of 5.1\,$M_\oplus$ (20\,$M_\oplus$). No contours are shown for the grid at $T_\mathrm{eq}=336$\,K because all models (except a handful at 2.8\,$M_\oplus$) have solid-rock MEB conditions. The transition from bridgmanite to post-perovksite is indicated with dotted lines. The contour closest to the zero-albedo equilibrium temperature of TOI-270 d is highlighted with a thicker line, and the corresponding maximum-probability set of inferred envelope mass fractions and metallicities for TOI-270 d are shown as a dashed line on top of the 5.1\,$M_\oplus$ model results (left panel). TOI-270 d's mantle is predicted to be molten at least up to an atmosphere metallicity of $\sim 150\times$solar. For higher planetary masses, a solid bridgmanite or post-perovksite states are reached at lower envelope mass fractions because of the higher MEB pressures (see Fig. \ref{fig:melt_illustrate_profiles}).
}
    \label{fig:melt_vs_Teq}
\end{figure}

We evaluate the conditions at the mantle-envelope boundary across our model grids. The impact of the competition between a greater greenhouse effect in the atmosphere and a shallower temperature gradient in the interior for increasing metallicity is illustrated in Fig. \ref{fig:melt_illustrate_profiles}, and the results over the entire grid of compositions for two planet masses are displayed on Figure \ref{fig:melt_vs_Teq}. 

\bibliography{references_clean}{}

@ARTICLE{NasaExoArchive2025,
       author = {{Christiansen}, Jessie L. and {McElroy}, Douglas L. and {Harbut}, Marcy and {Ciardi}, David R. and {Crane}, Megan and {Good}, John and {Hardegree-Ullman}, Kevin K. and {Kesseli}, Aurora Y. and {Lund}, Michael B. and {Lynn}, Meca and et al.},
        title = "{The NASA Exoplanet Archive and Exoplanet Follow-up Observing Program: Data, Tools, and Usage}",
      journal = {\psj},
         year = 2025,
        month = aug,
       volume = {6},
       number = {8},
          eid = {186},
        pages = {186},
          doi = {10.3847/PSJ/ade3c2},
archivePrefix = {arXiv},
       eprint = {2506.03299},
 primaryClass = {astro-ph.EP},
       adsurl = {https://ui.adsabs.harvard.edu/abs/2025PSJ.....6..186C}
}

@ARTICLE{NasaExoArchive2013,
       author = {{Akeson}, R.~L. and {Chen}, X. and {Ciardi}, D. and {Crane}, M. and {Good}, J. and {Harbut}, M. and {Jackson}, E. and {Kane}, S.~R. and {Laity}, A.~C. and {Leifer}, S. and et al.},
        title = "{The NASA Exoplanet Archive: Data and Tools for Exoplanet Research}",
      journal = {\pasp},
         year = 2013,
        month = aug,
       volume = {125},
       number = {930},
        pages = {989},
          doi = {10.1086/672273},
archivePrefix = {arXiv},
       eprint = {1307.2944},
 primaryClass = {astro-ph.IM},
       adsurl = {https://ui.adsabs.harvard.edu/abs/2013PASP..125..989A}
}

@misc{Thompson90,
	title        = {ANEOS--Analytic Equations of State for Shock Physics Codes},
	author       = {S. L. Thompson},
	year         = 1990,
	location     = {Albuquerque, NM},
	number       = {SAND89-2951},
	institution  = {Sandia National Laboratories},
	type         = {Technical Report}
}

@article{harris_array_2020,
	title        = {Array programming with {NumPy}},
	author       = {Harris, Charles R. and Millman, K. Jarrod and van der Walt, Stéfan J. and Gommers, Ralf and Virtanen, Pauli and Cournapeau, David and Wieser, Eric and Taylor, Julian and Berg, Sebastian and Smith, Nathaniel J. and Kern, Robert and Picus, Matti and Hoyer, Stephan and van Kerkwijk, Marten H. and Brett, Matthew and Haldane, Allan and del Río, Jaime Fernández and Wiebe, Mark and Peterson, Pearu and Gérard-Marchant, Pierre and Sheppard, Kevin and Reddy, Tyler and Weckesser, Warren and Abbasi, Hameer and Gohlke, Christoph and Oliphant, Travis E.},
	year         = 2020,
	month        = sep,
	journal      = {Nature},
	volume       = 585,
	pages        = {357--362},
	doi          = {10.1038/s41586-020-2649-2},
	issn         = {0028-0836},
	url          = {https://ui.adsabs.harvard.edu/abs/2020Natur.585..357H},
	urldate      = {2024-06-20},
	note         = {ADS Bibcode: 2020Natur.585..357H}
}

@article{virtanen_scipy_2020,
	title        = {{SciPy 1.0: fundamental algorithms for scientific computing in Python}},
	author       = {{Virtanen}, Pauli and {Gommers}, Ralf and {Oliphant}, Travis E. and {Haberland}, Matt and {Reddy}, Tyler and {Cournapeau}, David and {Burovski}, Evgeni and {Peterson}, Pearu and {Weckesser}, Warren and {Bright}, Jonathan and {van der Walt}, St{\'e}fan J. and {Brett}, Matthew and {Wilson}, Joshua and {Millman}, K. Jarrod and {Mayorov}, Nikolay and {Nelson}, Andrew R.~J. and {Jones}, Eric and {Kern}, Robert and {Larson}, Eric and {Carey}, C.~J. and {Polat}, {\.I}lhan and {Feng}, Yu and {Moore}, Eric W. and {VanderPlas}, Jake and {Laxalde}, Denis and {Perktold}, Josef and {Cimrman}, Robert and {Henriksen}, Ian and {Quintero}, E.~A. and {Harris}, Charles R. and {Archibald}, Anne M. and {Ribeiro}, Ant{\^o}nio H. and {Pedregosa}, Fabian and {van Mulbregt}, Paul and {SciPy 1. 0 Contributors}},
	year         = 2020,
	month        = feb,
	journal      = {Nature Methods},
	volume       = 17,
	pages        = {261--272},
	doi          = {10.1038/s41592-019-0686-2},
	archiveprefix = {arXiv},
	eprint       = {1907.10121},
	primaryclass = {cs.MS},
	adsurl       = {https://ui.adsabs.harvard.edu/abs/2020NatMe..17..261V}
}

@article{gupta_signatures_2020,
	title        = {Signatures of the core-powered mass-loss mechanism in the exoplanet population: dependence on stellar properties and observational predictions},
	shorttitle   = {Signatures of the core-powered mass-loss mechanism in the exoplanet population},
	author       = {Gupta, Akash and Schlichting, Hilke E.},
	year         = 2020,
	month        = mar,
	journal      = {Monthly Notices of the Royal Astronomical Society},
	volume       = 493,
	pages        = {792--806},
	doi          = {10.1093/mnras/staa315},
	url          = {http://adsabs.harvard.edu/abs/2020MNRAS.493..792G},
	urldate      = {2020-11-07}
}

@article{benneke_sub-neptune_2019,
	title        = {A sub-{Neptune} exoplanet with a low-metallicity methane-depleted atmosphere and {Mie}-scattering clouds},
	author       = {Benneke, Björn and Knutson, Heather A. and Lothringer, Joshua and Crossfield, Ian J. M. and Moses, Julianne I. and Morley, Caroline and Kreidberg, Laura and Fulton, Benjamin J. and Dragomir, Diana and Howard, Andrew W. and Wong, Ian and Désert, Jean-Michel and McCullough, Peter R. and Kempton, Eliza M.-R. and Fortney, Jonathan and Gilliland, Ronald and Deming, Drake and Kammer, Joshua},
	year         = 2019,
	month        = sep,
	journal      = {Nature Astronomy},
	publisher    = {Nature Publishing Group},
	volume       = 3,
	number       = 9,
	pages        = {813--821},
	doi          = {10.1038/s41550-019-0800-5},
	issn         = {2397-3366},
	url          = {https://www.nature.com/articles/s41550-019-0800-5},
	urldate      = {2024-03-14},
	copyright    = {2019 The Author(s), under exclusive licence to Springer Nature Limited},
	language     = {en}
}

@article{aguichine_mass-radius_2021,
	title        = {Mass-radius relationships for irradiated ocean planets},
	author       = {Aguichine, Artyom and Mousis, Olivier and Deleuil, Magali and Marcq, Emmanuel},
	year         = 2021,
	month        = may,
	journal      = {arXiv e-prints},
	volume       = 2105,
	pages        = {arXiv:2105.01102},
	url          = {http://adsabs.harvard.edu/abs/2021arXiv210501102A},
	urldate      = {2021-05-06}
}

@article{neil_evaluating_2022,
	title        = {Evaluating the {Evidence} for {Water} {World} {Populations} using {Mixture} {Models}},
	author       = {Neil, Andrew R. and Liston, Jessica and Rogers, Leslie A.},
	year         = 2022,
	month        = apr,
	journal      = {arXiv:2205.00006 [astro-ph]},
	url          = {http://arxiv.org/abs/2205.00006},
	urldate      = {2022-05-04},
	note         = {arXiv: 2205.00006}
}

@misc{pierrehumbert_runaway_2022,
	title        = {The runaway greenhouse on {subNeptune} waterworlds},
	author       = {Pierrehumbert, Raymond T.},
	year         = 2022,
	month        = dec,
	publisher    = {arXiv},
	doi          = {10.48550/arXiv.2212.02644},
	url          = {http://arxiv.org/abs/2212.02644},
	urldate      = {2022-12-07},
	note         = {arXiv:2212.02644 [astro-ph]}
}

@article{owen_evaporation_2017,
	title        = {The {Evaporation} {Valley} in the {Kepler} {Planets}},
	author       = {Owen, James E. and Wu, Yanqin},
	year         = 2017,
	month        = sep,
	journal      = {The Astrophysical Journal},
	volume       = 847,
	pages        = 29,
	doi          = {10.3847/1538-4357/aa890a},
	issn         = {0004-637X},
	url          = {http://adsabs.harvard.edu/abs/2017ApJ...847...29O},
	urldate      = {2020-02-05}
}

@misc{rigby_surface_2025,
	title        = {The {Surface} and {Interior} {Conditions} of {Temperate} {Sub}-{Neptune} {TOI}-270 d},
	author       = {Rigby, Frances E. and Madhusudhan, Nikku},
	year         = 2025,
	month        = nov,
	publisher    = {arXiv},
	doi          = {10.48550/arXiv.2511.16722},
	url          = {http://arxiv.org/abs/2511.16722},
	urldate      = {2025-11-24},
	note         = {arXiv:2511.16722 [astro-ph]}
}

@article{young_differentiation_2025,
	title        = {Differentiation, the exception not the rule -- {Evidence} for full miscibility in sub-{Neptune} interiors},
	author       = {Young, Edward D. and Werlen, Aaron and Marcum, Sarah P. and Dullemond, Cornelis P.},
	year         = 2025,
	month        = jul,
	publisher    = {arXiv},
	doi          = {10.48550/arXiv.2507.00947},
	url          = {http://arxiv.org/abs/2507.00947},
	urldate      = {2025-07-02},
	note         = {arXiv:2507.00947 [astro-ph]}
}

@article{bergermann_gibbs-ensemble_2021,
	title        = {Gibbs-ensemble {Monte} {Carlo} simulation of {H2}–{H2O} mixtures},
	author       = {Bergermann, Armin and French, Martin and Redmer, Ronald},
	year         = 2021,
	month        = jun,
	journal      = {Physical Chemistry Chemical Physics},
	publisher    = {The Royal Society of Chemistry},
	volume       = 23,
	number       = 22,
	pages        = {12637--12643},
	doi          = {10.1039/D1CP00515D},
	issn         = {1463-9084},
	url          = {https://pubs.rsc.org/en/content/articlelanding/2021/cp/d1cp00515d},
	urldate      = {2025-11-21},
	language     = {en}
}

@article{soubiran_miscibility_2015,
	title        = {{MISCIBILITY} {CALCULATIONS} {FOR} {WATER} {AND} {HYDROGEN} {IN} {GIANT} {PLANETS}},
	author       = {Soubiran, François and Militzer, Burkhard},
	year         = 2015,
	month        = jun,
	journal      = {The Astrophysical Journal},
	publisher    = {The American Astronomical Society},
	volume       = 806,
	number       = 2,
	pages        = 228,
	doi          = {10.1088/0004-637X/806/2/228},
	issn         = {0004-637X},
	url          = {https://dx.doi.org/10.1088/0004-637X/806/2/228},
	urldate      = {2024-05-16},
	language     = {en}
}

@article{peng_puffy_2024,
	title        = {Puffy {Venuses}: the {Mass}-{Radius} {Impact} of {Carbon}-{Rich} {Atmospheres} on {Lava} {Worlds}},
	shorttitle   = {Puffy {Venuses}},
	author       = {Peng, Bo and Valencia, Diana},
	year         = 2024,
	month        = may,
	publisher    = {arXiv},
	url          = {http://arxiv.org/abs/2405.08998},
	urldate      = {2024-05-16},
	note         = {arXiv:2405.08998 [astro-ph]}
}

@article{guillot_radiative_2010,
	title        = {On the radiative equilibrium of irradiated planetary atmospheres},
	author       = {Guillot, T.},
	year         = 2010,
	month        = sep,
	journal      = {Astronomy and Astrophysics},
	publisher    = {EDP},
	volume       = 520,
	pages        = {A27},
	doi          = {10.1051/0004-6361/200913396},
	issn         = {0004-6361},
	url          = {https://ui.adsabs.harvard.edu/abs/2010A&A...520A..27G},
	urldate      = {2025-11-21},
	note         = {ADS Bibcode: 2010A\&A...520A..27G}
}

@misc{rogers_redefining_2025,
	title        = {Redefining interiors and envelopes: hydrogen-silicate miscibility and its consequences for the structure and evolution of sub-{Neptunes}},
	shorttitle   = {Redefining interiors and envelopes},
	author       = {Rogers, James G. and Young, Edward D. and Schlichting, Hilke E.},
	year         = 2025,
	month        = sep,
	publisher    = {arXiv},
	doi          = {10.48550/arXiv.2509.13320},
	url          = {http://arxiv.org/abs/2509.13320},
	urldate      = {2025-09-17},
	note         = {arXiv:2509.13320 [astro-ph]}
}

@article{katz_new_2003,
	title        = {A new parameterization of hydrous mantle melting},
	author       = {Katz, Richard F. and Spiegelman, Marc and Langmuir, Charles H.},
	year         = 2003,
	journal      = {Geochemistry, Geophysics, Geosystems},
	volume       = 4,
	number       = 9,
	doi          = {10.1029/2002GC000433},
	issn         = {1525-2027},
	url          = {https://onlinelibrary.wiley.com/doi/abs/10.1029/2002GC000433},
	urldate      = {2024-06-10},
	copyright    = {Copyright 2003 by the American Geophysical Union.},
	note         = {\_eprint: https://onlinelibrary.wiley.com/doi/pdf/10.1029/2002GC000433},
	language     = {en}
}

@article{nettelmann_ab_2008,
	title        = {Ab {Initio} {Equation} of {State} {Data} for {Hydrogen}, {Helium}, and {Water} and the {Internal} {Structure} of {Jupiter}},
	author       = {Nettelmann, Nadine and Holst, Bastian and Kietzmann, André and French, Martin and Redmer, Ronald and Blaschke, David},
	year         = 2008,
	month        = aug,
	journal      = {The Astrophysical Journal},
	publisher    = {IOP},
	volume       = 683,
	pages        = {1217--1228},
	doi          = {10.1086/589806},
	issn         = {0004-637X},
	url          = {https://ui.adsabs.harvard.edu/abs/2008ApJ...683.1217N},
	urldate      = {2025-11-19},
	note         = {ADS Bibcode: 2008ApJ...683.1217N}
}

@phdthesis{thorngren_bayesian_2019,
	title        = {Bayesian {Statistical} {Inference} of {Giant} {Planet} {Physics}},
	author       = {Thorngren, Daniel P.},
	year         = 2019,
	month        = jan,
	url          = {https://ui.adsabs.harvard.edu/abs/2019PhDT........51T},
	urldate      = {2025-11-19},
	note         = {ADS Bibcode: 2019PhDT........51T},
	type         = {Ph.{D}. thesis}
}

@misc{paszke_pytorch_2019,
	title        = {{PyTorch}: {An} {Imperative} {Style}, {High}-{Performance} {Deep} {Learning} {Library}},
	shorttitle   = {{PyTorch}},
	author       = {Paszke, Adam and Gross, Sam and Massa, Francisco and Lerer, Adam and Bradbury, James and Chanan, Gregory and Killeen, Trevor and Lin, Zeming and Gimelshein, Natalia and Antiga, Luca and Desmaison, Alban and Köpf, Andreas and Yang, Edward and DeVito, Zach and Raison, Martin and Tejani, Alykhan and Chilamkurthy, Sasank and Steiner, Benoit and Fang, Lu and Bai, Junjie and Chintala, Soumith},
	year         = 2019,
	month        = dec,
	publisher    = {arXiv},
	doi          = {10.48550/arXiv.1912.01703},
	url          = {http://arxiv.org/abs/1912.01703},
	urldate      = {2025-11-19},
	note         = {arXiv:1912.01703 [cs]}
}

@article{toon_rapid_1989,
	title        = {Rapid calculation of radiative heating rates and photodissociation rates in inhomogeneous multiple scattering atmospheres},
	author       = {Toon, Owen B. and McKay, C. P. and Ackerman, T. P. and Santhanam, K.},
	year         = 1989,
	month        = nov,
	journal      = {Journal of Geophysical Research},
	publisher    = {Wiley},
	volume       = 94,
	pages        = {16287--16301},
	doi          = {10.1029/JD094iD13p16287},
	issn         = {0148-0227},
	url          = {https://ui.adsabs.harvard.edu/abs/1989JGR....9416287T},
	urldate      = {2025-11-19},
	note         = {ADS Bibcode: 1989JGR....9416287T}
}

@article{batalha_exoplanet_2019,
	title        = {Exoplanet {Reflected} {Light} {Spectroscopy} with {PICASO}},
	author       = {Batalha, Natasha E. and Marley, Mark S. and Lewis, Nikole K. and Fortney, Jonathan J.},
	year         = 2019,
	month        = jun,
	journal      = {The Astrophysical Journal},
	volume       = 878,
	number       = 1,
	pages        = 70,
	doi          = {10.3847/1538-4357/ab1b51},
	issn         = {0004-637X, 1538-4357},
	url          = {http://arxiv.org/abs/1904.09355},
	urldate      = {2025-11-19},
	note         = {arXiv:1904.09355 [astro-ph]}
}

@misc{constantinou_atmospheric_2025,
	title        = {The atmospheric composition of {TOI}-270 d},
	author       = {Constantinou, Savvas and Madhusudhan, Nikku and Holmberg, Måns},
	year         = 2025,
	month        = nov,
	publisher    = {arXiv},
	doi          = {10.48550/arXiv.2511.13830},
	url          = {http://arxiv.org/abs/2511.13830},
	urldate      = {2025-11-19},
	note         = {arXiv:2511.13830 [astro-ph]}
}

@article{sakai_experimental_2016,
	title        = {Experimental and theoretical thermal equations of state of {MgSiO3} post-perovskite at multi-megabar pressures},
	author       = {Sakai, Takeshi and Dekura, Haruhiko and Hirao, Naohisa},
	year         = 2016,
	month        = mar,
	journal      = {Scientific Reports},
	volume       = 6,
	pages        = 22652,
	doi          = {10.1038/srep22652},
	issn         = {2045-2322},
	url          = {https://ui.adsabs.harvard.edu/abs/2016NatSR...622652S},
	urldate      = {2025-11-11},
	note         = {ADS Bibcode: 2016NatSR...622652S}
}

@article{oganov_theoretical_2004,
	title        = {Theoretical and experimental evidence for a post-perovskite phase of {MgSiO3} in {Earth}'s {D}'' layer},
	author       = {Oganov, Artem R. and Ono, Shigeaki},
	year         = 2004,
	month        = jul,
	journal      = {Nature},
	volume       = 430,
	pages        = {445--448},
	doi          = {10.1038/nature02701},
	issn         = {0028-0836},
	url          = {https://ui.adsabs.harvard.edu/abs/2004Natur.430..445O},
	urldate      = {2025-11-11},
	note         = {ADS Bibcode: 2004Natur.430..445O}
}

@article{more_new_1988,
	title        = {A new quotidian equation of state ({QEOS}) for hot dense matter},
	author       = {More, R. M. and Warren, K. H. and Young, D. A. and Zimmerman, G. B.},
	year         = 1988,
	month        = oct,
	journal      = {Physics of Fluids},
	publisher    = {AIP},
	volume       = 31,
	pages        = {3059--3078},
	doi          = {10.1063/1.866963},
	issn         = {0899-82131070-6631},
	url          = {https://ui.adsabs.harvard.edu/abs/1988PhFl...31.3059M},
	urldate      = {2025-11-07},
	note         = {ADS Bibcode: 1988PhFl...31.3059M}
}

@article{eylen_masses_2021,
	title        = {Masses and compositions of three small planets orbiting the nearby {M} dwarf {L231}-32 ({TOI}-270) and the {M} dwarf radius valley},
	author       = {Eylen, Vincent Van and Astudillo-Defru, N. and Bonfils, X. and Livingston, J. and Hirano, T. and Luque, R. and Lam, K. W. F. and Justesen, A. B. and Winn, J. N. and Gandolfi, D. and Nowak, G. and Palle, E. and Albrecht, S. and Dai, F. and Estrada, B. Campos and Owen, J. E. and Foreman-Mackey, D. and Fridlund, M. and Korth, J. and Mathur, S. and Forveille, T. and Mikal-Evans, T. and Osborne, H. L. M. and Ho, C. S. K. and Almenara, J. M. and Artigau, E. and Barragán, O. and Barros, S. C. C. and Bouchy, F. and Cabrera, J. and Caldwell, D. A. and Charbonneau, D. and Chaturvedi, P. and Cochran, W. D. and Csizmadia, S. and Damasso, M. and Delfosse, X. and Medeiros, J. R. De and Díaz, R. F. and Doyon, R. and Esposito, M. and Fűrész, G. and Figueira, P. and Georgieva, I. and Goffo, E. and Grziwa, S. and Guenther, E. and Hatzes, A. P. and Jenkins, J. M. and Kabath, P. and Knudstrup, E. and Latham, D. W. and Lavie, B. and Lovis, C. and Mennickent, R. E. and Mullally, S. E. and Murgas, F. and Narita, N. and Pepe, F. A. and Persson, C. M. and Redfield, S. and Ricker, G. R. and Santos, N. C. and Seager, S. and Serrano, L. M. and Smith, A. M. S. and Mascareño, A. Suárez and Subjak, J. and Twicken, J. D. and Udry, S. and Vanderspek, R. and Osorio, M. R. Zapatero},
	year         = 2021,
	month        = aug,
	journal      = {Monthly Notices of the Royal Astronomical Society},
	pages        = {stab2143},
	doi          = {10.1093/mnras/stab2143},
	issn         = {0035-8711, 1365-2966},
	url          = {http://arxiv.org/abs/2101.01593},
	urldate      = {2025-11-07},
	note         = {arXiv:2101.01593 [astro-ph]}
}

@article{bergermann_ab_2024,
	title        = {Ab initio calculation of the miscibility diagram for mixtures of hydrogen and water},
	author       = {Bergermann, Armin and French, Martin and Redmer, Ronald},
	year         = 2024,
	month        = may,
	journal      = {Physical Review B},
	publisher    = {APS},
	volume       = 109,
	pages        = 174107,
	doi          = {10.1103/PhysRevB.109.174107},
	issn         = {0163-18291098-0121},
	url          = {https://ui.adsabs.harvard.edu/abs/2024PhRvB.109q4107B},
	urldate      = {2025-11-07},
	note         = {ADS Bibcode: 2024PhRvB.109q4107B}
}

@article{batalha_planetary_2013,
	title        = {{PLANETARY} {CANDIDATES} {OBSERVED} {BY} {KEPLER}. {III}. {ANALYSIS} {OF} {THE} {FIRST} 16 {MONTHS} {OF} {DATA}},
	author       = {Batalha, Natalie M. and Rowe, Jason F. and Bryson, Stephen T. and Barclay, Thomas and Burke, Christopher J. and Caldwell, Douglas A. and Christiansen, Jessie L. and Mullally, Fergal and Thompson, Susan E. and Brown, Timothy M. and Dupree, Andrea K. and Fabrycky, Daniel C. and Ford, Eric B. and Fortney, Jonathan J. and Gilliland, Ronald L. and Isaacson, Howard and Latham, David W. and Marcy, Geoffrey W. and Quinn, Samuel N. and Ragozzine, Darin and Shporer, Avi and Borucki, William J. and Ciardi, David R. and Gautier, Thomas N. and Haas, Michael R. and Jenkins, Jon M. and Koch, David G. and Lissauer, Jack J. and Rapin, William and Basri, Gibor S. and Boss, Alan P. and Buchhave, Lars A. and Carter, Joshua A. and Charbonneau, David and Christensen-Dalsgaard, Joergen and Clarke, Bruce D. and Cochran, William D. and Demory, Brice-Olivier and Desert, Jean-Michel and Devore, Edna and Doyle, Laurance R. and Esquerdo, Gilbert A. and Everett, Mark and Fressin, Francois and Geary, John C. and Girouard, Forrest R. and Gould, Alan and Hall, Jennifer R. and Holman, Matthew J. and Howard, Andrew W. and Howell, Steve B. and Ibrahim, Khadeejah A. and Kinemuchi, Karen and Kjeldsen, Hans and Klaus, Todd C. and Li, Jie and Lucas, Philip W. and Meibom, Søren and Morris, Robert L. and Prša, Andrej and Quintana, Elisa and Sanderfer, Dwight T. and Sasselov, Dimitar and Seader, Shawn E. and Smith, Jeffrey C. and Steffen, Jason H. and Still, Martin and Stumpe, Martin C. and Tarter, Jill C. and Tenenbaum, Peter and Torres, Guillermo and Twicken, Joseph D. and Uddin, Kamal and Van Cleve, Jeffrey and Walkowicz, Lucianne and Welsh, William F.},
	year         = 2013,
	month        = feb,
	journal      = {The Astrophysical Journal Supplement Series},
	publisher    = {The American Astronomical Society},
	volume       = 204,
	number       = 2,
	pages        = 24,
	doi          = {10.1088/0067-0049/204/2/24},
	issn         = {0067-0049},
	url          = {https://doi.org/10.1088/0067-0049/204/2/24},
	urldate      = {2025-11-07},
	language     = {en}
}

@article{howard_planet_2012,
	title        = {Planet {Occurrence} within 0.25 {AU} of {Solar}-type {Stars} from {Kepler}},
	author       = {Howard, Andrew W. and Marcy, Geoffrey W. and Bryson, Stephen T. and Jenkins, Jon M. and Rowe, Jason F. and Batalha, Natalie M. and Borucki, William J. and Koch, David G. and Dunham, Edward W. and Gautier, Thomas N. and Van Cleve, Jeffrey and Cochran, William D. and Latham, David W. and Lissauer, Jack J. and Torres, Guillermo and Brown, Timothy M. and Gilliland, Ronald L. and Buchhave, Lars A. and Caldwell, Douglas A. and Christensen-Dalsgaard, Jørgen and Ciardi, David and Fressin, Francois and Haas, Michael R. and Howell, Steve B. and Kjeldsen, Hans and Seager, Sara and Rogers, Leslie and Sasselov, Dimitar D. and Steffen, Jason H. and Basri, Gibor S. and Charbonneau, David and Christiansen, Jessie and Clarke, Bruce and Dupree, Andrea and Fabrycky, Daniel C. and Fischer, Debra A. and Ford, Eric B. and Fortney, Jonathan J. and Tarter, Jill and Girouard, Forrest R. and Holman, Matthew J. and Johnson, John Asher and Klaus, Todd C. and Machalek, Pavel and Moorhead, Althea V. and Morehead, Robert C. and Ragozzine, Darin and Tenenbaum, Peter and Twicken, Joseph D. and Quinn, Samuel N. and Isaacson, Howard and Shporer, Avi and Lucas, Philip W. and Walkowicz, Lucianne M. and Welsh, William F. and Boss, Alan and Devore, Edna and Gould, Alan and Smith, Jeffrey C. and Morris, Robert L. and Prsa, Andrej and Morton, Timothy D. and Still, Martin and Thompson, Susan E. and Mullally, Fergal and Endl, Michael and MacQueen, Phillip J.},
	year         = 2012,
	month        = aug,
	journal      = {ApJS},
	volume       = 201,
	number       = 2,
	pages        = 15,
	doi          = {10.1088/0067-0049/201/2/15},
	issn         = {0067-0049},
	url          = {https://ui.adsabs.harvard.edu/abs/2012ApJS..201...15H/abstract},
	urldate      = {2020-05-01},
	language     = {en}
}

@misc{hu_water-rich_2025,
	title        = {A water-rich interior in the temperate sub-{Neptune} {K2}-18 b revealed by {JWST}},
	author       = {Hu, Renyu and Bello-Arufe, Aaron and Tokadjian, Armen and Yang, Jeehyun and Damiano, Mario and Roy, Pierre-Alexis and Coulombe, Louis-Philippe and Madhusudhan, Nikku and Constantinou, Savvas and Benneke, Björn},
	year         = 2025,
	month        = jul,
	publisher    = {arXiv},
	doi          = {10.48550/arXiv.2507.12622},
	url          = {https://ui.adsabs.harvard.edu/abs/2025arXiv250712622H},
	urldate      = {2025-10-10},
	note         = {ADS Bibcode: 2025arXiv250712622H}
}

@misc{nixon_magma_2025,
	title        = {Magma ocean interactions can explain {JWST} observations of the sub-{Neptune} {TOI}-270 d},
	author       = {Nixon, Matthew C. and Somers, R. Sander and Savel, Arjun B. and Ih, Jegug and Kempton, Eliza M.-R. and Young, Edward D. and Schlichting, Hilke E. and Lichtenberg, Tim and Welbanks, Luis and Misener, William and Piette, Anjali A. A. and Wogan, Nicholas F.},
	year         = 2025,
	month        = oct,
	publisher    = {arXiv},
	doi          = {10.48550/arXiv.2510.07367},
	url          = {http://arxiv.org/abs/2510.07367},
	urldate      = {2025-10-10},
	note         = {arXiv:2510.07367 [astro-ph]}
}

@article{werlen_sub-neptunes_2025,
	title        = {Sub-{Neptunes} {Are} {Drier} than {They} {Seem}: {Rethinking} the {Origins} of {Water}-rich {Worlds}},
	shorttitle   = {Sub-{Neptunes} {Are} {Drier} than {They} {Seem}},
	author       = {Werlen, Aaron and Dorn, Caroline and Burn, Remo and Schlichting, Hilke E. and Grimm, Simon L. and Young, Edward D.},
	year         = 2025,
	month        = sep,
	journal      = {The Astrophysical Journal},
	publisher    = {IOP},
	volume       = 991,
	pages        = {L16},
	doi          = {10.3847/2041-8213/adff73},
	issn         = {0004-637X},
	url          = {https://ui.adsabs.harvard.edu/abs/2025ApJ...991L..16W},
	urldate      = {2025-10-05},
	note         = {ADS Bibcode: 2025ApJ...991L..16W}
}

@article{ahrer_escaping_2025,
	title        = {Escaping {Helium} and a {Highly} {Muted} {Spectrum} {Suggest} a {Metal}-{Enriched} {Atmosphere} on {Sub}-{Neptune} {GJ3090b} from {JWST} {Transit} {Spectroscopy}},
	author       = {Ahrer, Eva-Maria and Radica, Michael and Piaulet-Ghorayeb, Caroline and Raul, Eshan and Wiser, Lindsey S. and Welbanks, Luis and Acuna, Lorena and Allart, Romain and Coulombe, Louis-Philippe and Louca, Amy J. and MacDonald, Ryan J. and Saidel, Morgan and Evans-Soma, Thomas M. and Benneke, Björn and Christie, Duncan and Beatty, Thomas G. and Cadieux, Charles and Cloutier, Ryan and Doyon, René and Fortney, Jonathan J. and Gagnebin, Anna and Gapp, Cyril and Innes, Hamish and Knutson, Heather A. and Komacek, Thaddeus D. and Krissansen-Totton, Joshua E. and Miguel, Yamila and Pierrehumbert, Raymond T. and Roy, Pierre-Alexis and Schlichting, Hilke E.},
	year         = 2025,
	month        = apr,
	doi          = {10.3847/2041-8213/add010},
	url          = {http://arxiv.org/abs/2504.20428},
	urldate      = {2025-05-15},
	note         = {arXiv:2504.20428 [astro-ph]}
}

@article{felix_evidence_2025,
	title        = {Evidence for sulfur chemistry in the atmosphere of the warm sub-{Neptune} {TOI}-270 d},
	author       = {Felix, Lukas and Kitzmann, Daniel and Demory, Brice-Olivier and Mordasini, Christoph},
	year         = 2025,
	month        = apr,
	publisher    = {arXiv},
	doi          = {10.48550/arXiv.2504.13039},
	url          = {http://arxiv.org/abs/2504.13039},
	urldate      = {2025-04-18},
	note         = {arXiv:2504.13039 [astro-ph]}
}

@misc{breza_not_2025,
	title        = {Not {All} {Sub}-{Neptune} {Exoplanets} {Have} {Magma} {Oceans}},
	author       = {Breza, Bodie and Nixon, Matthew C. and Kempton, Eliza M.-R.},
	year         = 2025,
	month        = sep,
	publisher    = {arXiv},
	doi          = {10.48550/arXiv.2509.20429},
	url          = {http://arxiv.org/abs/2509.20429},
	urldate      = {2025-09-26},
	note         = {arXiv:2509.20429 [astro-ph]}
}

@article{seward_system_1981,
	title        = {The system hydrogen - water up to 440°{C} and 2500 bar pressure},
	author       = {Seward, T. M. and Franck, E. U.},
	year         = 1981,
	journal      = {Berichte der Bunsengesellschaft für physikalische Chemie},
	volume       = 85,
	number       = 1,
	pages        = {2--7},
	doi          = {10.1002/bbpc.19810850103},
	issn         = {0005-9021},
	url          = {https://onlinelibrary.wiley.com/doi/abs/10.1002/bbpc.19810850103},
	urldate      = {2025-08-07},
	note         = {\_eprint: https://onlinelibrary.wiley.com/doi/pdf/10.1002/bbpc.19810850103},
	language     = {en}
}

@misc{holmberg_possible_2024,
	title        = {Possible {Hycean} conditions in the sub-{Neptune} {TOI}-270 d},
	author       = {Holmberg, Måns and Madhusudhan, Nikku},
	year         = 2024,
	month        = mar,
	journal      = {Astronomy and Astrophysics},
	volume       = 683,
	pages        = {L2},
	doi          = {10.1051/0004-6361/202348238},
	issn         = {0004-6361},
	url          = {https://ui.adsabs.harvard.edu/abs/2024A&A...683L...2H},
	urldate      = {2024-05-13},
	note         = {ADS Bibcode: 2024A\&A...683L...2H}
}

@article{madhusudhan_habitability_2021,
	title        = {Habitability and {Biosignatures} of {Hycean} {Worlds}},
	author       = {Madhusudhan, Nikku and Piette, Anjali A. A. and Constantinou, Savvas},
	year         = 2021,
	month        = sep,
	journal      = {The Astrophysical Journal},
	volume       = 918,
	pages        = 1,
	doi          = {10.3847/1538-4357/abfd9c},
	issn         = {0004-637X},
	url          = {https://ui.adsabs.harvard.edu/abs/2021ApJ...918....1M},
	urldate      = {2021-11-18},
	note         = {ADS Bibcode: 2021ApJ...918....1M}
}

@article{roy_is_2022,
	title        = {Is the {Hot}, {Dense} {Sub}-{Neptune} {TOI}-824 b an {Exposed} {Neptune} {Mantle}? {Spitzer} {Detection} of the {Hot} {Dayside} and {Reanalysis} of the {Interior} {Composition}},
	shorttitle   = {Is the {Hot}, {Dense} {Sub}-{Neptune} {TOI}-824 b an {Exposed} {Neptune} {Mantle}?},
	author       = {Roy, Pierre-Alexis and Benneke, Björn and Piaulet, Caroline and Crossfield, Ian J. M. and Kreidberg, Laura and Dragomir, Diana and Deming, Drake and Werner, Michael W. and Parmentier, Vivien and Christiansen, Jessie L. and Dressing, Courtney D. and Kane, Stephen R. and Morales, Farisa Y.},
	year         = 2022,
	month        = dec,
	journal      = {The Astrophysical Journal},
	publisher    = {IOP},
	volume       = 941,
	pages        = 89,
	doi          = {10.3847/1538-4357/ac9f18},
	issn         = {0004-637X},
	url          = {https://ui.adsabs.harvard.edu/abs/2022ApJ...941...89R},
	urldate      = {2025-08-03},
	note         = {ADS Bibcode: 2022ApJ...941...89R}
}

@misc{werlen_atmospheric_2025,
	title        = {Atmospheric {C}/{O} {Ratios} of {Sub}-{Neptunes} with {Magma} {Oceans}: {Homemade} rather than {Inherited}},
	shorttitle   = {Atmospheric {C}/{O} {Ratios} of {Sub}-{Neptunes} with {Magma} {Oceans}},
	author       = {Werlen, Aaron and Dorn, Caroline and Schlichting, Hilke E. and Grimm, Simon L. and Young, Edward D.},
	year         = 2025,
	month        = apr,
	publisher    = {arXiv},
	doi          = {10.48550/arXiv.2504.20450},
	url          = {http://arxiv.org/abs/2504.20450},
	urldate      = {2025-04-30},
	note         = {arXiv:2504.20450 [astro-ph]}
}

@article{nettelmann_thermal_2011,
	title        = {{THERMAL} {EVOLUTION} {AND} {STRUCTURE} {MODELS} {OF} {THE} {TRANSITING} {SUPER}-{EARTH} {GJ} 1214b},
	author       = {Nettelmann, N. and Fortney, J. J. and Kramm, U. and Redmer, R.},
	year         = 2011,
	month        = apr,
	journal      = {The Astrophysical Journal},
	publisher    = {The American Astronomical Society},
	volume       = 733,
	number       = 1,
	pages        = 2,
	doi          = {10.1088/0004-637X/733/1/2},
	issn         = {0004-637X},
	url          = {https://dx.doi.org/10.1088/0004-637X/733/1/2},
	urldate      = {2025-07-29},
	language     = {en}
}

@misc{howard_possibility_2025,
	title        = {The {Possibility} of {Hydrogen}-{Water} {Demixing} in {Uranus}, {Neptune}, {K2}-18b and {TOI}-270d},
	author       = {Howard, Saburo and Helled, Ravit and Bergermann, Armin and Redmer, Ronald},
	year         = 2025,
	month        = jul,
	publisher    = {arXiv},
	doi          = {10.48550/arXiv.2507.06288},
	url          = {http://arxiv.org/abs/2507.06288},
	urldate      = {2025-07-10},
	note         = {arXiv:2507.06288 [astro-ph]}
}

@misc{arevalo_different_2025,
	title        = {Different {Inhomogeneous} {Evolutionary} {Histories} for {Uranus} and {Neptune}},
	author       = {Arevalo, Roberto Tejada},
	year         = 2025,
	month        = jun,
	publisher    = {arXiv},
	doi          = {10.48550/arXiv.2506.13857},
	url          = {http://arxiv.org/abs/2506.13857},
	urldate      = {2025-06-18},
	note         = {arXiv:2506.13857 [astro-ph]}
}

@article{schlichting_chemical_2022,
	title        = {Chemical {Equilibrium} between {Cores}, {Mantles}, and {Atmospheres} of {Super}-{Earths} and {Sub}-{Neptunes} and {Implications} for {Their} {Compositions}, {Interiors}, and {Evolution}},
	author       = {Schlichting, Hilke E. and Young, Edward D.},
	year         = 2022,
	month        = may,
	journal      = {The Planetary Science Journal},
	publisher    = {IOP},
	volume       = 3,
	pages        = 127,
	doi          = {10.3847/PSJ/ac68e6},
	issn         = {2632-3338},
	url          = {https://ui.adsabs.harvard.edu/abs/2022PSJ.....3..127S},
	urldate      = {2024-05-16},
	note         = {ADS Bibcode: 2022PSJ.....3..127S}
}

@misc{teske_jwst_2025,
	title        = {{JWST} {COMPASS}: {NIRSpec}/{G395H} {Transmission} {Observations} of {TOI}-776 c, a 2 {Rearth} {M} {Dwarf} {Planet}},
	shorttitle   = {{JWST} {COMPASS}},
	author       = {Teske, Johanna and Batalha, Natasha E. and Wallack, Nicole L. and Kirk, James and Wogan, Nicholas F. and Gordon, Tyler A. and Alam, Munazza K. and Aguichine, Artyom and Wolfgang, Angie and Wakeford, Hannah R. and Scarsdale, Nicholas and Adams Redai, Jea and Moran, Sarah E. and López-Morales, Mercedes and Meech, Annabella and Gao, Peter and Batalha, Natalie M. and Alderson, Lili and Gagnebin, Anna},
	year         = 2025,
	month        = feb,
	doi          = {10.48550/arXiv.2502.20501},
	url          = {https://ui.adsabs.harvard.edu/abs/2025arXiv250220501T},
	urldate      = {2025-03-06},
	note         = {Publication Title: arXiv e-prints ADS Bibcode: 2025arXiv250220501T}
}

@misc{aguichine_evolution_2024,
	title        = {Evolution of steam worlds: energetic aspects},
	shorttitle   = {Evolution of steam worlds},
	author       = {Aguichine, Artyom and Batalha, Natalie and Fortney, Jonathan J. and Nettelmann, Nadine and Owen, James E. and Kempton, Eliza M.-R.},
	year         = 2024,
	month        = dec,
	publisher    = {arXiv},
	doi          = {10.48550/arXiv.2412.17945},
	url          = {http://arxiv.org/abs/2412.17945},
	urldate      = {2025-02-25},
	note         = {arXiv:2412.17945 [astro-ph]}
}

@article{piaulet-ghorayeb_jwstniriss_2024,
	title        = {{JWST}/{NIRISS} reveals the water-rich "steam world" atmosphere of {GJ} 9827 d},
	author       = {Piaulet-Ghorayeb, Caroline and Benneke, Bjorn and Radica, Michael and Raul, Eshan and Coulombe, Louis-Philippe and Ahrer, Eva-Maria and Kubyshkina, Daria and Howard, Ward S. and Krissansen-Totton, Joshua and MacDonald, Ryan and Roy, Pierre-Alexis and Louca, Amy and Christie, Duncan and Fournier-Tondreau, Marylou and Allart, Romain and Miguel, Yamila and Schlichting, Hilke E. and Welbanks, Luis and Cadieux, Charles and Dorn, Caroline and Evans-Soma, Thomas M. and Fortney, Jonathan J. and Pierrehumbert, Raymond and Lafreniere, David and Acuna, Lorena and Komacek, Thaddeus and Innes, Hamish and Beatty, Thomas G. and Cloutier, Ryan and Doyon, Rene and Gagnebin, Anna and Gapp, Cyril and Knutson, Heather A.},
	year         = 2024,
	month        = oct,
	journal      = {The Astrophysical Journal Letters},
	volume       = 974,
	number       = 1,
	pages        = {L10},
	doi          = {10.3847/2041-8213/ad6f00},
	issn         = {2041-8205, 2041-8213},
	url          = {http://arxiv.org/abs/2410.03527},
	urldate      = {2025-01-20},
	note         = {arXiv:2410.03527 [astro-ph]}
}

@article{alibert_formation_2017,
	title        = {Formation and composition of planets around very low mass stars},
	author       = {Alibert, Yann and Benz, Willy},
	year         = 2017,
	month        = feb,
	journal      = {Astronomy \& Astrophysics},
	volume       = 598,
	pages        = {L5},
	doi          = {10.1051/0004-6361/201629671},
	issn         = {0004-6361, 1432-0746},
	url          = {http://arxiv.org/abs/1610.03460},
	urldate      = {2024-06-09},
	note         = {arXiv:1610.03460 [astro-ph]},
	language     = {en}
}

@misc{tang_reassessing_2024,
	title        = {Reassessing {Sub}-{Neptune} {Structure}, {Radii}, and {Thermal} {Evolution}},
	author       = {Tang, Yao and Fortney, Jonathan J. and Nimmo, Francis and Thorngren, Daniel and Ohno, Kazumasa and Murray-Clay, Ruth},
	year         = 2024,
	month        = oct,
	publisher    = {arXiv},
	url          = {http://arxiv.org/abs/2410.21584},
	urldate      = {2024-11-19},
	note         = {arXiv:2410.21584}
}

@misc{wallack_jwst_2024,
	title        = {{JWST} {COMPASS}: {A} {NIRSpec}/{G395H} {Transmission} {Spectrum} of the {Sub}-{Neptune} {TOI}-836c},
	shorttitle   = {{JWST} {COMPASS}},
	author       = {Wallack, Nicole L. and Batalha, Natasha E. and Alderson, Lili and Scarsdale, Nicholas and Redai, Jea I. Adams and Aguichine, Artyom and Alam, Munazza K. and Gao, Peter and Wolfgang, Angie and Batalha, Natalie M. and Kirk, James and López-Morales, Mercedes and Moran, Sarah E. and Teske, Johanna and Wakeford, Hannah R. and Wogan, Nicholas F.},
	year         = 2024,
	month        = apr,
	publisher    = {arXiv},
	url          = {http://arxiv.org/abs/2404.01264},
	urldate      = {2024-08-01},
	note         = {arXiv:2404.01264 [astro-ph]},
	language     = {en}
}

@misc{amoros_h_2-h_2o_2024,
	title        = {H\$\_\{2\}\$-{H}\$\_\{2\}\${O} demixing in {Uranus} and {Neptune}: {Adiabatic} structure models},
	shorttitle   = {H\$\_\{2\}\$-{H}\$\_\{2\}\${O} demixing in {Uranus} and {Neptune}},
	author       = {Amoros, Marina Cano and Nettelmann, Nadine and Tosi, Nicola and Baumeister, Philipp and Rauer, Heike},
	year         = 2024,
	month        = oct,
	publisher    = {arXiv},
	url          = {http://arxiv.org/abs/2410.21099},
	urldate      = {2024-10-29},
	note         = {arXiv:2410.21099 [astro-ph]},
	language     = {en}
}

@misc{gupta_miscibility_2024,
	title        = {The miscibility of hydrogen and water in planetary atmospheres and interiors},
	author       = {Gupta, Akash and Stixrude, Lars and Schlichting, Hilke E.},
	year         = 2024,
	month        = jul,
	doi          = {10.48550/arXiv.2407.04685},
	url          = {https://ui.adsabs.harvard.edu/abs/2024arXiv240704685G},
	urldate      = {2024-09-24},
	note         = {Publication Title: arXiv e-prints ADS Bibcode: 2024arXiv240704685G}
}

@misc{nixon_new_2024,
	title        = {New insights into the internal structure of {GJ} 1214 b informed by {JWST}},
	author       = {Nixon, Matthew C. and Piette, Anjali A. A. and Kempton, Eliza M.-R. and Gao, Peter and Bean, Jacob L. and Steinrueck, Maria E. and Mahajan, Alexandra S. and Eastman, Jason D. and Zhang, Michael and Rogers, Leslie A.},
	year         = 2024,
	month        = jul,
	publisher    = {arXiv},
	url          = {http://arxiv.org/abs/2407.12079},
	urldate      = {2024-07-18},
	note         = {arXiv:2407.12079 [astro-ph]}
}

@article{lichtenberg_bifurcation_2021,
	title        = {Bifurcation of planetary building blocks during {Solar} {System} formation},
	author       = {Lichtenberg, Tim and Drążkowska, Joanna and Schönbächler, Maria and Golabek, Gregor J. and Hands, Thomas O.},
	year         = 2021,
	month        = jan,
	journal      = {Science},
	volume       = 371,
	pages        = {365--370},
	doi          = {10.1126/science.abb3091},
	issn         = {0036-8075},
	url          = {https://ui.adsabs.harvard.edu/abs/2021Sci...371..365L},
	urldate      = {2024-06-18},
	note         = {ADS Bibcode: 2021Sci...371..365L}
}

@article{lichtenberg_redox_2021,
	title        = {Redox {Hysteresis} of {Super}-{Earth} {Exoplanets} from {Magma} {Ocean} {Circulation}},
	author       = {Lichtenberg, Tim},
	year         = 2021,
	month        = jun,
	journal      = {The Astrophysical Journal},
	publisher    = {IOP},
	volume       = 914,
	pages        = {L4},
	doi          = {10.3847/2041-8213/ac0146},
	issn         = {0004-637X},
	url          = {https://ui.adsabs.harvard.edu/abs/2021ApJ...914L...4L},
	urldate      = {2024-06-10},
	note         = {ADS Bibcode: 2021ApJ...914L...4L}
}

@misc{vazan_how_2024,
	title        = {How planets form by pebble accretion {V}. {Silicate} rainout delays contraction of sub-{Neptunes}},
	author       = {Vazan, A. and Ormel, C. W. and Brouwers, M. G.},
	year         = 2024,
	month        = may,
	publisher    = {arXiv},
	doi          = {10.48550/arXiv.2405.09900},
	url          = {http://arxiv.org/abs/2405.09900},
	urldate      = {2024-05-20},
	note         = {arXiv:2405.09900 [astro-ph]}
}

@misc{tulekeyev_constraints_2024,
	title        = {Constraints on the long-term existence of dilute cores in giant planets},
	author       = {Tulekeyev, A. and Garaud, P. and Idini, B. and Fortney, J. J.},
	year         = 2024,
	month        = may,
	publisher    = {arXiv},
	url          = {http://arxiv.org/abs/2405.06790},
	urldate      = {2024-05-14},
	note         = {arXiv:2405.06790 [astro-ph]}
}

@misc{sur_apple_2024,
	title        = {{APPLE}: {An} {Evolution} {Code} for {Modeling} {Giant} {Planets}},
	shorttitle   = {{APPLE}},
	author       = {Sur, Ankan and Su, Yubo and Arevalo, Roberto Tejada and Chen, Yi-Xian and Burrows, Adam},
	year         = 2024,
	month        = apr,
	publisher    = {arXiv},
	url          = {http://arxiv.org/abs/2404.14483},
	urldate      = {2024-04-24},
	note         = {arXiv:2404.14483 [astro-ph]}
}

@article{piaulet_evidence_2023,
	title        = {Evidence for the volatile-rich composition of a 1.5-{Earth}-radius planet},
	author       = {Piaulet, Caroline and Benneke, Björn and Almenara, Jose M. and Dragomir, Diana and Knutson, Heather A. and Thorngren, Daniel and Peterson, Merrin S. and Crossfield, Ian J. M. and Kempton, Eliza M. -R. and Kubyshkina, Daria and Howard, Andrew W. and Angus, Ruth and Isaacson, Howard and Weiss, Lauren M. and Beichman, Charles A. and Fortney, Jonathan J. and Fossati, Luca and Lammer, Helmut and McCullough, P. R. and Morley, Caroline V. and Wong, Ian},
	year         = 2023,
	month        = feb,
	journal      = {Nature Astronomy},
	volume       = 7,
	pages        = {206--222},
	doi          = {10.1038/s41550-022-01835-4},
	issn         = {2397-3366},
	url          = {https://ui.adsabs.harvard.edu/abs/2023NatAs...7..206P},
	urldate      = {2024-04-19},
	note         = {ADS Bibcode: 2023NatAs...7..206P}
}

@article{kempton_where_2023,
	title        = {Where are the {Water} {Worlds}?: {Self}-consistent {Models} of {Water}-rich {Exoplanet} {Atmospheres}},
	shorttitle   = {Where are the {Water} {Worlds}?},
	author       = {Kempton, Eliza M. -R. and Lessard, Madeline and Malik, Matej and Rogers, Leslie A. and Futrowsky, Kate E. and Ih, Jegug and Marounina, Nadejda and Muñoz-Romero, Carlos E.},
	year         = 2023,
	month        = aug,
	journal      = {The Astrophysical Journal},
	volume       = 953,
	pages        = 57,
	doi          = {10.3847/1538-4357/ace10d},
	issn         = {0004-637X},
	url          = {https://ui.adsabs.harvard.edu/abs/2023ApJ...953...57K},
	urldate      = {2024-03-28},
	note         = {ADS Bibcode: 2023ApJ...953...57K}
}

@misc{benneke_jwst_2024,
	title        = {{JWST} {Reveals} {CH}\$\_4\$, {CO}\$\_2\$, and {H}\$\_2\${O} in a {Metal}-rich {Miscible} {Atmosphere} on a {Two}-{Earth}-{Radius} {Exoplanet}},
	author       = {Benneke, Björn and Roy, Pierre-Alexis and Coulombe, Louis-Philippe and Radica, Michael and Piaulet, Caroline and Ahrer, Eva-Maria and Pierrehumbert, Raymond and Krissansen-Totton, Joshua and Schlichting, Hilke E. and Hu, Renyu and Yang, Jeehyun and Christie, Duncan and Thorngren, Daniel and Young, Edward D. and Pelletier, Stefan and Knutson, Heather A. and Miguel, Yamila and Evans-Soma, Thomas M. and Dorn, Caroline and Gagnebin, Anna and Fortney, Jonathan J. and Komacek, Thaddeus and MacDonald, Ryan and Raul, Eshan and Cloutier, Ryan and Acuna, Lorena and Lafrenière, David and Cadieux, Charles and Doyon, René and Welbanks, Luis and Allart, Romain},
	year         = 2024,
	month        = mar,
	publisher    = {arXiv},
	url          = {http://arxiv.org/abs/2403.03325},
	urldate      = {2024-03-08},
	note         = {arXiv:2403.03325 [astro-ph]}
}

@article{innes_runaway_2023,
	title        = {The {Runaway} {Greenhouse} {Effect} on {Hycean} {Worlds}},
	author       = {Innes, Hamish and Tsai, Shang-Min and Pierrehumbert, Raymond T.},
	year         = 2023,
	month        = aug,
	journal      = {The Astrophysical Journal},
	publisher    = {The American Astronomical Society},
	volume       = 953,
	number       = 2,
	pages        = 168,
	doi          = {10.3847/1538-4357/ace346},
	issn         = {0004-637X},
	url          = {https://dx.doi.org/10.3847/1538-4357/ace346},
	urldate      = {2024-03-05},
	language     = {en}
}

@misc{burn_radius_2024,
	title        = {A radius valley between migrated steam worlds and evaporated rocky cores},
	author       = {Burn, Remo and Mordasini, Christoph and Mishra, Lokesh and Haldemann, Jonas and Venturini, Julia and Emsenhuber, Alexandre and Henning, Thomas},
	year         = 2024,
	month        = jan,
	publisher    = {arXiv},
	url          = {http://arxiv.org/abs/2401.04380},
	urldate      = {2024-02-13},
	note         = {arXiv:2401.04380 [astro-ph]}
}

@article{thorngren_bayesian_2018,
	title        = {Bayesian {Analysis} of {Hot}-{Jupiter} {Radius} {Anomalies}: {Evidence} for {Ohmic} {Dissipation}?},
	shorttitle   = {Bayesian {Analysis} of {Hot}-{Jupiter} {Radius} {Anomalies}},
	author       = {Thorngren, Daniel P. and Fortney, Jonathan J.},
	year         = 2018,
	month        = may,
	journal      = {The Astronomical Journal},
	volume       = 155,
	number       = 5,
	pages        = 214,
	doi          = {10.3847/1538-3881/aaba13},
	issn         = {0004-6256, 1538-3881},
	url          = {https://iopscience.iop.org/article/10.3847/1538-3881/aaba13},
	urldate      = {2024-02-02},
	language     = {en}
}

@misc{rogers_conclusive_2023,
	title        = {Conclusive evidence for a population of water-worlds around {M}-dwarfs remains elusive},
	author       = {Rogers, James G. and Schlichting, Hilke E. and Owen, James E.},
	year         = 2023,
	month        = jan,
	publisher    = {arXiv},
	url          = {http://arxiv.org/abs/2301.04321},
	urldate      = {2023-01-12},
	note         = {arXiv:2301.04321 [astro-ph]}
}

@article{luque_density_2022,
	title        = {Density, not radius, separates rocky and water-rich small planets orbiting {M} dwarf stars},
	author       = {Luque, Rafael and Pallé, Enric},
	year         = 2022,
	month        = sep,
	journal      = {Science},
	volume       = 377,
	pages        = {1211--1214},
	doi          = {10.1126/science.abl7164},
	issn         = {0036-8075},
	url          = {https://ui.adsabs.harvard.edu/abs/2022Sci...377.1211L},
	urldate      = {2022-09-26},
	note         = {ADS Bibcode: 2022Sci...377.1211L}
}

@article{husser_new_2013,
	title        = {A new extensive library of {PHOENIX} stellar atmospheres and synthetic spectra},
	author       = {Husser, T.-O. and Wende-von Berg, S. and Dreizler, S. and Homeier, D. and Reiners, A. and Barman, T. and Hauschildt, P. H.},
	year         = 2013,
	month        = may,
	journal      = {Astronomy \&amp; Astrophysics, Volume 553, id.A6, {\textless}NUMPAGES{\textgreater}9{\textless}/NUMPAGES{\textgreater} pp.},
	volume       = 553,
	pages        = {A6},
	doi          = {10.1051/0004-6361/201219058},
	issn         = {0004-6361},
	url          = {https://ui.adsabs.harvard.edu/abs/2013A%26A...553A...6H/abstract},
	urldate      = {2022-07-07},
	language     = {en}
}

@article{turbet_revised_2020,
	title        = {Revised mass-radius relationships for water-rich rocky planets more irradiated than the runaway greenhouse limit},
	author       = {Turbet, Martin and Bolmont, Emeline and Ehrenreich, David and Gratier, Pierre and Leconte, Jérémy and Selsis, Franck and Hara, Nathan and Lovis, Christophe},
	year         = 2020,
	month        = jun,
	journal      = {Astronomy \&amp; Astrophysics, Volume 638, id.A41, {\textless}NUMPAGES{\textgreater}10{\textless}/NUMPAGES{\textgreater} pp.},
	volume       = 638,
	pages        = {A41},
	doi          = {10.1051/0004-6361/201937151},
	issn         = {0004-6361},
	url          = {https://ui.adsabs.harvard.edu/abs/2020A%26A...638A..41T/abstract},
	urldate      = {2022-06-05},
	language     = {en}
}

@article{rogers_photoevaporation_2021,
	title        = {Photoevaporation versus core-powered mass-loss: model comparison with the {3D} radius gap},
	shorttitle   = {Photoevaporation versus core-powered mass-loss},
	author       = {Rogers, James G. and Gupta, Akash and Owen, James E. and Schlichting, Hilke E.},
	year         = 2021,
	month        = dec,
	journal      = {Monthly Notices of the Royal Astronomical Society},
	volume       = 508,
	pages        = {5886--5902},
	doi          = {10.1093/mnras/stab2897},
	issn         = {0035-8711},
	url          = {https://ui.adsabs.harvard.edu/abs/2021MNRAS.508.5886R},
	urldate      = {2021-12-17},
	note         = {ADS Bibcode: 2021MNRAS.508.5886R}
}

@techreport{schlichting_chemical_2021,
	title        = {Chemical equilibrium between {Cores}, {Mantles}, and {Atmospheres} of {Super}-{Earths} and {Sub}-{Neptunes}, and {Implications} for their {Compositions}, {Interiors} and {Evolution}},
	author       = {Schlichting, Hilke E. and Young, Edward D.},
	year         = 2021,
	month        = jul,
	url          = {https://ui.adsabs.harvard.edu/abs/2021arXiv210710405S},
	urldate      = {2021-11-25},
	note         = {Publication Title: arXiv e-prints ADS Bibcode: 2021arXiv210710405S Type: article}
}

@article{kite_water_2021,
	title        = {Water on {Hot} {Rocky} {Exoplanets}},
	author       = {Kite, Edwin S. and Schaefer, Laura},
	year         = 2021,
	month        = mar,
	journal      = {The Astrophysical Journal},
	volume       = 909,
	pages        = {L22},
	doi          = {10.3847/2041-8213/abe7dc},
	issn         = {0004-637X},
	url          = {https://ui.adsabs.harvard.edu/abs/2021ApJ...909L..22K},
	urldate      = {2021-11-01},
	note         = {ADS Bibcode: 2021ApJ...909L..22K}
}

@article{lopez_understanding_2014,
	title        = {Understanding the {Mass}-{Radius} {Relation} for {Sub}-{Neptunes}: {Radius} as a {Proxy} for {Composition}},
	shorttitle   = {Understanding the {Mass}-{Radius} {Relation} for {Sub}-{Neptunes}},
	author       = {Lopez, Eric D. and Fortney, Jonathan J.},
	year         = 2014,
	month        = aug,
	journal      = {The Astrophysical Journal},
	volume       = 792,
	number       = 1,
	pages        = 1,
	doi          = {10.1088/0004-637X/792/1/1},
	issn         = {1538-4357},
	url          = {http://arxiv.org/abs/1311.0329},
	urldate      = {2019-05-05},
	note         = {arXiv: 1311.0329}
}

@article{miller-ricci_atmospheric_2009,
	title        = {The {Atmospheric} {Signatures} of {Super}-{Earths}: {How} to {Distinguish} {Between} {Hydrogen}-{Rich} and {Hydrogen}-{Poor} {Atmospheres}},
	shorttitle   = {The {Atmospheric} {Signatures} of {Super}-{Earths}},
	author       = {Miller-Ricci, Eliza and Seager, Sara and Sasselov, Dimitar},
	year         = 2009,
	month        = jan,
	journal      = {The Astrophysical Journal},
	volume       = 690,
	pages        = {1056--1067},
	doi          = {10.1088/0004-637X/690/2/1056},
	issn         = {0004-637X},
	url          = {http://adsabs.harvard.edu/abs/2009ApJ...690.1056M},
	urldate      = {2020-05-14}
}

@article{mazevet_ab_2019,
	title        = {Ab initio based equation of state of dense water for planetary and exoplanetary modeling},
	author       = {Mazevet, S. and Licari, A. and Chabrier, G. and Potekhin, A. Y.},
	year         = 2019,
	month        = jan,
	journal      = {Astronomy and Astrophysics},
	volume       = 621,
	pages        = {A128},
	doi          = {10.1051/0004-6361/201833963},
	issn         = {0004-6361},
	url          = {http://adsabs.harvard.edu/abs/2019A%26A...621A.128M},
	urldate      = {2021-03-23}
}

@article{chabrier_new_2019,
	title        = {A {New} {Equation} of {State} for {Dense} {Hydrogen}-{Helium} {Mixtures}},
	author       = {Chabrier, G. and Mazevet, S. and Soubiran, F.},
	year         = 2019,
	month        = feb,
	journal      = {The Astrophysical Journal},
	volume       = 872,
	pages        = 51,
	doi          = {10.3847/1538-4357/aaf99f},
	issn         = {0004-637X},
	url          = {http://adsabs.harvard.edu/abs/2019ApJ...872...51C},
	urldate      = {2021-03-23}
}

@article{piaulet_wasp-107bs_2021,
	title        = {{WASP}-107b’s {Density} {Is} {Even} {Lower}: {A} {Case} {Study} for the {Physics} of {Planetary} {Gas} {Envelope} {Accretion} and {Orbital} {Migration}},
	shorttitle   = {{WASP}-107b’s {Density} {Is} {Even} {Lower}},
	author       = {Piaulet, Caroline and Benneke, Björn and Rubenzahl, Ryan A. and Howard, Andrew W. and Lee, Eve J. and Thorngren, Daniel and Angus, Ruth and Peterson, Merrin and Schlieder, Joshua E. and Werner, Michael and Kreidberg, Laura and Jaouni, Tareq and Crossfield, Ian J. M. and Ciardi, David R. and Petigura, Erik A. and Livingston, John and Dressing, Courtney D. and Fulton, Benjamin J. and Beichman, Charles and Christiansen, Jessie L. and Gorjian, Varoujan and Hardegree-Ullman, Kevin K. and Krick, Jessica and Sinukoff, Evan},
	year         = 2021,
	month        = jan,
	journal      = {The Astronomical Journal},
	publisher    = {IOP Publishing},
	volume       = 161,
	number       = 2,
	pages        = 70,
	doi          = {10.3847/1538-3881/abcd3c},
	issn         = {1538-3881},
	url          = {https://iopscience.iop.org/article/10.3847/1538-3881/abcd3c/meta},
	urldate      = {2021-01-21},
	language     = {en}
}

@article{kempton_framework_2018,
	title        = {A {Framework} for {Prioritizing} the {TESS} {Planetary} {Candidates} {Most} {Amenable} to {Atmospheric} {Characterization}},
	author       = {Kempton, Eliza M.-R. and Bean, Jacob L. and Louie, Dana R. and Deming, Drake and Koll, Daniel D. B. and Mansfield, Megan and Christiansen, Jessie L. and López-Morales, Mercedes and Swain, Mark R. and Zellem, Robert T. and Ballard, Sarah and Barclay, Thomas and Barstow, Joanna K. and Batalha, Natasha E. and Beatty, Thomas G. and Berta-Thompson, Zach and Birkby, Jayne and Buchhave, Lars A. and Charbonneau, David and Cowan, Nicolas B. and Crossfield, Ian and de Val-Borro, Miguel and Doyon, René and Dragomir, Diana and Gaidos, Eric and Heng, Kevin and Hu, Renyu and Kane, Stephen R. and Kreidberg, Laura and Mallonn, Matthias and Morley, Caroline V. and Narita, Norio and Nascimbeni, Valerio and Pallé, Enric and Quintana, Elisa V. and Rauscher, Emily and Seager, Sara and Shkolnik, Evgenya L. and Sing, David K. and Sozzetti, Alessandro and Stassun, Keivan G. and Valenti, Jeff A. and von Essen, Carolina},
	year         = 2018,
	month        = nov,
	journal      = {Publications of the Astronomical Society of the Pacific},
	volume       = 130,
	pages        = 114401,
	doi          = {10.1088/1538-3873/aadf6f},
	issn         = {0004-6280},
	url          = {http://adsabs.harvard.edu/abs/2018PASP..130k4401K},
	urldate      = {2020-11-23}
}

@article{fulton_california-kepler_2017,
	title        = {The {California}-{Kepler} {Survey}. {III}. {A} {Gap} in the {Radius} {Distribution} of {Small} {Planets}},
	author       = {Fulton, Benjamin J. and Petigura, Erik A. and Howard, Andrew W. and Isaacson, Howard and Marcy, Geoffrey W. and Cargile, Phillip A. and Hebb, Leslie and Weiss, Lauren M. and Johnson, John Asher and Morton, Timothy D. and Sinukoff, Evan and Crossfield, Ian J. M. and Hirsch, Lea A.},
	year         = 2017,
	month        = sep,
	journal      = {The Astronomical Journal},
	volume       = 154,
	pages        = 109,
	doi          = {10.3847/1538-3881/aa80eb},
	issn         = {0004-6256},
	url          = {http://adsabs.harvard.edu/abs/2017AJ....154..109F},
	urldate      = {2020-05-03}
}

@article{astropy_collaboration_astropy_2013,
	title        = {Astropy: {A} community {Python} package for astronomy},
	shorttitle   = {Astropy},
	author       = {{Astropy Collaboration} and Robitaille, Thomas P. and Tollerud, Erik J. and Greenfield, Perry and Droettboom, Michael and Bray, Erik and Aldcroft, Tom and Davis, Matt and Ginsburg, Adam and Price-Whelan, Adrian M. and Kerzendorf, Wolfgang E. and Conley, Alexander and Crighton, Neil and Barbary, Kyle and Muna, Demitri and Ferguson, Henry and Grollier, Frédéric and Parikh, Madhura M. and Nair, Prasanth H. and Unther, Hans M. and Deil, Christoph and Woillez, Julien and Conseil, Simon and Kramer, Roban and Turner, James E. H. and Singer, Leo and Fox, Ryan and Weaver, Benjamin A. and Zabalza, Victor and Edwards, Zachary I. and Azalee Bostroem, K. and Burke, D. J. and Casey, Andrew R. and Crawford, Steven M. and Dencheva, Nadia and Ely, Justin and Jenness, Tim and Labrie, Kathleen and Lim, Pey Lian and Pierfederici, Francesco and Pontzen, Andrew and Ptak, Andy and Refsdal, Brian and Servillat, Mathieu and Streicher, Ole},
	year         = 2013,
	month        = oct,
	journal      = {Astronomy and Astrophysics},
	volume       = 558,
	pages        = {A33},
	doi          = {10.1051/0004-6361/201322068},
	issn         = {0004-6361},
	url          = {http://adsabs.harvard.edu/abs/2013A%26A...558A..33A},
	urldate      = {2020-10-09}
}

@article{hunter_matplotlib_2007,
	title        = {Matplotlib: {A} {2D} {Graphics} {Environment}},
	shorttitle   = {Matplotlib},
	author       = {Hunter, John D.},
	year         = 2007,
	month        = may,
	journal      = {Computing in Science and Engineering},
	volume       = 9,
	pages        = {90--95},
	doi          = {10.1109/MCSE.2007.55},
	url          = {http://adsabs.harvard.edu/abs/2007CSE.....9...90H},
	urldate      = {2020-10-09}
}

@article{benneke_atmospheric_2012,
	title        = {Atmospheric {Retrieval} for {Super}-{Earths}: {Uniquely} {Constraining} the {Atmospheric} {Composition} with {Transmission} {Spectroscopy}},
	shorttitle   = {Atmospheric {Retrieval} for {Super}-{Earths}},
	author       = {Benneke, Bjoern and Seager, Sara},
	year         = 2012,
	month        = jul,
	journal      = {The Astrophysical Journal},
	volume       = 753,
	pages        = 100,
	doi          = {10.1088/0004-637X/753/2/100},
	issn         = {0004-637X},
	url          = {http://adsabs.harvard.edu/abs/2012ApJ...753..100B},
	urldate      = {2020-05-14}
}

@article{zeng_growth_2019,
	title        = {Growth model interpretation of planet size distribution},
	author       = {Zeng, Li and Jacobsen, Stein B. and Sasselov, Dimitar D. and Petaev, Michail I. and Vanderburg, Andrew and Lopez-Morales, Mercedes and Perez-Mercader, Juan and Mattsson, Thomas R. and Li, Gongjie and Heising, Matthew Z. and Bonomo, Aldo S. and Damasso, Mario and Berger, Travis A. and Cao, Hao and Levi, Amit and Wordsworth, Robin D.},
	year         = 2019,
	month        = may,
	journal      = {Proceedings of the National Academy of Science},
	volume       = 116,
	pages        = {9723--9728},
	doi          = {10.1073/pnas.1812905116},
	issn         = {0027-8424},
	url          = {http://adsabs.harvard.edu/abs/2019PNAS..116.9723Z},
	urldate      = {2020-05-27}
}

@article{kuchner_volatile-rich_2003,
	title        = {Volatile-rich {Earth}-{Mass} {Planets} in the {Habitable} {Zone}},
	author       = {Kuchner, Marc J.},
	year         = 2003,
	month        = oct,
	journal      = {The Astrophysical Journal Letters},
	volume       = 596,
	pages        = {L105--L108},
	doi          = {10.1086/378397},
	issn         = {0004-637X},
	url          = {http://adsabs.harvard.edu/abs/2003ApJ...596L.105K},
	urldate      = {2020-05-27}
}

@article{rogers_framework_2010,
	title        = {A {Framework} for {Quantifying} the {Degeneracies} of {Exoplanet} {Interior} {Compositions}},
	author       = {Rogers, L. A. and Seager, S.},
	year         = 2010,
	month        = apr,
	journal      = {The Astrophysical Journal},
	volume       = 712,
	pages        = {974--991},
	doi          = {10.1088/0004-637X/712/2/974},
	issn         = {0004-637X},
	url          = {http://adsabs.harvard.edu/abs/2010ApJ...712..974R},
	urldate      = {2020-04-10}
}

@article{madhusudhan_interior_2020,
	title        = {The {Interior} and {Atmosphere} of the {Habitable}-zone {Exoplanet} {K2}-18b},
	author       = {Madhusudhan, Nikku and Nixon, Matthew C. and Welbanks, Luis and Piette, Anjali A. A. and Booth, Richard A.},
	year         = 2020,
	month        = mar,
	journal      = {The Astrophysical Journal Letters},
	volume       = 891,
	pages        = {L7},
	doi          = {10.3847/2041-8213/ab7229},
	issn         = {0004-637X},
	url          = {http://adsabs.harvard.edu/abs/2020ApJ...891L...7M},
	urldate      = {2020-05-11}
}

@article{fulton_california-kepler_2018,
	title        = {The {California}-{Kepler} {Survey}. {VII}. {Precise} {Planet} {Radii} {Leveraging} {Gaia} {DR2} {Reveal} the {Stellar} {Mass} {Dependence} of the {Planet} {Radius} {Gap}},
	author       = {Fulton, Benjamin J. and Petigura, Erik A.},
	year         = 2018,
	month        = dec,
	journal      = {The Astronomical Journal},
	volume       = 156,
	pages        = 264,
	doi          = {10.3847/1538-3881/aae828},
	issn         = {0004-6256},
	url          = {http://adsabs.harvard.edu/abs/2018AJ....156..264F},
	urldate      = {2020-03-18}
}

@article{zeng_mass-radius_2016,
	title        = {Mass-{Radius} {Relation} for {Rocky} {Planets} {Based} on {PREM}},
	author       = {Zeng, Li and Sasselov, Dimitar D. and Jacobsen, Stein B.},
	year         = 2016,
	month        = mar,
	journal      = {The Astrophysical Journal},
	volume       = 819,
	pages        = 127,
	doi          = {10.3847/0004-637X/819/2/127},
	issn         = {0004-637X},
	url          = {http://adsabs.harvard.edu/abs/2016ApJ...819..127Z},
	urldate      = {2020-03-17}
}

@article{thorngren_connecting_2019,
	title        = {Connecting {Giant} {Planet} {Atmosphere} and {Interior} {Modeling}: {Constraints} on {Atmospheric} {Metal} {Enrichment}},
	shorttitle   = {Connecting {Giant} {Planet} {Atmosphere} and {Interior} {Modeling}},
	author       = {Thorngren, Daniel and Fortney, Jonathan J.},
	year         = 2019,
	month        = apr,
	journal      = {The Astrophysical Journal},
	volume       = 874,
	number       = 2,
	pages        = {L31},
	doi          = {10.3847/2041-8213/ab1137},
	url          = {https://ui.adsabs.harvard.edu/abs/2019ApJ...874L..31T/abstract},
	urldate      = {2019-05-01},
	language     = {en}
}

@article{fortney_planetary_2007,
	title        = {Planetary {Radii} across {Five} {Orders} of {Magnitude} in {Mass} and {Stellar} {Insolation}: {Application} to {Transits}},
	shorttitle   = {Planetary {Radii} across {Five} {Orders} of {Magnitude} in {Mass} and {Stellar} {Insolation}},
	author       = {Fortney, J. J. and Marley, M. S. and Barnes, J. W.},
	year         = 2007,
	month        = apr,
	journal      = {The Astrophysical Journal},
	volume       = 659,
	pages        = {1661--1672},
	doi          = {10.1086/512120},
	issn         = {0004-637X},
	url          = {http://adsabs.harvard.edu/abs/2007ApJ...659.1661F},
	urldate      = {2020-01-27}
}

@article{benneke_characterizing_2013,
	title        = {Characterizing {Super}-{Earth} {Exoplanets} {Using} {Transmission} {Spectroscopy} - {Application} to {GJ} 1214b},
	author       = {Benneke, Bjoern},
	year         = 2013,
	month        = jan,
	journal      = {American Astronomical Society Meeting Abstracts \#221},
	volume       = 221,
	pages        = {224.04},
	url          = {https://ui.adsabs.harvard.edu/abs/2013AAS...22122404B/abstract},
	urldate      = {2019-08-07},
	language     = {en}
}

@article{seager_mass-radius_2007,
	title        = {Mass-{Radius} {Relationships} for {Solid} {Exoplanets}},
	author       = {Seager, S. and Kuchner, M. and Hier-Majumder, C. and Militzer, B.},
	year         = 2007,
	month        = nov,
	journal      = {The Astrophysical Journal},
	volume       = 669,
	number       = 2,
	pages        = {1279--1297},
	doi          = {10.1086/521346},
	issn         = {0004-637X, 1538-4357},
	url          = {http://arxiv.org/abs/0707.2895},
	urldate      = {2019-05-01},
	note         = {arXiv: 0707.2895}
}

@article{lodders_solar_2010,
	title        = {Solar {System} {Abundances} of the {Elements}},
	author       = {Lodders, Katharina},
	year         = 2010,
	journal      = {arXiv:1010.2746 [astro-ph]},
	pages        = {379--417},
	doi          = {10.1007/978-3-642-10352-0_8},
	url          = {http://arxiv.org/abs/1010.2746},
	urldate      = {2019-03-25},
	note         = {arXiv: 1010.2746}
}

@article{thorngren_mass-metallicity_2016,
	title        = {The {Mass}-{Metallicity} {Relation} for {Giant} {Planets}},
	author       = {Thorngren, Daniel P. and Fortney, Jonathan J. and Murray-Clay, Ruth A. and Lopez, Eric D.},
	year         = 2016,
	month        = nov,
	journal      = {The Astrophysical Journal},
	volume       = 831,
	number       = 1,
	pages        = 64,
	doi          = {10.3847/0004-637X/831/1/64},
	url          = {https://ui.adsabs.harvard.edu/#abs/2016ApJ...831...64T/abstract},
	urldate      = {2019-01-08},
	language     = {en}
}

@article{vazan_rocky_2023,
	title        = {Rocky sub-{Neptunes} formed by pebble accretion: {Rain} of rock from polluted envelopes},
	shorttitle   = {Rocky sub-{Neptunes} formed by pebble accretion},
	author       = {Vazan, Allona and Ormel, Chris W.},
	year         = 2023,
	month        = aug,
	journal      = {Astronomy \& Astrophysics},
	publisher    = {EDP Sciences},
	volume       = 676,
	pages        = {L8},
	doi          = {10.1051/0004-6361/202346574},
	issn         = {0004-6361, 1432-0746},
	url          = {https://www.aanda.org/articles/aa/abs/2023/08/aa46574-23/aa46574-23.html},
	urldate      = {2026-03-13},
	copyright    = {© The Authors 2023},
	language     = {en}
}

@article{cmiel_characterizing_2025,
	title        = {Characterizing the {Radiative}–{Convective} {Structure} of {Dense} {Rocky} {Planet} {Atmospheres}},
	author       = {Cmiel, Jessica and Wordsworth, Robin and Seeley, Jacob T.},
	year         = 2025,
	month        = may,
	journal      = {The Planetary Science Journal},
	publisher    = {IOP},
	volume       = 6,
	pages        = 123,
	doi          = {10.3847/PSJ/adcd5f},
	issn         = {2632-3338},
	url          = {https://ui.adsabs.harvard.edu/abs/2025PSJ.....6..123C},
	urldate      = {2026-03-13},
	note         = {ADS Bibcode: 2025PSJ.....6..123C}
}

@misc{selsis_steam_2024,
	title        = {Steam atmospheres and the implications for {Venus} and {Venus}-like planets},
	author       = {Selsis, Franck and Leconte, Jérémy and Turbet, Martin and Chaverot, Guillaume and Bolmont, Emeline},
	year         = 2024,
	month        = apr,
	pages        = 22187,
	doi          = {10.5194/egusphere-egu24-22187},
	url          = {https://ui.adsabs.harvard.edu/abs/2024EGUGA..2622187S},
	urldate      = {2026-03-13},
	note         = {ADS Bibcode: 2024EGUGA..2622187S}
}

@article{werlen_effects_2026,
	title        = {The {Effects} of {Non}-ideal {Mixing} in {Planetary} {Magma} {Oceans} and {Atmospheres}},
	author       = {Werlen, Aaron and Young, Edward D. and Schlichting, Hilke E. and Dorn, Caroline and Shahar, Anat},
	year         = 2026,
	month        = mar,
	journal      = {The Astrophysical Journal},
	publisher    = {The American Astronomical Society},
	volume       = 999,
	number       = 2,
	pages        = 178,
	doi          = {10.3847/1538-4357/ae434d},
	issn         = {0004-637X},
	url          = {https://doi.org/10.3847/1538-4357/ae434d},
	urldate      = {2026-03-13},
	language     = {en}
}

@article{militzer_ab_2013,
	title        = {Ab {Initio} {Equation} of {State} for {Hydrogen}-{Helium} {Mixtures} with {Recalibration} of the {Giant}-planet {Mass}-{Radius} {Relation}},
	author       = {Militzer, B. and Hubbard, W. B.},
	year         = 2013,
	month        = sep,
	journal      = {The Astrophysical Journal},
	publisher    = {IOP},
	volume       = 774,
	pages        = 148,
	doi          = {10.1088/0004-637X/774/2/148},
	issn         = {0004-637X},
	url          = {https://ui.adsabs.harvard.edu/abs/2013ApJ...774..148M},
	urldate      = {2026-03-13},
	note         = {ADS Bibcode: 2013ApJ...774..148M}
}

@article{chabrier_heat_2007,
	title        = {Heat transport in giant (exo)planets: a new perspective},
	shorttitle   = {Heat transport in giant (exo)planets},
	author       = {Chabrier, Gilles and Baraffe, Isabelle},
	year         = 2007,
	month        = may,
	journal      = {The Astrophysical Journal},
	volume       = 661,
	number       = 1,
	pages        = {L81--L84},
	doi          = {10.1086/518473},
	issn         = {0004-637X, 1538-4357},
	url          = {http://arxiv.org/abs/astro-ph/0703755},
	urldate      = {2026-03-01},
	note         = {arXiv:astro-ph/0703755}
}

@article{korre_convective_2019,
	title        = {Convective overshooting and penetration in a {Boussinesq} spherical shell},
	author       = {Korre, L and Garaud, P and Brummell, N H},
	year         = 2019,
	month        = mar,
	journal      = {Monthly Notices of the Royal Astronomical Society},
	volume       = 484,
	number       = 1,
	pages        = {1220--1237},
	doi          = {10.1093/mnras/stz047},
	issn         = {0035-8711},
	url          = {https://doi.org/10.1093/mnras/stz047},
	urldate      = {2026-03-01}
}

@article{anders_convective_2023,
	title        = {Convective {Boundary} {Mixing} in {Main}-{Sequence} {Stars}: {Theory} and {Empirical} {Constraints}},
	shorttitle   = {Convective {Boundary} {Mixing} in {Main}-{Sequence} {Stars}},
	author       = {Anders, Evan H. and Pedersen, May G.},
	year         = 2023,
	month        = apr,
	journal      = {Galaxies},
	publisher    = {Multidisciplinary Digital Publishing Institute},
	volume       = 11,
	number       = 2,
	pages        = 56,
	doi          = {10.3390/galaxies11020056},
	issn         = {2075-4434},
	url          = {https://www.mdpi.com/2075-4434/11/2/56},
	urldate      = {2026-03-01},
	copyright    = {http://creativecommons.org/licenses/by/3.0/},
	language     = {en}
}

@article{astropy_collaboration_astropy_2018,
	title        = {The {Astropy} {Project}: {Building} an {Open}-science {Project} and {Status} of the v2.0 {Core} {Package}},
	shorttitle   = {The {Astropy} {Project}},
	author       = {{Astropy Collaboration} and Price-Whelan, A. M. and Sipőcz, B. M. and Günther, H. M. and Lim, P. L. and Crawford, S. M. and Conseil, S. and Shupe, D. L. and Craig, M. W. and Dencheva, N. and Ginsburg, A. and VanderPlas, J. T. and Bradley, L. D. and Pérez-Suárez, D. and de Val-Borro, M. and Aldcroft, T. L. and Cruz, K. L. and Robitaille, T. P. and Tollerud, E. J. and Ardelean, C. and Babej, T. and Bach, Y. P. and Bachetti, M. and Bakanov, A. V. and Bamford, S. P. and Barentsen, G. and Barmby, P. and Baumbach, A. and Berry, K. L. and Biscani, F. and Boquien, M. and Bostroem, K. A. and Bouma, L. G. and Brammer, G. B. and Bray, E. M. and Breytenbach, H. and Buddelmeijer, H. and Burke, D. J. and Calderone, G. and Cano Rodríguez, J. L. and Cara, M. and Cardoso, J. V. M. and Cheedella, S. and Copin, Y. and Corrales, L. and Crichton, D. and D'Avella, D. and Deil, C. and Depagne, É. and Dietrich, J. P. and Donath, A. and Droettboom, M. and Earl, N. and Erben, T. and Fabbro, S. and Ferreira, L. A. and Finethy, T. and Fox, R. T. and Garrison, L. H. and Gibbons, S. L. J. and Goldstein, D. A. and Gommers, R. and Greco, J. P. and Greenfield, P. and Groener, A. M. and Grollier, F. and Hagen, A. and Hirst, P. and Homeier, D. and Horton, A. J. and Hosseinzadeh, G. and Hu, L. and Hunkeler, J. S. and Ivezić, Ž. and Jain, A. and Jenness, T. and Kanarek, G. and Kendrew, S. and Kern, N. S. and Kerzendorf, W. E. and Khvalko, A. and King, J. and Kirkby, D. and Kulkarni, A. M. and Kumar, A. and Lee, A. and Lenz, D. and Littlefair, S. P. and Ma, Z. and Macleod, D. M. and Mastropietro, M. and McCully, C. and Montagnac, S. and Morris, B. M. and Mueller, M. and Mumford, S. J. and Muna, D. and Murphy, N. A. and Nelson, S. and Nguyen, G. H. and Ninan, J. P. and Nöthe, M. and Ogaz, S. and Oh, S. and Parejko, J. K. and Parley, N. and Pascual, S. and Patil, R. and Patil, A. A. and Plunkett, A. L. and Prochaska, J. X. and Rastogi, T. and Reddy Janga, V. and Sabater, J. and Sakurikar, P. and Seifert, M. and Sherbert, L. E. and Sherwood-Taylor, H. and Shih, A. Y. and Sick, J. and Silbiger, M. T. and Singanamalla, S. and Singer, L. P. and Sladen, P. H. and Sooley, K. A. and Sornarajah, S. and Streicher, O. and Teuben, P. and Thomas, S. W. and Tremblay, G. R. and Turner, J. E. H. and Terrón, V. and van Kerkwijk, M. H. and de la Vega, A. and Watkins, L. L. and Weaver, B. A. and Whitmore, J. B. and Woillez, J. and Zabalza, V. and {Astropy Contributors}},
	year         = 2018,
	month        = sep,
	journal      = {The Astronomical Journal},
	publisher    = {IOP},
	volume       = 156,
	pages        = 123,
	doi          = {10.3847/1538-3881/aabc4f},
	issn         = {0004-6256},
	url          = {https://ui.adsabs.harvard.edu/abs/2018AJ....156..123A},
	urldate      = {2026-03-01},
	note         = {ADS Bibcode: 2018AJ....156..123A}
}

@article{astropy_collaboration_astropy_2022,
	title        = {The {Astropy} {Project}: {Sustaining} and {Growing} a {Community}-oriented {Open}-source {Project} and the {Latest} {Major} {Release} (v5.0) of the {Core} {Package}},
	shorttitle   = {The {Astropy} {Project}},
	author       = {{Astropy Collaboration} and Price-Whelan, Adrian M. and Lim, Pey Lian and Earl, Nicholas and Starkman, Nathaniel and Bradley, Larry and Shupe, David L. and Patil, Aarya A. and Corrales, Lia and Brasseur, C. E. and Nöthe, Maximilian and Donath, Axel and Tollerud, Erik and Morris, Brett M. and Ginsburg, Adam and Vaher, Eero and Weaver, Benjamin A. and Tocknell, James and Jamieson, William and van Kerkwijk, Marten H. and Robitaille, Thomas P. and Merry, Bruce and Bachetti, Matteo and Günther, H. Moritz and Aldcroft, Thomas L. and Alvarado-Montes, Jaime A. and Archibald, Anne M. and Bódi, Attila and Bapat, Shreyas and Barentsen, Geert and Bazán, Juanjo and Biswas, Manish and Boquien, Médéric and Burke, D. J. and Cara, Daria and Cara, Mihai and Conroy, Kyle E. and Conseil, Simon and Craig, Matthew W. and Cross, Robert M. and Cruz, Kelle L. and D'Eugenio, Francesco and Dencheva, Nadia and Devillepoix, Hadrien A. R. and Dietrich, Jörg P. and Eigenbrot, Arthur Davis and Erben, Thomas and Ferreira, Leonardo and Foreman-Mackey, Daniel and Fox, Ryan and Freij, Nabil and Garg, Suyog and Geda, Robel and Glattly, Lauren and Gondhalekar, Yash and Gordon, Karl D. and Grant, David and Greenfield, Perry and Groener, Austen M. and Guest, Steve and Gurovich, Sebastian and Handberg, Rasmus and Hart, Akeem and Hatfield-Dodds, Zac and Homeier, Derek and Hosseinzadeh, Griffin and Jenness, Tim and Jones, Craig K. and Joseph, Prajwel and Kalmbach, J. Bryce and Karamehmetoglu, Emir and Kałuszyński, Mikołaj and Kelley, Michael S. P. and Kern, Nicholas and Kerzendorf, Wolfgang E. and Koch, Eric W. and Kulumani, Shankar and Lee, Antony and Ly, Chun and Ma, Zhiyuan and MacBride, Conor and Maljaars, Jakob M. and Muna, Demitri and Murphy, N. A. and Norman, Henrik and O'Steen, Richard and Oman, Kyle A. and Pacifici, Camilla and Pascual, Sergio and Pascual-Granado, J. and Patil, Rohit R. and Perren, Gabriel I. and Pickering, Timothy E. and Rastogi, Tanuj and Roulston, Benjamin R. and Ryan, Daniel F. and Rykoff, Eli S. and Sabater, Jose and Sakurikar, Parikshit and Salgado, Jesús and Sanghi, Aniket and Saunders, Nicholas and Savchenko, Volodymyr and Schwardt, Ludwig and Seifert-Eckert, Michael and Shih, Albert Y. and Jain, Anany Shrey and Shukla, Gyanendra and Sick, Jonathan and Simpson, Chris and Singanamalla, Sudheesh and Singer, Leo P. and Singhal, Jaladh and Sinha, Manodeep and Sipőcz, Brigitta M. and Spitler, Lee R. and Stansby, David and Streicher, Ole and Šumak, Jani and Swinbank, John D. and Taranu, Dan S. and Tewary, Nikita and Tremblay, Grant R. and de Val-Borro, Miguel and Van Kooten, Samuel J. and Vasović, Zlatan and Verma, Shresth and de Miranda Cardoso, José Vinícius and Williams, Peter K. G. and Wilson, Tom J. and Winkel, Benjamin and Wood-Vasey, W. M. and Xue, Rui and Yoachim, Peter and Zhang, Chen and Zonca, Andrea and {Astropy Project Contributors}},
	year         = 2022,
	month        = aug,
	journal      = {The Astrophysical Journal},
	publisher    = {IOP},
	volume       = 935,
	pages        = 167,
	doi          = {10.3847/1538-4357/ac7c74},
	issn         = {0004-637X},
	url          = {https://ui.adsabs.harvard.edu/abs/2022ApJ...935..167A},
	urldate      = {2026-03-01},
	note         = {ADS Bibcode: 2022ApJ...935..167A}
}

@article{nettelmann_uranus_2016,
	title        = {Uranus evolution models with simple thermal boundary layers},
	author       = {Nettelmann, N. and Wang, K. and Fortney, J. J. and Hamel, S. and Yellamilli, S. and Bethkenhagen, M. and Redmer, R.},
	year         = 2016,
	month        = sep,
	journal      = {Icarus},
	volume       = 275,
	pages        = {107--116},
	doi          = {10.1016/j.icarus.2016.04.008},
	issn         = {0019-1035},
	url          = {https://www.sciencedirect.com/science/article/pii/S0019103516300537},
	urldate      = {2026-03-01}
}

@article{bethkenhagen_planetary_2017,
	title        = {Planetary {Ices} and the {Linear} {Mixing} {Approximation}},
	author       = {Bethkenhagen, M. and Meyer, E. R. and Hamel, S. and Nettelmann, N. and French, M. and Scheibe, L. and Ticknor, C. and Collins, L. A. and Kress, J. D. and Fortney, J. J. and Redmer, R.},
	year         = 2017,
	month        = oct,
	journal      = {The Astrophysical Journal},
	publisher    = {IOP},
	volume       = 848,
	pages        = 67,
	doi          = {10.3847/1538-4357/aa8b14},
	issn         = {0004-637X},
	url          = {https://ui.adsabs.harvard.edu/abs/2017ApJ...848...67B},
	urldate      = {2026-03-01},
	note         = {ADS Bibcode: 2017ApJ...848...67B}
}

@article{chabrier_new_2021,
	title        = {A {New} {Equation} of {State} for {Dense} {Hydrogen}-{Helium} {Mixtures}. {II}. {Taking} into {Account} {Hydrogen}-{Helium} {Interactions}},
	author       = {Chabrier, Gilles and Debras, Florian},
	year         = 2021,
	month        = aug,
	journal      = {The Astrophysical Journal},
	publisher    = {IOP},
	volume       = 917,
	pages        = 4,
	doi          = {10.3847/1538-4357/abfc48},
	issn         = {0004-637X},
	url          = {https://ui.adsabs.harvard.edu/abs/2021ApJ...917....4C},
	urldate      = {2026-02-27},
	note         = {ADS Bibcode: 2021ApJ...917....4C}
}

@article{valencia_internal_2006,
	title        = {Internal structure of massive terrestrial planets},
	author       = {Valencia, Diana and O'Connell, Richard J. and Sasselov, Dimitar},
	year         = 2006,
	month        = apr,
	journal      = {Icarus},
	publisher    = {Elsevier},
	volume       = 181,
	pages        = {545--554},
	doi          = {10.1016/j.icarus.2005.11.021},
	issn         = {0019-1035},
	url          = {https://ui.adsabs.harvard.edu/abs/2006Icar..181..545V},
	urldate      = {2026-02-27},
	note         = {ADS Bibcode: 2006Icar..181..545V}
}

@article{li_soot_2026,
	title        = {Soot {Planets} {Instead} of {Water} {Worlds}},
	author       = {Li, Jie and Bergin, Edwin A. and Hirschmann, Marc M. and Blake, Geoffrey A. and Ciesla, Fred J. and Kempton, Eliza M.-R.},
	year         = 2026,
	month        = jan,
	journal      = {The Astrophysical Journal},
	publisher    = {IOP},
	volume       = 997,
	pages        = {L29},
	doi          = {10.3847/2041-8213/ae29a6},
	issn         = {0004-637X},
	url          = {https://ui.adsabs.harvard.edu/abs/2026ApJ...997L..29L},
	urldate      = {2026-02-06},
	note         = {ADS Bibcode: 2026ApJ...997L..29L}
}

@misc{tejada_arevalo_sub-neptune_2025,
	title        = {Sub-{Neptune} {Memories} {I}: {Implications} of {Inefficient} {Mantle} {Cooling} and {Silicate} {Rain}},
	shorttitle   = {Sub-{Neptune} {Memories} {I}},
	author       = {Tejada Arevalo, Roberto and Gupta, Akash and Burrows, Adam and Zheng, Donghao and Tang, Yao and Deng, Jie},
	year         = 2025,
	month        = dec,
	publisher    = {arXiv},
	doi          = {10.48550/arXiv.2601.00059},
	url          = {https://ui.adsabs.harvard.edu/abs/2026arXiv260100059T},
	urldate      = {2026-01-12},
	note         = {ADS Bibcode: 2026arXiv260100059T}
}
\bibliographystyle{aasjournal}



\end{document}